\documentclass[12pt]{article}
\pdfoutput=1

\usepackage{cite}
\usepackage{amsmath,amsfonts,amssymb}
\usepackage{psfrag}
\usepackage{enumerate}
\usepackage{mathrsfs}
\usepackage{graphicx}
\usepackage{wrapfig}
\usepackage{xcolor}
\usepackage{caption}
\newcommand\blfootnote[1]{%
  \begingroup
  \renewcommand\thefootnote{}\footnote{\hspace{-6mm}#1}%
  \addtocounter{footnote}{-1}%
  \endgroup
}

\makeatletter\renewcommand\section{\@startsection {section}{1}{\z@}%
                                   {-3.5ex \@plus -1ex \@minus -.2ex}
                                   {2.3ex \@plus.2ex}%
                                   {\normalfont\large\bfseries}}
\renewcommand\subsection{\@startsection{subsection}{2}{\z@}%
                                     {-3.25ex\@plus -1ex \@minus -.2ex}%
                                     {1.5ex \@plus .2ex}%
                                     {\normalfont\bfseries}}

\newcommand{\be}{\begin{equation}}
\newcommand{\ee}{\end{equation}}
\newcommand{\beq}{\begin{eqnarray}}
\newcommand{\eeq}{\end{eqnarray}}

\def\[{\left [}
\def\]{\right ]}
\def\({\left (}
\def\){\right )}

\def\r2{\sqrt{2}}

\newcommand{\bbibitem}[1]{\bibitem{#1}\marginpar{#1}}

\newcommand{\myeq}[1]{\begin{equation} #1 \end{equation}}

\newtheorem{mydef}{Definition}

\def\Label#1{\label{#1}%
  \smash{\hbox to0pt{\raise1ex\hbox{\tiny[#1]}\hss}}}
\def\noLabels{\let\Label=\label}
\def\nobbibitem{\let\bbibitem=\bibitem}

\begin{document}
\noLabels 
\nobbibitem 

\clearpage\thispagestyle{empty}
\begin{center}
{\Large \bf  Nuts and Bolts for Creating Space}

\vspace{7mm}

Bart{\l}omiej Czech and Lampros Lamprou

\blfootnote{\tt czech, llamprou -AT- stanford -DOT- edu}


\bigskip\centerline{\it Stanford Institute for Theoretical Physics, Stanford University}
\smallskip\centerline{\it 382 Via Pueblo Mall, Stanford, CA 94305-4060, USA}
\end{center}

\vspace{5mm}

\begin{abstract}
\noindent
We discuss the way in which field theory quantities assemble the spatial geometry of three-dimensional anti-de Sitter space (AdS$_3$). The field theory ingredients are the entanglement entropies of boundary intervals. A point in AdS$_3$ corresponds to a collection of boundary intervals, which is selected by a variational principle we discuss. Coordinates in AdS$_3$ are integration constants of the resulting equation of motion. We propose a distance function for this collection of points, which obeys the triangle inequality as a consequence of the strong subadditivity of entropy. Our construction correctly reproduces the static slice of AdS$_3$ and the Ryu-Takayanagi relation between geodesics and entanglement entropies. We discuss how these results extend to quotients of AdS$_3$ -- the conical defect and the BTZ geometries. In these cases, the set of entanglement entropies must be supplemented by other field theory quantities, which can carry the information about lengths of non-minimal geodesics. 
\end{abstract}

\setcounter{footnote}{0}
\newpage
\clearpage
\setcounter{page}{1}

\tableofcontents

\section{Introduction}

The gravitational force is the most familiar force of Nature. It was the first force to become an object of scientific inquiry \cite{aristotle, galileo} and the earliest one to be understood quantitatively, at both medium \cite{newton} and large distance scales \cite{einstein}. Yet at small scales, it remains the only known interaction still shrouded in mystery. There are, of course, good reasons for this: since gravity is the dynamical theory of space and time, it is not a priori clear what it would mean to understand gravity at arbitrarily small distances. Indeed, the very notion of a point in space -- one that underlies our microscopic understanding of other physical interactions captured by the Standard Model -- may well be a semiclassical construct, which on the level of the fundamental theory gives way to other, less familiar objects.

The present paper collects some lessons, which can be drawn about the microscopic nature of space from the holographic duality \cite{maldacena97}. The holographic setting is convenient for several reasons. First, it translates gravitational problems into the language of gauge theory, which is in principle understood at all scales. In this way, holography grants access to and control over both the semiclassical spacetime and the microscopic degrees of freedom from which it is built. In particular, classical geometric objects on the gravity side -- minimal surfaces -- are conjecturally related to a set of fundamentally quantum quantities -- entanglement entropies of spatial regions in the field theory \cite{rt1, rt2}. This relation allows one to go beyond gravitational perturbation theory and to study the fundamental theory of gravity in a background independent way. 
Finally, in the AdS$_3$/CFT$_2$ correspondence the minimal surfaces that compute entanglement entropies are spacelike geodesics -- simple objects, which in many cases are known analytically. Although the converse is not true -- not all spatial geodesics compute boundary entanglement entropies \cite{strongsub, fernando, plateau} -- much is known about their interpretation in the dual field theory \cite{geodapprox, geodapprox2, excursions, entwinement}. We will use these facts to explain how the spatial slice of AdS$_3$ arises as a geometric description of the vacuum state of the dual two-dimensional conformal field theory (Secs.~\ref{basicobjects}-\ref{pureads}). We will further discuss how our reasoning extends to static quotients of AdS$_3$ -- the conical defect geometry (Sec.~\ref{condef}) and the non-rotating BTZ black holes (Sec.~\ref{btzthermal}).

Our starting point is the Ryu-Takayanagi (RT) proposal \cite{rt1, rt2}. Its profound consequences for the emergence of spacetime were first pointed out in \cite{briansessay, marksessay} (see also \cite{VanRaamsdonk:2010pw, Swingle:2012wq, bianchimyers, hartmanmaldacena, Myers:2013lva, holeent, erepr, xiaoliang, complexity}), which argued that reducing the entanglement between complementary half-spaces in the field theory is dual to lengthening a bridge in the spacetime until the latter pinches off into disconnected components. Extrapolating from this reasoning, it was conjectured that a holographic spacetime arises as a geometrization of the entanglement structure of a quantum state \cite{bianchimyers} (see also \cite{erepr}). The present paper relies on a quantitative version of this statement, which appeared in \cite{holeography} (see also \cite{roblast, Hubeny:2014qwa, xijamieandi, robproof, wien}). It generalized the Ryu-Takayanagi proposal from spacelike geodesics to arbitrary differentiable curves on a spatial slice of AdS$_3$. Instead of entanglement entropy, the length of a non-geodesic curve computes a novel, UV-finite combination of field theory entanglement entropies, called differential entropy:
\begin{equation}
\label{sdiff}
E[\alpha(\theta)] = \frac{1}{2} \int_0^{2\pi} d\theta\, \frac{dS(\alpha)}{d\alpha}\Big|_{\alpha = \alpha(\theta)} = \frac{\rm circumference}{4G} 
\end{equation}
The differential entropy is a functional of a family of boundary intervals determined by one function $\alpha(\theta)$, which implicitly picks out a differentiable curve on a spatial slice of AdS$_3$. We review the derivation of and useful facts about eq.~(\ref{sdiff}) in Sec.~\ref{rev}. Eq.~(\ref{sdiff}) will allow us to understand quantitatively how two-dimensional hyperbolic space -- the spatial slice of AdS$_3$ in static coordinates -- emerges as a geometric encoding of entanglement entropies in the boundary field theory.

What does it mean for a spatial manifold to emerge? We answer this question and state our main results in Sec.~\ref{basicobjects}. In essence, a geometry is a set of points and a distance function, which satisfies the triangle inequality. Traditionally, we relate spacetime points to boundary quantities using the famous HKLL construction \cite{hkll, hkll2, kll}, but that relation is limited to bulk low energy effective field theory. Here we wish to understand the emergence of space, so our goal is a non-perturbative definition of a bulk point, which assumes no prior knowledge of the metric. Sec.~\ref{basicobjects} states just that: a physically motivated, abstract definition of a bulk point in field theory. We then use eq.~(\ref{sdiff}) to define a distance function on our abstractly defined bulk points and verify that its obeys the triangle inequality.\footnote{Our {\it ab initio} construction of points and distances should be distinguished from \cite{recon1, recon2, recon3, recon4}, who use entanglement entropies / minimal surfaces to reconstruct numerically a metric assuming a certain ansatz.} 

In Sec.~\ref{pureads} we explain the physical and geometric reasons why our definition of points and distances correctly reproduces the static spatial slice of AdS$_3$. To do so, we detail several aspects of the differential entropy formula, which have not been previously discussed in the literature. The discussion is geometric and shows that bulk points are essentially limits of differentiable bulk curves shrunk to zero size. But the geometric statements have an intimate connection with information theory: for example, the strong subadditivity of entropy is an essential premise of the construction.

We discuss how our results extend to static quotients of AdS$_3$, the conical defect geometry and the non-rotating BTZ black holes, in Secs.~\ref{condef} and Sec.~\ref{btzthermal}. On the geometry side the construction is identical as in pure AdS$_3$, but on the field theory side it requires new ingredients. This is related to the fact that away from pure AdS$_3$ not all geodesics compute entanglement entropies. The results raise several interesting issues, such as the emergence of locality on sub-AdS scales, which we briefly discuss. Comparing how our construction is applied to pure AdS$_3$ and to its quotients highlights a salient fact: that the boundary definition of a spacetime point is state-dependent. For example, the boundary object, which in the field theory vacuum defines a bulk point in pure AdS$_3$, in the thermal state with $T \sim L_{\rm AdS}$ defines a curve, which wraps around the black hole horizon with a berth of order $L_{\rm AdS}$. In the final section we discuss the significance of our findings and how they might be extended to holographic spacetimes beyond AdS$_3$ and its quotients.


\section{Review of hole-ography}
\label{rev}
We work in the context of the AdS$_3$ / CFT$_2$ duality, though the results of \cite{holeography} extend to more general setups \cite{roblast, xijamieandi, robproof}. Here and in Secs.~\ref{basicobjects} and \ref{pureads} we concentrate on pure AdS$_3$, which corresponds to the vacuum of the boundary CFT. We represent pure AdS$_3$ in static coordinates:
\begin{equation}
ds^2 = - \left(1 + \frac{R^2}{L^2}\right) dT^2 + \left(1 + \frac{R^2}{L^2}\right)^{-1} dR^2 + R^2\, d\tilde\theta^2.
\label{ads3metric}
\end{equation} 
We will distinguish the bulk coordinate $\tilde\theta$ from the angular coordinate on the boundary, denoted with $\theta$.

We will make frequent use of spacelike geodesics. In coordinates (\ref{ads3metric}), a spacelike geodesic centered at $\theta$ takes the form:
\begin{equation}
\tan^2(\tilde\theta - \theta)= \frac{R^2 \tan^2\alpha - L^2}{R^2 + L^2} 
\label{adsgeodesic}
\end{equation}
The regulated length of this geodesic in Planck units is
\begin{equation}
S(\alpha) = \frac{L}{2G} \log \frac{2L \sin \alpha}{\mu}\,,
\label{entinterval}
\end{equation}
where $L^2/\mu$ is a gravitational infrared cutoff. According to the Ryu-Takayanagi proposal \cite{rt1, rt2}, this quantity also computes the entanglement entropy of an interval of length $2\alpha$ in the vacuum of the dual CFT. On the field theory side, $\mu$ is an ultraviolet cutoff and $L/2G = c/3$ \cite{brownhenneaux}, where $c$ is the central charge of the CFT.

\begin{figure}[t]
\centering
\includegraphics[width=.6 \textwidth]{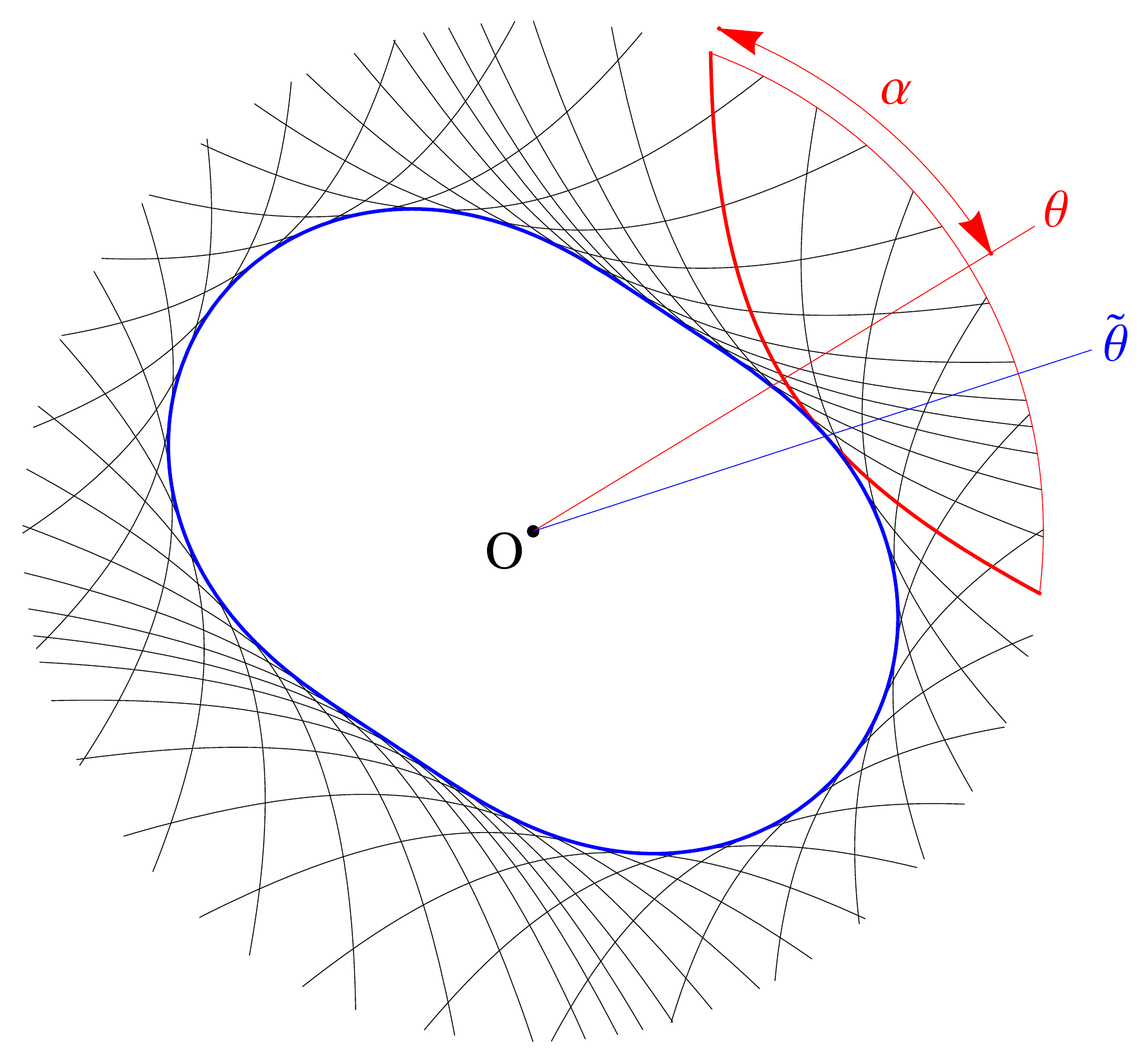}
\caption{The notation introduced in Sec.~\ref{rev}. We consider the set of spacelike geodesics tangent to a given curve $R = R(\tilde\theta)$, where $\tilde\theta$ is the bulk angular coordinate. The geodesics are centered at $\theta(\tilde\theta)$ and have width $\alpha(\tilde\theta)$. The variable $\theta$ is reserved for the boundary coordinate.}
\label{notation}
\end{figure}

\paragraph{Bulk curves} Consider a closed, smooth bulk curve $R = R(\tilde\theta)$ on the $T=0$ slice. For every point on the curve, there is a unique geodesic tangent to the curve at that point. This geodesic, which also lives on the $T=0$ slice, has both endpoints on the asymptotic boundary of AdS. We denote the angular coordinates of the endpoints with $\theta(\tilde\theta) \pm \alpha(\tilde\theta)$, with $\tilde\theta$ marking the tangency point on the bulk curve $R = R(\tilde\theta)$; see Fig.~\ref{notation}. In pure AdS$_3$ the parameters defining the geodesics are:
\begin{eqnarray}
\tan \alpha(\tilde\theta) & = & \frac{L}{R}\, \sqrt{1 + \frac{L^2}{R^2 + L^2} \left( \frac{d \log R}{d\tilde\theta}\right)^2} \label{tanalpha} \\
\tan\big(\tilde\theta - \theta(\tilde\theta)\big) & = & \frac{L^2}{R^2 + L^2}  \frac{d \log R}{d\tilde\theta} \label{tantheta}
\end{eqnarray}
Eqs.~(\ref{tanalpha}-\ref{tantheta}) can be inverted \cite{robproof, wien}. The inverse mapping, which reconstructs the bulk curve $R = R(\tilde\theta)$ from the boundary function $\alpha(\theta)$, is:
\begin{eqnarray}
R(\theta) & = & L \cot \alpha(\theta)\, \sqrt{\frac{1 + \alpha'(\theta)^2 \tan^2\! \alpha(\theta)}{1-\alpha'(\theta)^2}} \label{invr} \\
\tan\big(\theta - \tilde\theta(\theta)\big) & = & \alpha'(\theta) \tan\alpha(\theta)  
\label{invtheta}
\end{eqnarray}
Here and throughout this paper primes represent derivatives with respect to the boundary coordinate $\theta$:
\begin{equation}
\alpha'(\theta) = \frac{d\alpha}{d\theta} \qquad{\rm and}\qquad \alpha''(\theta) = \frac{d^2\alpha}{d\theta^2}
\end{equation}
Note that a boundary function $\alpha(\theta)$ defines a bulk curve only when: 
\begin{equation}
-1 < \alpha'(\theta) < 1.\label{consist}
\end{equation}

\paragraph{Differential entropy}
The main result of \cite{holeography} is that the circumference of the bulk curve $R = R(\tilde\theta)$ is computed on the field theory side by a novel quantity $E$ called differential entropy, which is a combination of (derivatives of) boundary entanglement entropies:
\begin{equation}
\label{sdiff0}
E[\alpha(\theta)] 
= \frac{1}{2} \int_0^{2\pi} d\theta\, \frac{dS(\alpha)}{d\alpha}\Big|_{\alpha = \alpha(\theta)} 
= \frac{\rm circumference}{4G} 
\end{equation}
Importantly, the measure of integration is uniform not in the bulk coordinate $\tilde\theta$, but in the boundary coordinate $\theta$.

This result is easily verified for a central circle $R = const.$, but in other cases its proof is nontrivial. Eq.~(\ref{sdiff0}) represents an equality of two closed integrals for all choices of contours. Thus, the difference between the integrands
\begin{equation*}
\frac{d{\rm (length)}}{4G} = \frac{d\tilde\theta}{4G}\, 
\sqrt{\left(1 + \frac{R^2}{L^2}\right)^{-1} \left(\frac{dR}{d\tilde\theta}\right)^2 + R^2}
\qquad {\rm and} \qquad 
\frac{d\theta}{2} \frac{dS(\alpha)}{d\alpha}\Big|_{\alpha = \alpha(\theta)}
\end{equation*}
must be an exact form $df$. The correct choice of $f$ is the length of the geodesic contained in the angular wedge between $\theta(\tilde\theta)$ and $\tilde\theta$, which in pure AdS$_3$ takes the form:
\begin{equation}
f(\tilde\theta) = \frac{L}{8G} \log\frac{\sin \big( \alpha(\tilde\theta) + (\tilde\theta - \theta(\tilde\theta))\big)}{\sin \big( \alpha(\tilde\theta) - (\tilde\theta - \theta(\tilde\theta))\big)}
\label{deff}
\end{equation}
We will interpret $f(\tilde\theta)$ in a different way in Sec.~\ref{endpt}. 
The total derivative term vanishes when $dR/d\tilde\theta = 0$, which is why for a circle the bulk formula~(\ref{sdiff0}) can be confirmed directly by inspection. An immediate corollary is a relation between the differential entropy and lengths of open curves in the bulk:
\begin{equation}
\frac{\rm length}{4G} = \frac{1}{4G} \int_{\tilde\theta_i}^{\tilde\theta_f} d\tilde\theta\, 
\sqrt{\left(1 + \frac{R^2}{L^2}\right)^{-1} \left(\frac{dR}{d\tilde\theta}\right)^2 + R^2} =
\frac{1}{2} \int_{\theta_i}^{\theta_f} d\theta\, \frac{dS(\alpha)}{d\alpha}\Big|_{\alpha = \alpha(\theta)} + f(\tilde\theta_f) - f(\tilde\theta_i)
\label{open}
\end{equation}
Eq.~(\ref{open}) computes the length of an open differentiable curve that stretches in the bulk angular wedge $\tilde\theta_i \leq \tilde\theta \leq \tilde\theta_f$. The boundary integral extends from $\theta_i$ to $\theta_f$ -- the midpoints of the intervals subtended by those geodesics, which are tangent to the curve at its endpoints.

\begin{figure}[t!]
\centering
\begin{tabular}{ccc}
\includegraphics[width=.31\textwidth]{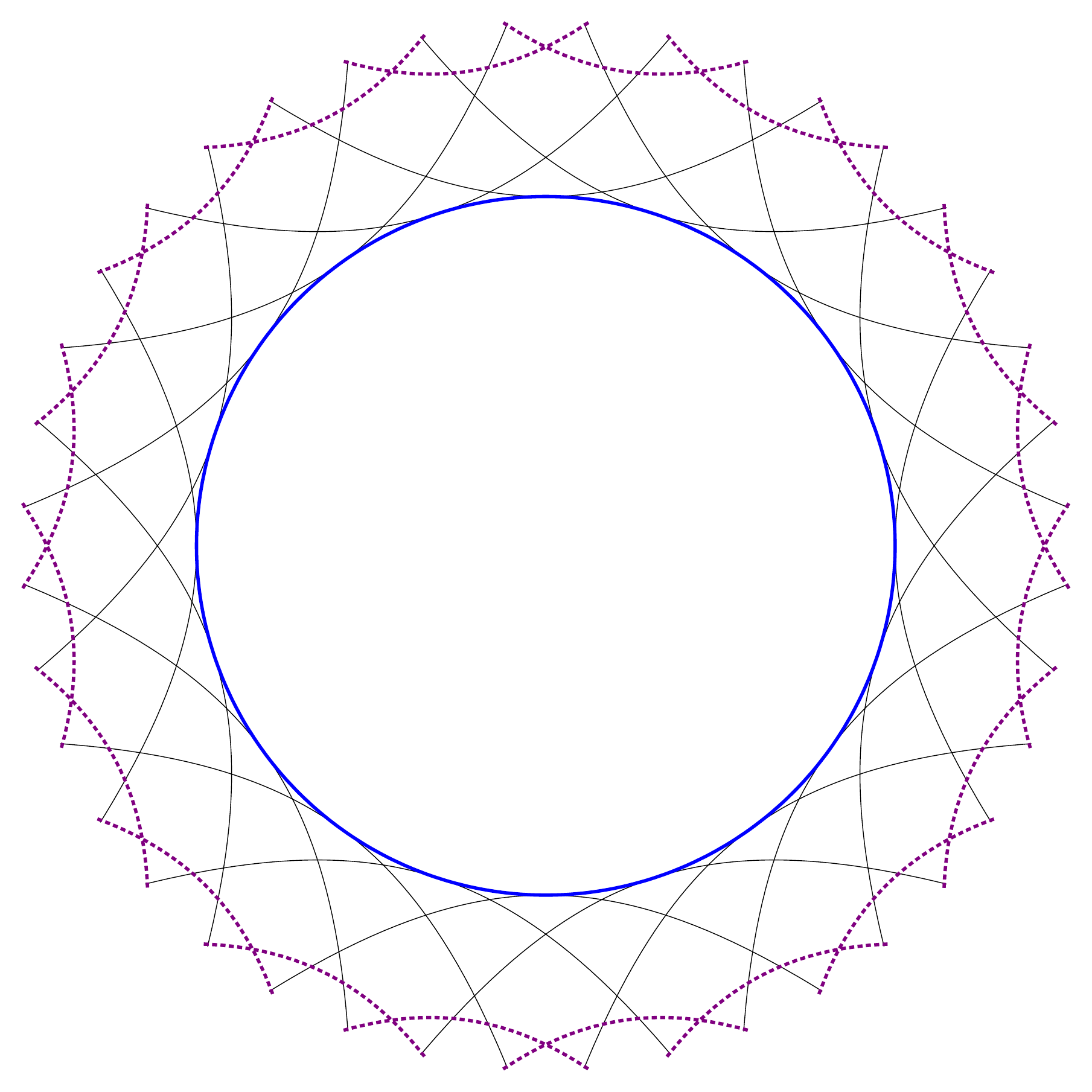} & 
\includegraphics[width=.31\textwidth]{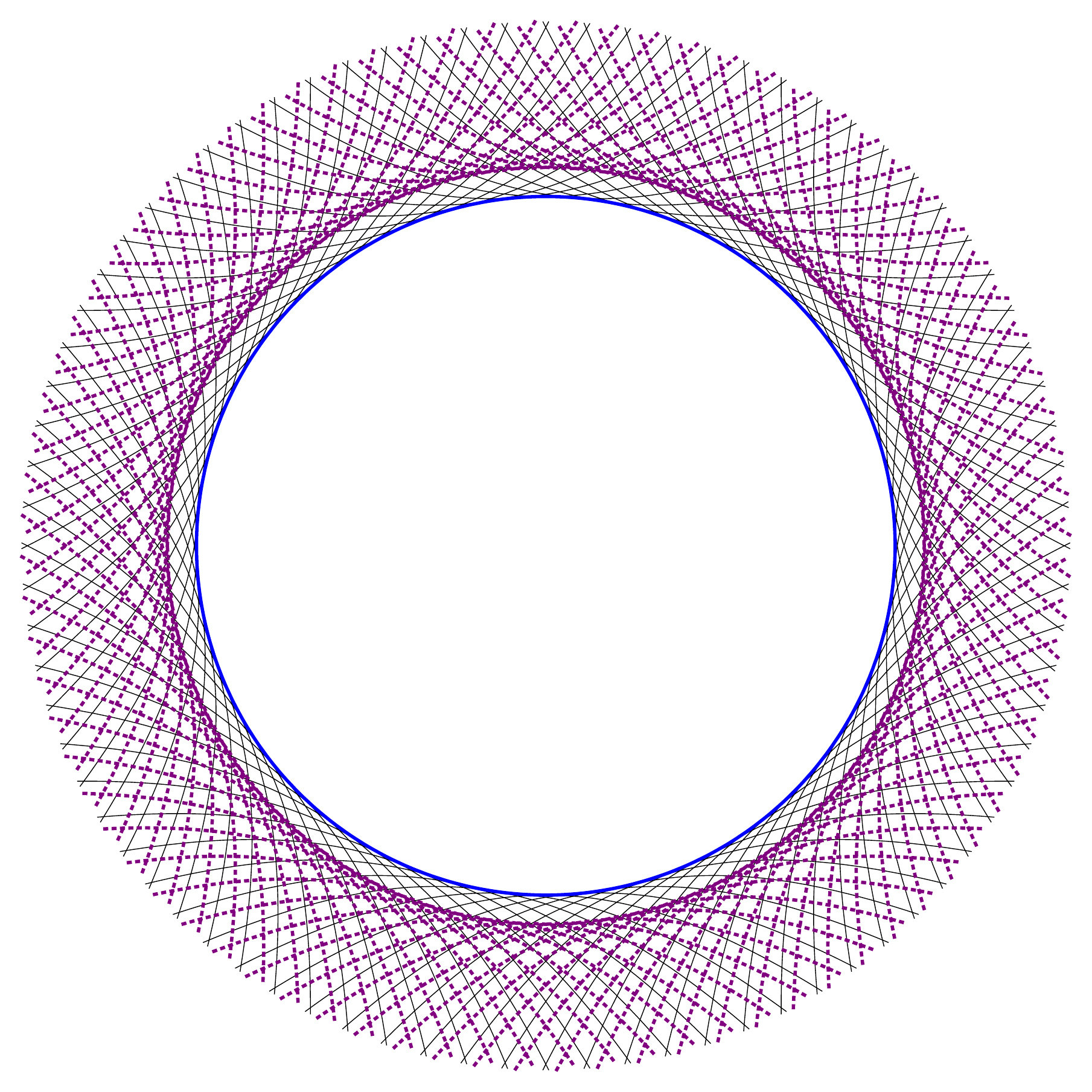} & 
\includegraphics[width=.31\textwidth]{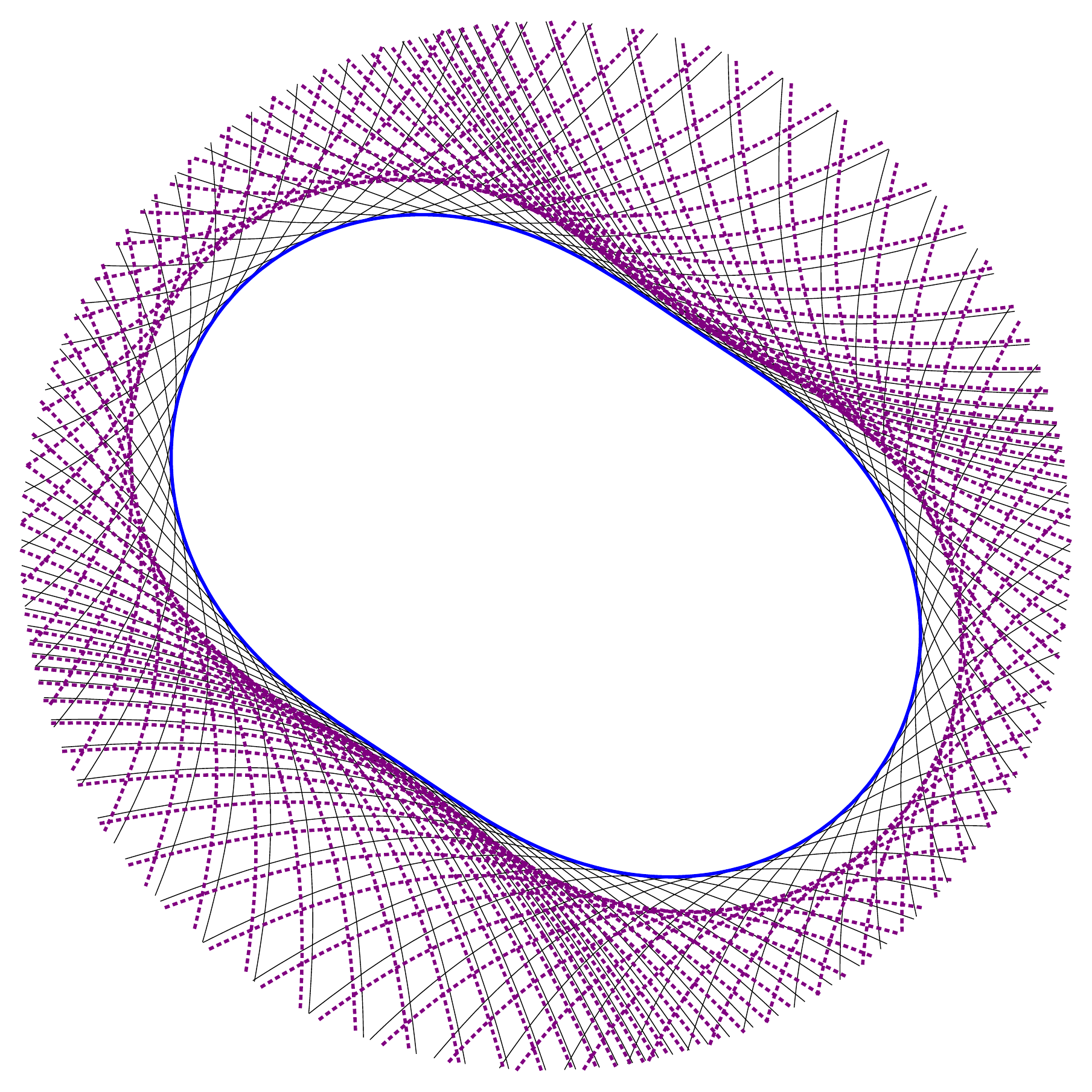}
\end{tabular}
\caption{A graphical representation of eq.~(\ref{differentials}). The black, continuous geodesics are tangent to the bulk curve; their lengths are $S(I_k)$. The purple, dotted geodesics subtend $I_k \cap I_{k+1}$. We illustrate the effect of the limit $N \to \infty$ by displaying finite combinations of geodesics at $N = 20$ and $N = 80$ for a circle. We also display the $N = 80$ finite sum for the curve shown in Fig.~\ref{notation}.}
\label{differences}
\end{figure}

\paragraph{A graphical explanation} Formula~(\ref{sdiff0}) can be motivated with the following heuristic argument. Rewrite the differential entropy by adding yet another total derivative term
\begin{equation}
E[\alpha(\theta)] = \frac{1}{2} \int_0^{2\pi} d\theta \left(1 - \frac{d\alpha}{d\theta}\right)
\frac{dS(\alpha)}{d\alpha}\Big|_{\alpha = \alpha(\theta)}
\end{equation}
and ``discretize'' the resulting integrand by Taylor expanding to first order in $d\theta$:
\begin{equation}
\frac{d\theta}{2} \left(1 - \frac{d\alpha}{d\theta}\right)
\frac{dS(\alpha)}{d\alpha}\Big|_{\alpha = \alpha(\theta)}
\approx S(\alpha(\theta)) - S\big(\alpha(\theta) + d\alpha/2 - d\theta/2\big).
\label{discretized}
\end{equation}
Let $I_\theta$ be the boundary interval subtended by the geodesic that is tangent to the bulk curve and centered at $\theta$. Explicitly, we have:
\begin{eqnarray}
I_\theta & = & \big(\theta - \alpha(\theta), \theta + \alpha(\theta)\big) \\
I_{\theta+d\theta} & = & \big(\theta + d\theta - \alpha(\theta) - d\alpha, \theta + d\theta + \alpha(\theta) + d\alpha \big)
\end{eqnarray}
Thus, eq.~(\ref{discretized}) computes the difference between the entanglement entropies of $I_\theta$ and of the overlap interval $I_\theta \cap I_{\theta+d\theta}$. In this way, the differential entropy can be expressed as
\begin{equation}
E = \lim_{N \to \infty} \sum_{k=1}^N S(I_k) - S(I_k \cap I_{k+1})
\label{differentials}
\end{equation}
for a family of $N$ intervals with uniformly distributed centers, each defining a geodesic that is tangent to the target curve in the bulk.\footnote{In this form, the argument applies only when the bulk curve is convex. See \cite{roblast} and Sec.~\ref{nonconvex} for a more general discussion.} A graphical representation of eq.~(\ref{differentials}) is shown in Fig.~\ref{differences}. 

A corollary of this graphical argument is that the bulk curve localizes on intersections of infinitesimally separated geodesics. This observation, which was emphasized in \cite{binchen}, gives the simplest method to obtain eqs.~(\ref{invr}-\ref{invtheta}). It also explains the necessary condition (\ref{consist}) for a boundary function $\alpha(\theta)$ to describe a bulk curve: when (\ref{consist}) is not satisfied, infinitesimally separated geodesics do not intersect \cite{roblast}.


\section{Points and distances}
\label{basicobjects}
We now explain how in AdS$_3$ the most basic geometric objects -- points and geodesics connecting them -- emerge from field theory data. The input from field theory is the complete set of entanglement entropies of spatial regions. We show how to use this data to construct a Euclidean geometry. The resulting space will by construction obey the relations reviewed in Sec.~\ref{rev}, including the Ryu-Takayanagi relation (\ref{entinterval}). In other words, the construction reproduces the spatial slice of AdS$_3$ in static coordinates.

In this section we emphasize the conceptual basis of our construction. A more technical discussion, including proofs of some results, is relegated to Sec.~\ref{pureads}.



\subsection{Bulk Points}
\label{secpoints}
What is a bulk point? We present three answers to this question that are supplied by hole-ography, proceeding in order of increasing abstraction and robustness. The first two answers are motivational; we include them for pedagogical purposes. The final answer is a \emph{constructive} definition of a bulk point -- a definition stated entirely in the boundary theory.

\subsubsection{Points as limits of curves}
\label{ptlimitcurves}
Consider a circle of radius $r$ centered at an arbitrary point on the $T=0$ slice of AdS$_3$. We can compute its corresponding boundary function $\alpha(\theta)$ using eqs.~(\ref{tanalpha}-\ref{tantheta}). From the bulk point of view it is then straightforward to take the limit $r\rightarrow 0$ and shrink the circle to a point -- a curve of zero length located at the center of the circle. By taking the same limit for the boundary function $\alpha(\theta)$ we obtain a function, which depends parametrically only on the bulk coordinates of the limiting point. It is a simple exercise to verify that for an arbitrary bulk point $A$ with coordinates $R$ and $\tilde\theta$ we obtain:
\begin{equation}
\label{point}
\alpha_A(\theta)= \cos^{-1} \frac{R \cos (\theta-\tilde{\theta})}{\sqrt{L^2 + R^2}}
\end{equation}
Inserting $\alpha_A(\theta)$ into the differential entropy formula (\ref{sdiff}) we get identically zero. This is consistent with what happens to the length of a bulk circle as it is shrunk to zero size. 


In light of this discussion one may be tempted to identify the set of bulk points with the set of zero differential entropy functions $\alpha(\theta)$ on the boundary. The rational would be to select curves of zero circumference, that is -- points. We will explain in Sec.~\ref{intkink} that this would be incorrect, because eq.~(\ref{sdiff}) computes \emph{signed} lengths of \emph{oriented} curves (see also \cite{robproof}). Consequently, there are infinitely many curves of finite length but changing orientation, whose differential entropy evaluates to zero. This fact means that a hole-ographic definition of a bulk point must be more subtle.

\subsubsection{Points from intersections of geodesics}
\label{intersectpoint}
The function $\alpha_A(\theta)$ in eq.~(\ref{point}) selects geodesics, which pass though the bulk point $A$. We may attempt to turn this observation into a definition of a bulk point. To do so, we would need to formulate a condition that two distinct geodesics intersect at a given point. 
It is sufficient to state this condition for two infinitesimally separated geodesics (centered at $\theta$ and $\theta + d\theta$) and then impose it over the full range of $\theta$. 

Two infinitesimally separated geodesics intersect at the bulk angle $\tilde\theta$ if they satisfy
\begin{equation} 
\frac{d R(\tilde{\theta};\, \alpha(\theta),\theta)}{d \theta}=0\,,
\end{equation}
where $R(\tilde\theta;  \alpha,\theta)$ is given by eq.~(\ref{adsgeodesic}). We solve this equation with respect to $\tilde{\theta}$ and then demand that the intersection point remain constant as a function of $\theta$. The latter condition is a second order differential equation, which selects $\alpha_A(\theta)$, the boundary functions corresponding to bulk points:
\myeq{ \label{pointdiffeq} 
\frac{d\tilde{\theta}(\alpha_A(\theta),\theta)}{d\theta}=0\qquad\Rightarrow\qquad 
-\sin \alpha_A(\theta) \alpha_A''(\theta)+\cos \alpha_A(\theta) \big(1-\alpha_A'(\theta)^2\big) = 0}
After a redefinition $\cos{\alpha(\theta)}=g(\theta)$ eq.~(\ref{pointdiffeq}) becomes a harmonic oscillator differential equation for $g(\theta)$. The solutions are precisely the functions (\ref{point}) derived in the previous section. It is not surprising that eq.~(\ref{pointdiffeq}) is second order: the integration constants are the coordinates $R$ and $\tilde\theta$ of the bulk point, which lives in a two-dimensional spatial slice of AdS$_3$. 

Eq.~(\ref{pointdiffeq}) uniquely picks the boundary projections of bulk points. It was derived using the explicit form of AdS$_3$ geodesics, but it is a useful guide toward a more general boundary definition of a bulk point. Consider the reciprocal of the left hand side of eq.~(\ref{pointdiffeq}): $d\theta/d\tilde{\theta}$ tracks the evolution of the boundary image of a point traveling along the curve, which is parameterized using the bulk angular coordinate $\tilde\theta$. This means that its sign agrees with the sign of the extrinsic curvature of the curve. In this way, eq.~(\ref{pointdiffeq}) diagnoses the orientation of the curve: when it is positive the curve is convex and when it is negative it is concave. Eq.~(\ref{pointdiffeq}) tells us that a point is a ``curve'' which is neither convex nor concave: its orientation is undefined. In the next subsection we will see that this last statement can be cast directly in the language of the field theory.

\subsubsection{Points as extrema of extrinsic curvature}
\label{pointsextrK}
In a negatively curved space the Gauss-Bonnet theorem states that:
\begin{equation}
\oint_C d\tau \,\sqrt{h} \,K = 2\pi - \int_A R\, dA \geq 2\pi. \label{gbineq}
\end{equation}
On the left hand side $d\tau \sqrt{h}$ is the length element along a curve and $K$ is its extrinsic curvature; the integral is taken over the length of the curve. In the middle expression, the Ricci scalar is integrated over the area enclosed by the curve. As a consequence, if a differentiable curve in a negatively curve space is shrunk to a point, the integral in (\ref{gbineq}) approaches $2\pi$. This means that if we extremize the left hand side of (\ref{gbineq}) over the set of all curves, we will have found points -- ``curves,'' which circle around a point infinitesimally.

In Appendix~\ref{extrinsic} we have calculated the extrinsic curvature of a bulk curve $R = R(\tilde\theta)$. Using eqs.~(\ref{tanalpha}-\ref{tantheta}) the result can be re-expressed as:\footnote{We remind the reader that in our notation $\alpha'(\theta) = d\alpha/d\theta$.}
\begin{equation}
d\tau\, \sqrt{h}\, K = \frac{d\theta \, \sqrt{1 - \alpha'(\theta)^2}}{\sin\alpha(\theta)}
\label{lagrangian}
\end{equation}
The numerator of eq.~(\ref{lagrangian}) can be rationalized in a simple way: it is the unique reparameterization invariant expression that vanishes when $\alpha(\theta)$ is proportional to $\pm \theta$ -- which, as we shall see in Sec.~\ref{diamonds}, is the case for $K=0$ curves, i.e. geodesics. We comment on the meaning of the denominator below.

If points extremize the integral in eq.~(\ref{gbineq}), we can treat eq.~(\ref{lagrangian}) as a Lagrangian. The extrema of the resulting action will be bulk points. Indeed, the Euler-Lagrange equation of the action
\begin{equation}
I = \int_0^{2\pi} d\theta\, \frac{\sqrt{1 - \alpha'(\theta)^2}}{\sin\alpha(\theta)}
\label{action0}
\end{equation}
is precisely eq.~(\ref{pointdiffeq}) and its complete set of solutions are the $\alpha_A(\theta)$ given in eq.~(\ref{point}). 

Eq.~(\ref{action0}) and its generalization (\ref{action}) below are the main results of our paper. Action~(\ref{action0}) defines bulk points purely in terms of the boundary theory: it can be stated and extremized without reference to the bulk, even though we arrived at it by studying curvatures in the bulk using eqs.~(\ref{tanalpha}-\ref{tantheta}).

\subsubsection{Toward a background-independent definition}
\label{pointsent}
Action (\ref{action0}) correctly picks out points in pure AdS$_3$, but it does not work in other asymptotically AdS$_3$ spacetimes such as the conical defect or the BTZ black hole. In particular, the denominator of eq.~(\ref{lagrangian}) was specifically chosen to reproduce the extrinsic curvature of a bulk curve in AdS$_3$ and not in other asymptotically AdS$_3$ spacetimes. We can fix this problem, at least for the conical defect and BTZ geometry, by observing that:
\begin{equation}
\frac{1}{\sin^2\alpha} = - \frac{2G}{L} \frac{d^2 S(\alpha)}{d\alpha^2}
\end{equation}
Substituting this in eq.~(\ref{action0}) yields our final boundary definition of a bulk point: it is a ``curve,'' which extremizes the action:
\begin{equation}
I = \int d\theta\, \sqrt{-\frac{d^2S}{d\alpha^2}\big(1 - \alpha'(\theta)^2\big)}
\label{action}
\end{equation}
Note that so long as the consistency condition (\ref{consist}) is satisfied, the expression under the square root is positive as a consequence of the strong subadditivity of entropy. The Euler-Lagrange equation reads:
\begin{equation}
(2\alpha_A'') \,\frac{d^2S}{d\alpha^2}\Big|_{\alpha_A} 
+ \big(1-\alpha_A'^2\big)\,\frac{d^3S}{d\alpha^3}\Big|_{\alpha_A} = 0
\label{eom}
\end{equation}
It reduces to eq.~(\ref{point}) in the case of pure AdS$_3$. We shall see in Secs.~\ref{condef} and \ref{btzthermal} that eq.~(\ref{eom}) correctly selects points in the bulk of the conical defect and BTZ spacetime.

\paragraph{Action (\ref{action}) as a length in an auxiliary space} Eq.~(\ref{action}) has a DBI form: it computes the length of a curve in an auxiliary Lorentzian space coordinatized by $\theta$ and $\alpha$ with metric
\begin{equation}
d\sigma^2 = - \frac{d^2S}{d\alpha^2}\, (d\theta^2 - d\alpha^2)\,.
\label{auxmetric}
\end{equation}
This auxiliary space is well known in integral geometry:\footnote{We would like to thank Michael Freedman for explaining to us the connection with integral geometry at the Aspen Center for Physics.}
its volume element is called the kinematic measure. Points in the kinematic space are geodesics on a spatial slice of the bulk. Eq.~(\ref{eom}) tells us that the ``inverse'' statement also holds: geodesics in the kinematic space are points in the bulk. Metric~(\ref{auxmetric}) is special, because it is the unique metric (up to overall scale) on the space of geodesics, which is invariant under rotations and boosts in the bulk. In the case of AdS$_3$, the kinematic space is de Sitter space. We comment briefly on the relevance of integral geometry to holography in the Discussion.

\subsection{Bulk Distances}
\label{secdistance}
\subsubsection{Definition of geodesic distance}
Given two points $A$ and $B$ with associated boundary functions $\alpha_A(\theta)$ and $\alpha_B(\theta)$, we give the following boundary definition of the geodesic distance between them:
\begin{mydef}
\label{defdistance}
Let $\gamma(\theta)= \min \{ \alpha_A(\theta), \alpha_B(\theta) \}$ for all $\theta$. The distance between $A$ and $B$ (measured in units of $4G$) is:
\begin{equation}
\label{defdist}
d(A,B)= \frac{1}{2}\, E[\gamma(\theta)]
\end{equation}
\end{mydef}
To explain why this formula is correct, we will need several additional results, which we present in Sec.~\ref{pureads}. In a nutshell, given two closed, convex bulk curves with boundary functions $\alpha(\theta)$ and $\beta(\theta)$, their pointwise minimum $\gamma(\theta) \equiv \min \{ \alpha(\theta), \beta(\theta) \}$ is the boundary image of their joint convex cover in the bulk. This result is developed in Sec.~\ref{ptmin} and illustrated in Fig.~\ref{convexcover} therein. When the two bulk ``curves'' are points $A$ and $B$, their convex cover is the ``closed curve,'' which runs from $A$ to $B$ along a geodesic and then returns from $B$ to $A$. The circumference of the cover, which is double the distance between $A$ and $B$, is then given by $E[\gamma(\theta)]$.

The simplest example of eq.~(\ref{defdist}) is the distance of any point on the $T=0$ slice of AdS$_3$ from the origin. The origin has $\alpha_O = \pi/2$ and its $dS/d\alpha$ vanishes everywhere. For a point $A$ located on $\tilde\theta = 0$, $\alpha_A(\theta)$ in eq.~(\ref{point}) is greater than $\pi/2$ outside the interval $-\pi/2 \leq \theta \leq \pi/2$. Thus, formula~(\ref{defdist}) and eq.~(\ref{point}) give $d(O,A)$ equal to:
\begin{equation}
\frac{1}{4} \int_{-\pi/2}^{\pi/2} d\theta\, \frac{dS(\alpha)}{d\alpha}\Big|_{\alpha = \alpha_A(\theta)} =
\frac{L}{8G} \int_{-\pi/2}^{\pi/2} d\theta\, \frac{R \cos\theta}{\sqrt{L^2 + R^2 \sin^2 \theta}} 
= \frac{1}{4G} \int_0^R dR \left(1 + \frac{R^2}{L^2}\right)^{-1/2}
\end{equation}
We will present another example of formula~(\ref{defdist}) at work in Sec.~\ref{diamonds}.

\subsubsection{Properties of distance}
\label{secproperties}
\begin{figure}[t!]
\centering
\begin{tabular}{ccc}
\includegraphics[width=.31\textwidth]{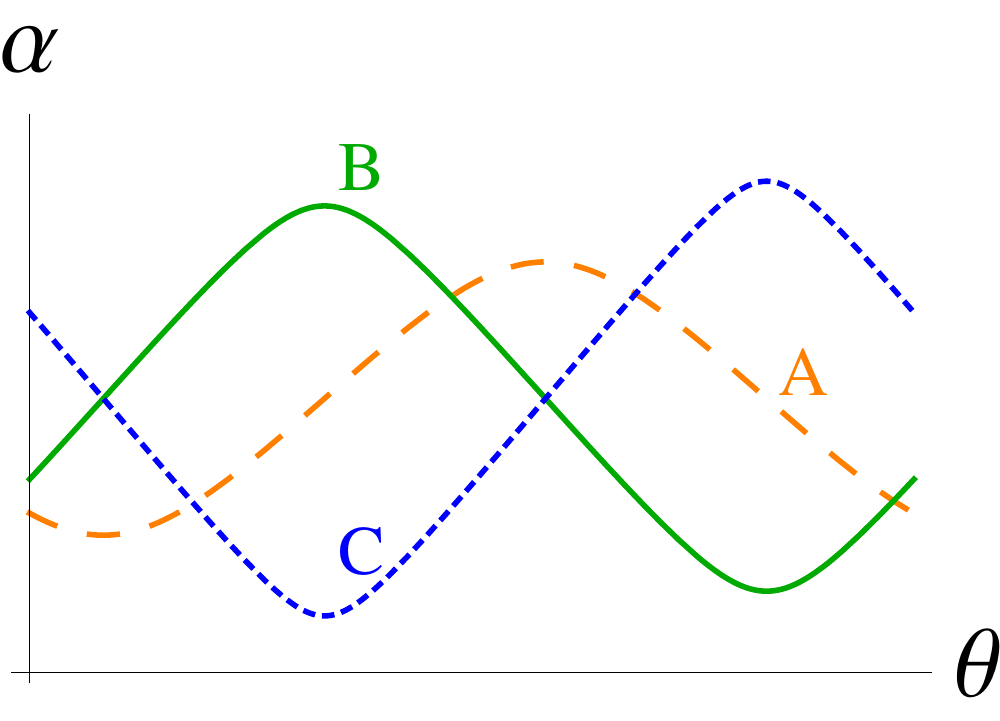} & 
\includegraphics[width=.31\textwidth]{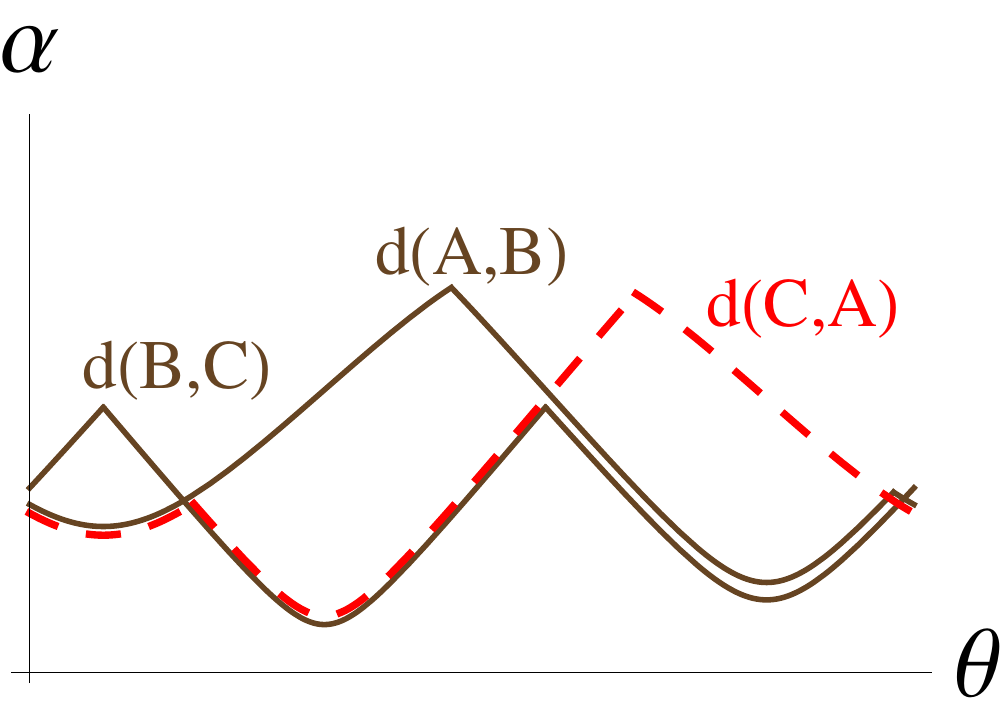} &
\includegraphics[width=.31\textwidth]{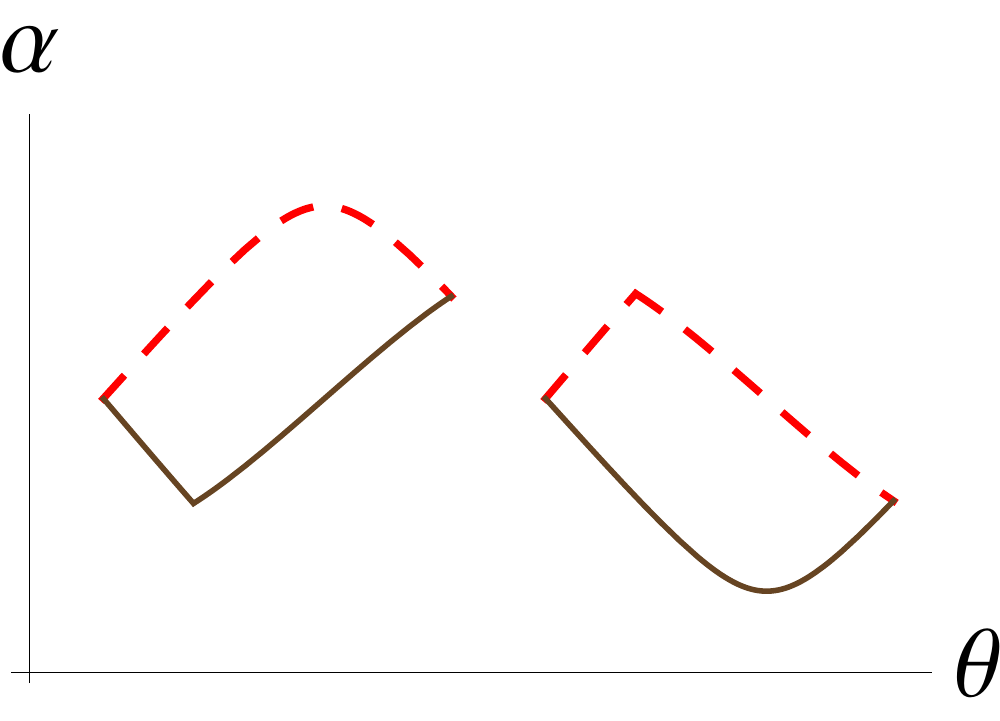}
\end{tabular}
\caption{The differential entropy proof of the triangle inequality (\ref{triineq}). Terms on the left hand side are drawn in continuous brown while the right hand side is in dashed red. See text below for more explanation.}
\label{triangle}
\end{figure}

Let us confirm that Definition~\ref{defdistance} obeys the axioms of a distance function. We remarked below eq.~(\ref{point}) that $E[\alpha_A(\theta)] = 0$, so $d(A,A) = 0$. Reflexivity $d(A,B) = d(B,A)$ follows directly from the definition. To confirm positivity $d(A,B)\geq 0$, write
\begin{equation}
d(A,B) = d(A,B) - d(A,A) = 
\frac{1}{4} \int_0^{2\pi} d\theta\, \frac{dS(\alpha)}{d\alpha}\Big|_{\alpha = \min \{ \alpha_A(\theta), \alpha_B(\theta) \}} - 
\frac{1}{4} \int_0^{2\pi} d\theta\, \frac{dS(\alpha)}{d\alpha}\Big|_{\alpha = \alpha_A(\theta)}
\end{equation}
and note that the integrand in the first term is greater than or equal to the integrand in the second term, because $d^2 S / d\alpha^2 < 0$. It remains to prove:

\paragraph{The triangle inequality.} We wish to verify that Definition~\ref{defdistance} satisfies:
\begin{equation}
d(A,B) + d(B, C) \geq d(A, C).
\label{triineq}
\end{equation}
We have illustrated the computation for a generic triple of points in Fig.~\ref{triangle}. The functions $\alpha_A(\theta), \alpha_B(\theta), \alpha_C(\theta)$ that define the bulk points are shown in the left panel. In the middle panel, we present the boundary functions that compute the pairwise distances between $A, B$ and $C$ according to eq.~(\ref{defdist}). The terms on the left hand side of~(\ref{triineq}) are marked in continuous brown while the right hand side is drawn in dashed red. In many places the dashed red overlaps with a continuous brown line and we have a direct cancellation. However, there is one interval over which we have two identical contributions from the left hand side: it is the interval where $\alpha_B(\theta)$ is smaller than $\alpha_A(\theta)$ and $\alpha_C(\theta)$. Using the fact that $d(B,B) = E[\alpha_B(\theta)] = 0$ we can trade one of these two contributions for the differential entropy of $\alpha_B(\theta)$ over the complementary interval, dressed with an extra minus sign. This leads to further cancellations, with the final result displayed in the right panel of Fig.~\ref{triangle}.

After the cancellations, both sides of the inequality are differential entropies computed over the same two intervals, but for different boundary functions:
\begin{equation}
\frac{1}{4} \int d\theta\, \frac{dS(\alpha)}{d\alpha}\Bigg|_{\alpha = LHS(\theta)} 
>\,\,\,
\frac{1}{4} \int d\theta\, \frac{dS(\alpha)}{d\alpha}\Bigg|_{\alpha = RHS(\theta)}
\end{equation}
Because the boundary function on the left hand side is strictly smaller than the boundary function the right hand side, the result again follows from the strong subadditivity of entropy: $d^2S/d\alpha^2 < 0$.

The proof is considerably simpler if we exploit conformal symmetry to place $B$ in the center of AdS$_3$. We chose not to do this and presented instead a proof, which relies only on two robust facts: the zero differential entropy of the point functions $d(B,B) = 0$ and the concavity of entropy $d^2S/d\alpha^2<0$. In this form, the proof carries over to other holographic spacetimes discussed in the present paper -- the conical defect geometry (Sec.~\ref{condef}) and the static BTZ black hole (Sec.~\ref{btzthermal}).


\section{Hole-ography of curves, points and distances}
\label{pureads}

In this section we discuss aspects of hole-ography, which lie at the root of the material in Sec.~\ref{basicobjects}. The first three subsections explain our definition of geodesic distance given in eq.~(\ref{defdist}). The results also make it convenient to discuss nonconvex curves, which we do in Sec.~\ref{nonconvex}. The last two subsections are relevant to understanding our definition of a point as an extremum of action~(\ref{action0}).

\subsection{Recovering the Ryu-Takayanagi formula}
\label{diamonds}
\begin{figure}[t]
\centering
\includegraphics[width=.4 \textwidth]{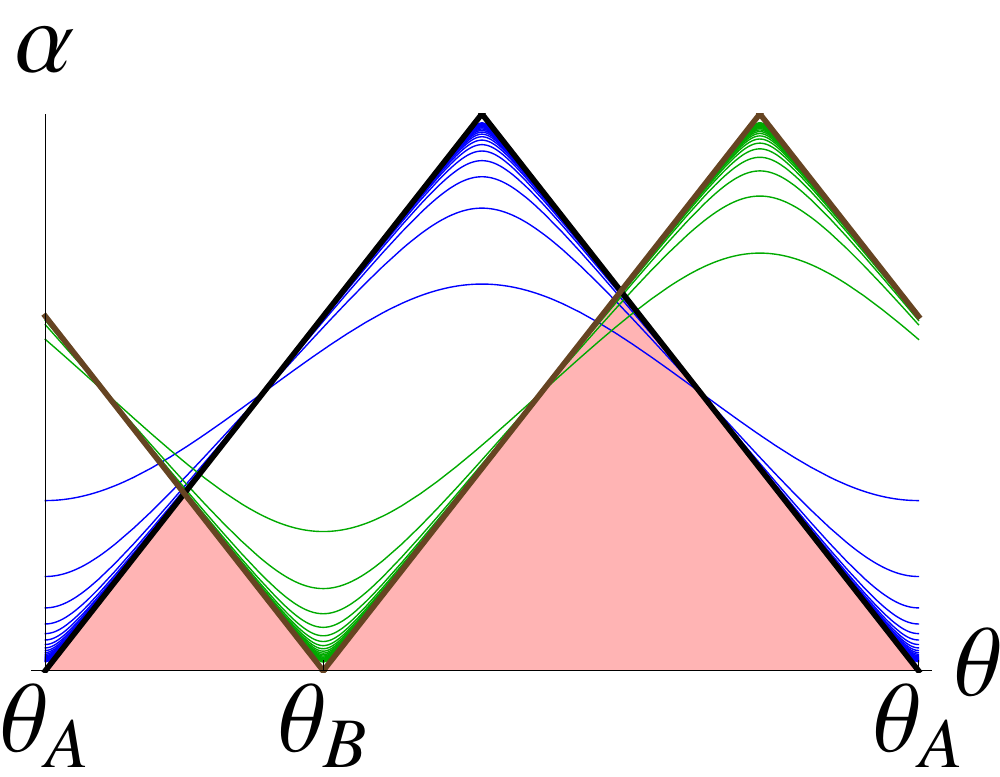}
\caption{The functions $\alpha_{A,B}(\theta)$ of two points $A$ and $B$, which approach the boundary at fixed angular coordinates $\tilde\theta = \theta_{A,B}$. The pointwise minimum $\gamma(\theta) = \min\{\alpha_A(\theta), \alpha_B(\theta)\}$ approaches the causal diamond (shaded) of the interval $(\theta_A, \theta_B)$ and of its complement.}
\label{figdiamonds}
\end{figure}

Another consistency check on the distance function~(\ref{defdist}) is that it correctly reproduces the lengths of geodesics anchored at the boundary. Consider two points $A$ and $B$ at some fixed angular locations $\tilde\theta = \theta_{A,B}$ and at equal radial coordinate $R_A = R_B$. Sending it to infinity, the distance $d(A, B)$ should approach the entanglement entropy (\ref{entinterval}) for the interval $(\theta_A, \theta_B)$ on the boundary. To corroborate this, examine the behavior of $\alpha_A(\theta)$ as $R_A$ approaches infinity:
\begin{equation}
\lim_{R_A \to \infty} \alpha(\theta) = \lim_{R_A \to \infty} \cos^{-1} \frac{R_A \cos(\theta-\theta_A)}{\sqrt{L^2 + R_A^2}} = |\theta - \theta_A|
\label{boundarylimit}
\end{equation}
This limit is illustrated in Fig.~\ref{figdiamonds}. Recall that our definition (\ref{defdist}) of geodesic distance involves the pointwise minimum of $\alpha_A(\theta)$ and $\alpha_B(\theta)$. In the limit (\ref{boundarylimit}) this pointwise minimum becomes the boundary causal diamond of the interval $(\theta_A, \theta_B)$ and of its complement. To make contact with formula (\ref{entinterval}), we must therefore examine the differential entropy (\ref{sdiff0}) of a boundary causal diamond.

This appears problematic, because eq.~(\ref{boundarylimit}) does not satisfy the consistency condition (\ref{consist}). But it is useful to include boundary causal diamonds as a special case. Although formulas~(\ref{invr}-\ref{invtheta}) do not recover the correct bulk curve (the geodesic connecting the boundary points at $\theta = \theta_{A,B}$), the differential entropy of a causal diamond can still be evaluated. Let $\alpha_0 = (\theta_B - \theta_A)/2$ and $\theta_0 = (\theta_B + \theta_A)/2$. For the causal diamond of the interval $(\theta_A, \theta_B)$, the differential entropy is:
\begin{equation}
\label{sdiffdiam}
E = \frac{1}{2} \int_{\theta_0 - \alpha_0}^{\theta_0 + \alpha_0} d\theta\, \frac{dS(\alpha)}{d\alpha}\Big|_{\alpha = \alpha_0 - |\theta - \theta_0|} =
\frac{1}{2} \int_0^{\alpha_0} d\alpha\, \frac{dS(\alpha)}{d\alpha} - 
\frac{1}{2} \int_{\alpha_0}^0 d\alpha\, \frac{dS(\alpha)}{d\alpha} = S(\alpha_0).
\end{equation}
Thus, in AdS$_3$ the differential entropy of a causal diamond equals the entanglement entropy of the boundary interval supporting it. Because we are working in a geometry dual to a pure state of the field theory, the causal diamond of $(\theta_B, 2\pi + \theta_A) = \overline{(\theta_A, \theta_B)}$ gives a contribution equal to (\ref{sdiffdiam}). Applying this conclusion to $\gamma(\theta) = \min\{\alpha_A(\theta), \alpha_B(\theta)\}$, we get $E[\gamma(\theta)] = 2 S(\alpha_0)$, so that the distance between boundary points $A$ and $B$ correctly reproduces the entanglement entropy of interval $(\theta_A, \theta_B)$.

\subsection{Discontinuity in $\alpha'(\theta)$ and finite pieces of bulk geodesics}
\label{kinkina}
\begin{figure}[t!]
\centering
\begin{tabular}{cc}
\raisebox{-.5cm}{\includegraphics[width=.48\textwidth]{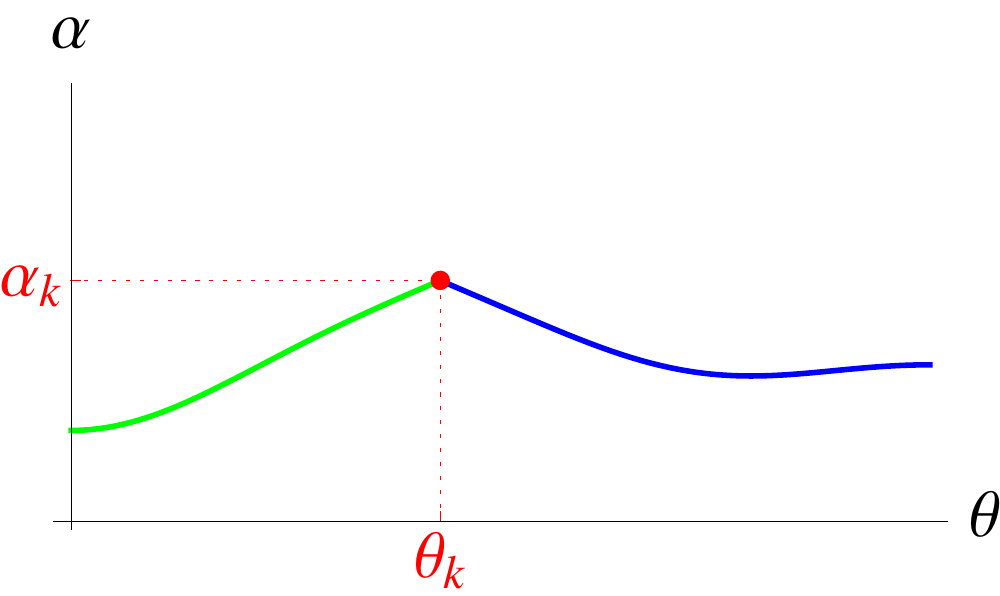}} &
\includegraphics[width=.48\textwidth]{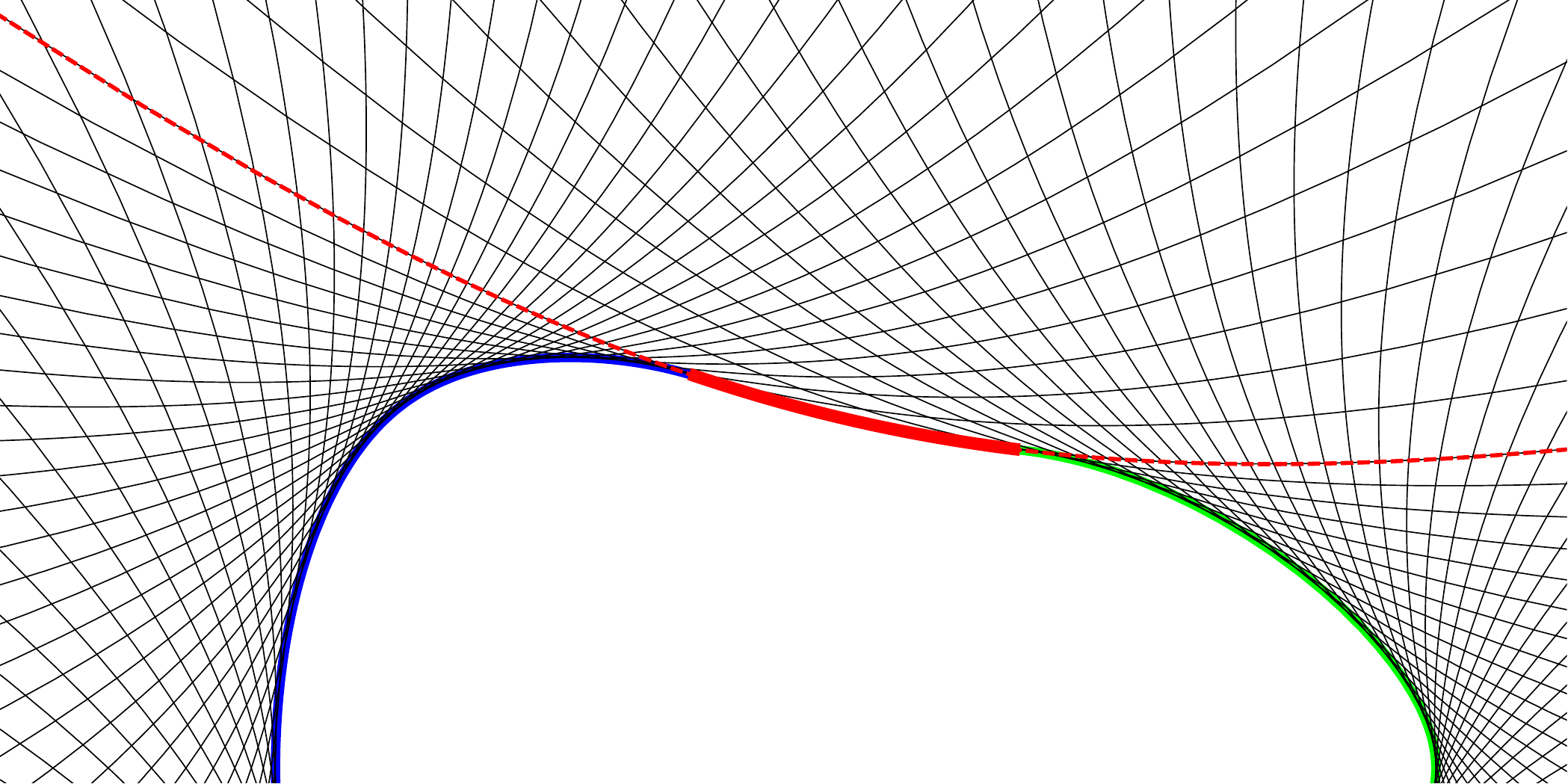} \\
\includegraphics[width=.48\textwidth]{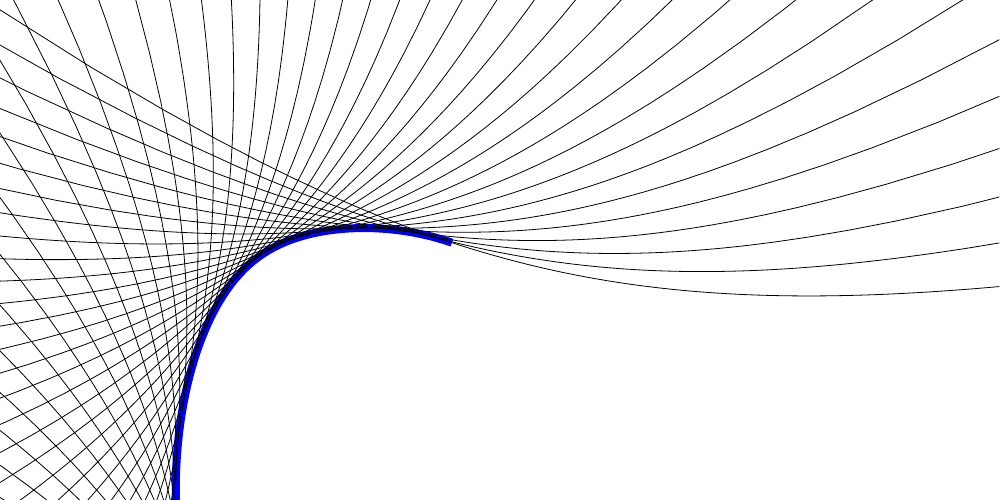} & 
\includegraphics[width=.48\textwidth]{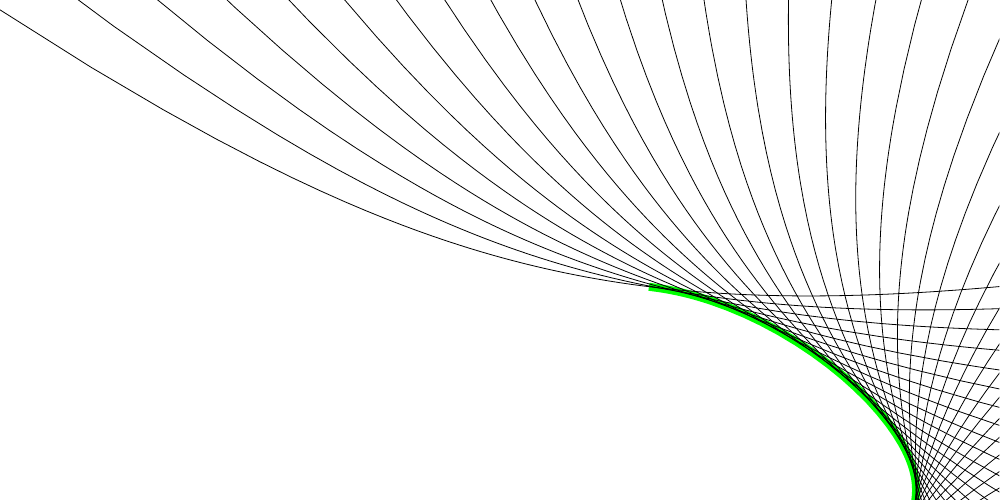}
\end{tabular}
\caption{Upper left: $\alpha(\theta)$ (green and blue), whose derivative is discontinuous at a point $(\theta_k, \alpha_k)$ (red). Upper right: the geodesics defined by $\alpha(\theta)$ and their (color coded) joint outer envelope. The two bulk open curves have the same tangent geodesic at their endpoints (red), because $\alpha(\theta)$ is continuous. The thickened part of this geodesic stretches between the endpoints of the open curves; its length is $f(\alpha_k, \theta_k, \tilde\theta_{\rm blue})-f(\alpha_k, \theta_k, \tilde\theta_{\rm green})$, viz. eq.~(\ref{kinkinaexpl}). The differential entropy of $\alpha(\theta)$ in the upper left panel is the length of this outer envelope. Lower panels: the individual bulk open curves and their tangent geodesics.}
\label{intersect}
\end{figure}

As a step toward understanding eq.~(\ref{defdist}), consider a boundary function $\alpha(\theta)$, which is only piecewise differentiable; see Fig.~\ref{intersect}. Suppose $\alpha'(\theta)$ jumps at some $\theta_k$. Let us examine $\alpha(\theta)$ separately in the ranges $\theta \gtrless \theta_k$, where it is differentiable. Each range defines a bulk curve with an endpoint. Both endpoints live on a common geodesic given by $\theta_k$ and $\alpha_k \equiv \alpha(\theta_k)$. But looking at eqs.~(\ref{invr}-\ref{invtheta}), we see that the two endpoints are a finite distance apart, because their locations depend on $\alpha'(\theta)$. In particular, eq.~(\ref{invtheta}) gives us the bulk angular coordinates of the endpoints $\tilde\theta_L$ and $\tilde\theta_R$, which are necessarily distinct.

We would like to understand the differential entropy formula evaluated on our piecewise differentiable $\alpha(\theta)$. To this end, we use eq.~(\ref{open}) for $\theta < \theta_k$ and $\theta > \theta_k$ individually:
\begin{align}
E[\alpha(\theta)] & = \frac{1}{2} \int_0^{\theta_k} d\theta\, \frac{dS(\alpha)}{d\alpha}\Big|_{\alpha = \alpha(\theta)} \quad\qquad +\frac{1}{2} \int_{\theta_k}^{2\pi} d\theta\, \frac{dS(\alpha)}{d\alpha}\Big|_{\alpha = \alpha(\theta)} \nonumber \\
& = \frac{\textrm{length on left}}{4G} - f(\alpha_k, \theta_k, \tilde\theta_L) +
\frac{\textrm{length on right}}{4G} + f(\alpha_k, \theta_k, \tilde\theta_R) 
\label{kinkinaexpl}\\
& = \frac{\textrm{length on left}}{4G}  + \frac{\textrm{length on right}}{4G} + 
\frac{\textrm{length of $\theta_k$-centered geodesic in $\tilde\theta_L < \tilde\theta < \tilde\theta_R$}}{4G} \nonumber
\end{align}
In the final line we used the fact that the term $f(\alpha_k, \theta_k, \tilde\theta_R)$ computes the length of the geodesic with opening angle $\alpha_k$ centered at $\theta_k$, which is contained in the angular wedge $\theta_k < \tilde\theta < \tilde\theta_R$. Likewise, $-f(\alpha_k, \theta_k, \tilde\theta_L)$ is the length of the same geodesic contained in $\tilde\theta_L < \tilde\theta < \theta_k$; the minus sign arises to give a negative contribution if $\theta_k < \tilde\theta_L$. Thus, the two $f$-terms combine to form the length of the geodesic that connects the endpoints of the two open curves. Overall, $E[\alpha(\theta)]$ computes the length of a continuously differentiable bulk curve, which is formed by joining the left and right open curves with the geodesic tangent at their endpoints; see the top right panel of Fig.~\ref{intersect}.

The final conclusion is that a discontinuity in $\alpha'(\theta)$ corresponds to a finite stretch of the bulk curve following one geodesic. The length over which the geodesic is followed depends on the jump in $\alpha'(\theta)$ via eq.~(\ref{invtheta}). We encountered one example of this finding in the previous subsection, where we saw that the differential entropy of a causal diamond (whose $\alpha'(\theta)$ jumps at the top) computes the full length of a geodesic.

\subsection{Pointwise minimum $\gamma(\theta) = \min\{\alpha(\theta), \beta(\theta)\}$ and the convex cover of curves}
\label{ptmin}
\begin{figure}[t]
\centering
\begin{tabular}{lcr}
\includegraphics[width=.47\textwidth]{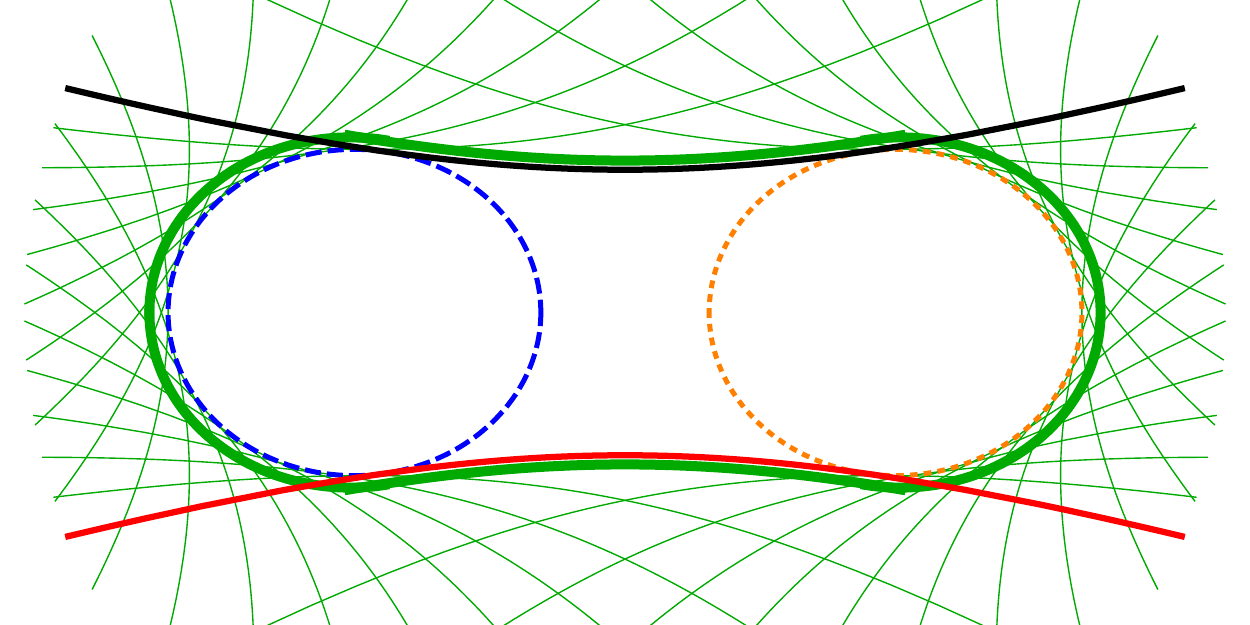} & \,\,\, &
\includegraphics[width=.47\textwidth]{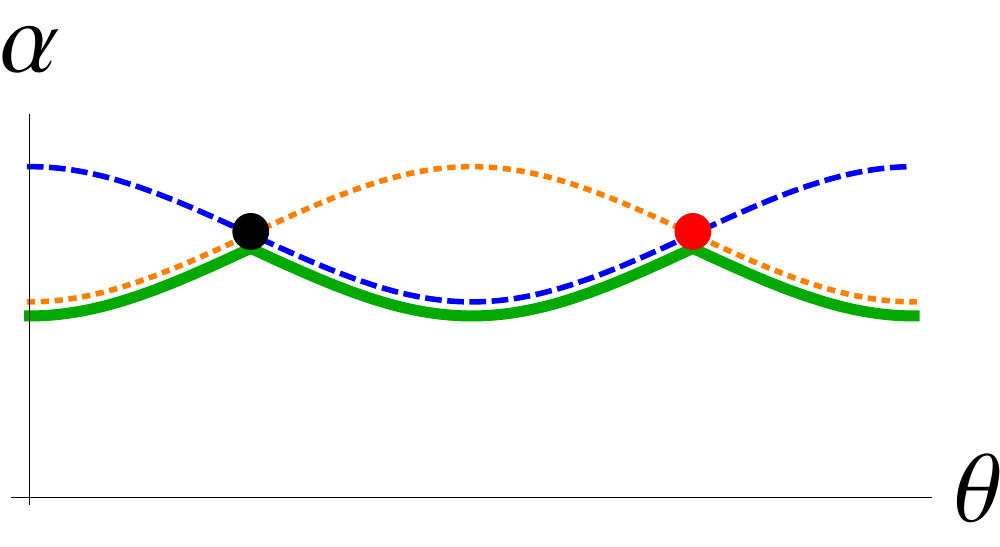}
\end{tabular}
\caption{Left: two closed bulk curves $a$ and $b$ (dashed blue and dotted orange) and their joint convex cover (thick green). We have drawn all geodesics tangent to the convex cover; the special geodesics that are tangent to both curves are drawn in black and red. Right: the functions $\alpha(\theta)$ and $\beta(\theta)$ corresponding to the two bulk curves (dashed blue and dotted orange), their pointwise minimum $\gamma(\theta)$ (thick green) and intersection points (black and red.)}
\label{convexcover}
\end{figure}

The discussion above has an interesting consequence. Consider two closed, convex, differentiable curves $a$ and $b$ in the bulk; see Fig.~\ref{convexcover}. Let their boundary functions be $\alpha(\theta)$ and $\beta(\theta)$ and call their pointwise minimum $\gamma(\theta)$. First, suppose that $\alpha(\theta)$ and $\beta(\theta)$ intersect. We discussed this situation in Sec.~\ref{kinkina} and illustrated it in Fig.~\ref{intersect}. Generically, the derivative $\gamma'(\theta)$ jumps at the intersection point and the bulk curve defined by $\gamma(\theta)$ first follows curve $a$, then follows a geodesic which is tangent to both $a$ and $b$ (this is the geodesic selected by the intersection of $\alpha(\theta)$ and $\beta(\theta)$), then follows curve $b$. This is nothing but the convex cover of $a$ and $b$. A second, special case is if $\alpha(\theta)<\beta(\theta)$ everywhere, but then $a$ properly encloses $b$ in the bulk. 
The final conclusion is that the operation of taking a pointwise minimum on the boundary corresponds to taking the convex cover in the bulk. 

This means, in particular, that if $\alpha_A(\theta)$ and $\alpha_B(\theta)$ describe two points $A$ and $B$ in the bulk, then $\gamma(\theta) = \min\{\alpha_A(\theta), \alpha_B(\theta)\}$ defines the closed curve, which circumscribes the convex cover of the set $\{A, B\}$. This closed curve follows the geodesic from $A$ to $B$ and then returns the same way from $B$ to $A$, which agrees with our prescription (\ref{defdist}) for geodesic distance. Fig.~\ref{convexcover} illustrates this construction, though one must imagine shrinking the bulk circles to points, as described in Sec.~\ref{ptlimitcurves}.


\subsection{Inflection points and nonconvex curves}
\label{nonconvex}

\begin{figure}[t!]
\centering
\begin{tabular}{ccc}
\includegraphics[width=.31\textwidth]{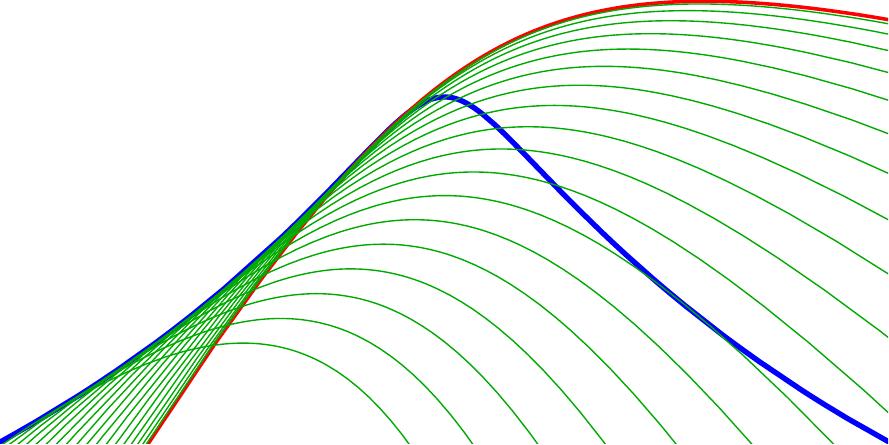} & 
\includegraphics[width=.31\textwidth]{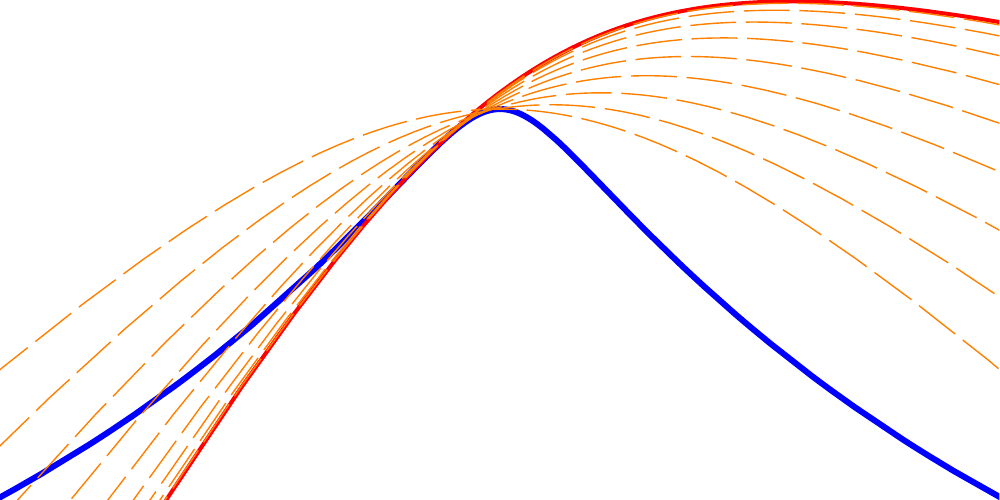} &
\raisebox{-.23cm}{\includegraphics[width=.24\textwidth]{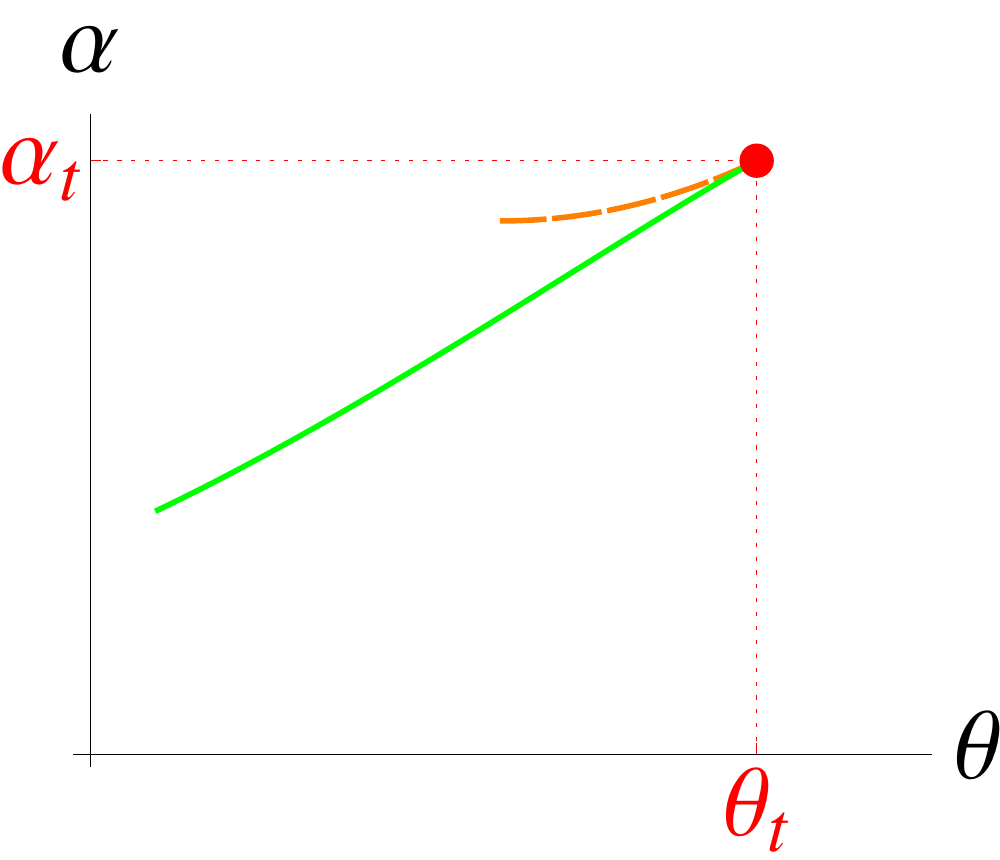}} \\ && \\
\includegraphics[width=.31\textwidth]{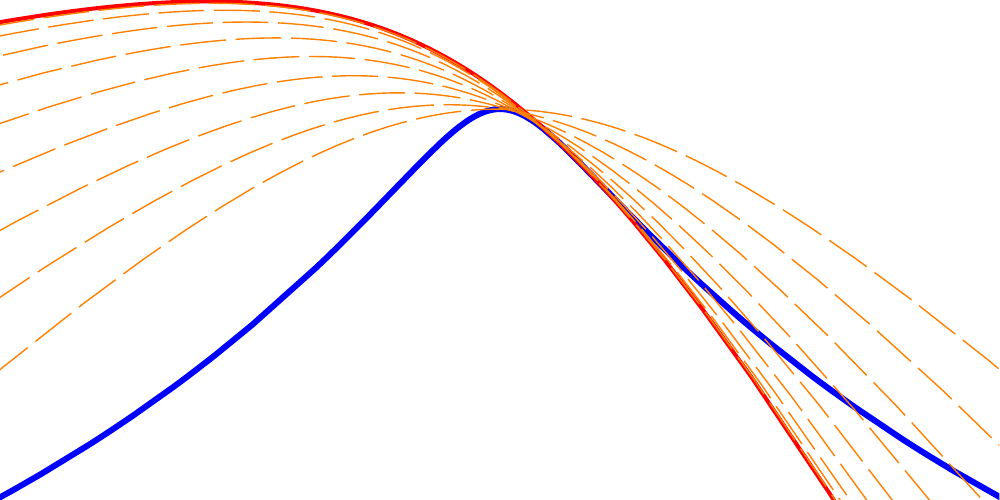} & 
\includegraphics[width=.31\textwidth]{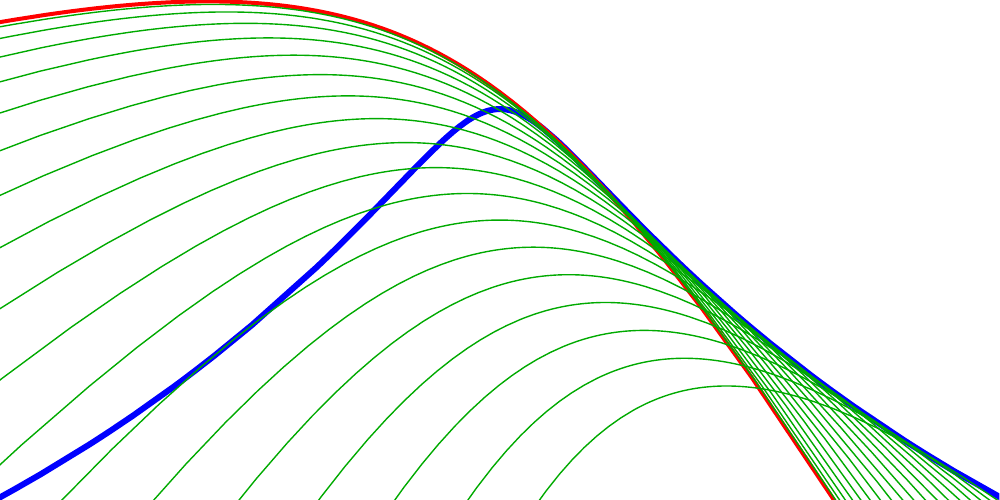} &
\raisebox{-.23cm}{\includegraphics[width=.24\textwidth]{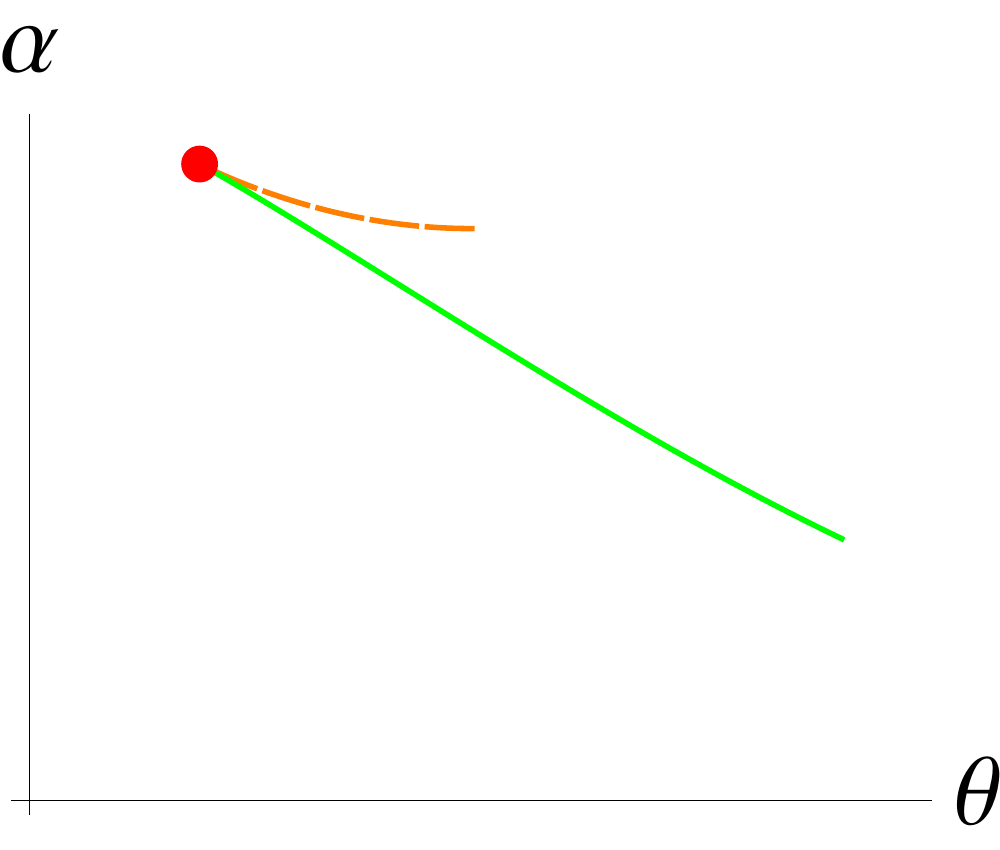}} \\ && \\
\includegraphics[width=.31\textwidth]{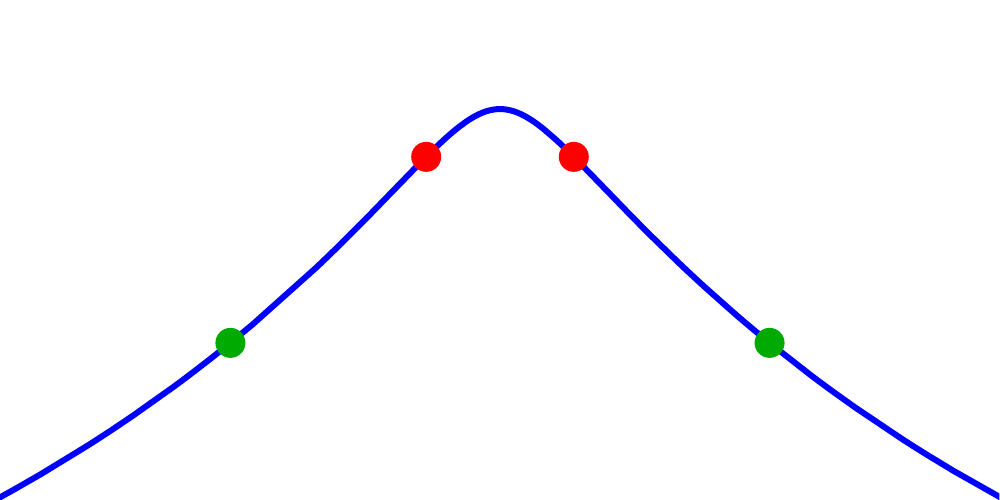} & 
\includegraphics[width=.31\textwidth]{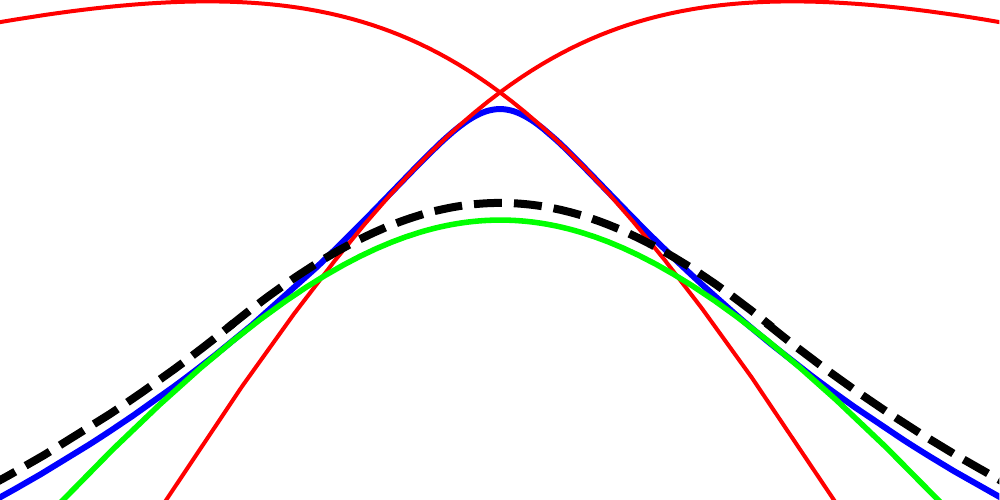} &
\raisebox{-.23cm}{\includegraphics[width=.24\textwidth]{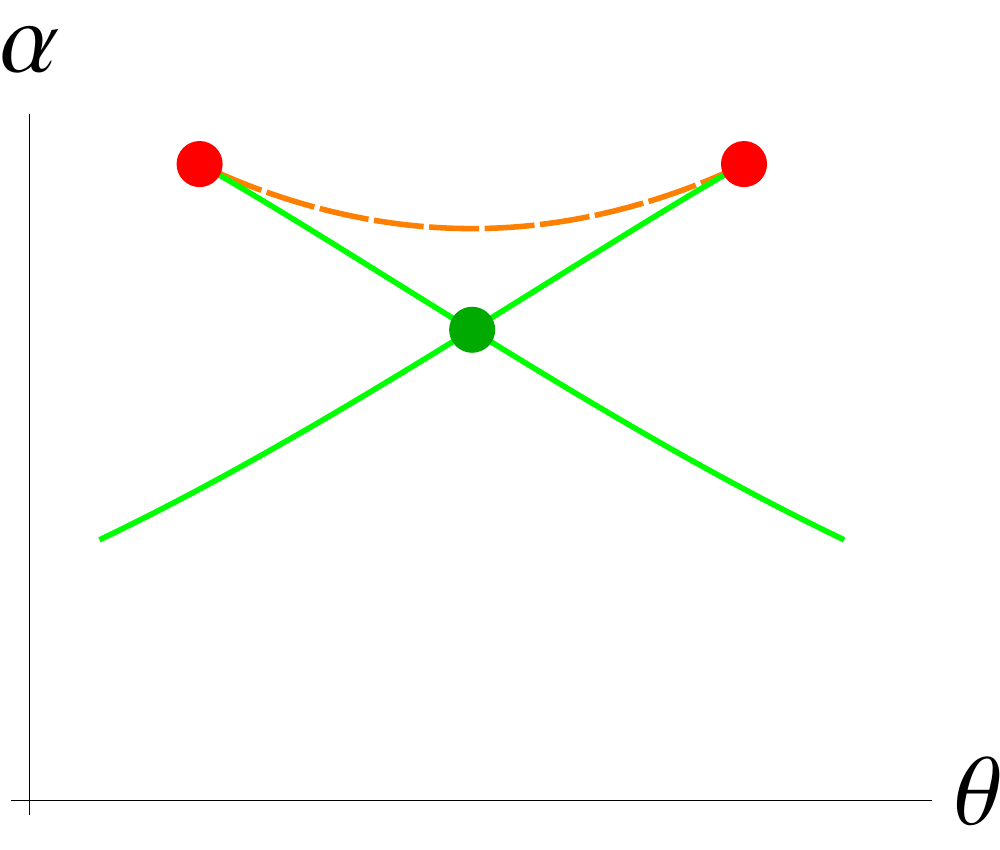}}
\end{tabular}
\caption{Top: As we approach the concavity from the left, the associated boundary intervals move to the far right (green continuous plots) until the first inflection point (red), where they turn back (orange dashed plots). Middle row: As we leave the concavity, the intervals shift to the left (orange dashed plots) until the second inflection point (red), whereafter they return to moving forward (green continuous plots). Bottom: An aggregate plot of the bulk curve and of its $\alpha(\theta(\tilde\theta))$, with three special geodesics and their tangency points singled out. The two red ones are tangent at the two inflection points; the green one occurs at the self-intersection of the plot of $\alpha(\theta(\tilde\theta))$ and is tangent to the curve at two distinct points. This geodesic traces the convex cover of the curve (black dashed; see also Sec.~\ref{ptmin}.)}
\label{concave}
\end{figure}

In the preceding subsection we assumed that the curves were convex. In the context of AdS$_3$, convexity means that the curve does not cross its tangent geodesics. We devote this subsection to consider a nonconvex curve -- one that crosses its own geodesic in at least one point; see Fig.~\ref{concave}. In analogy to flat space geometry, such a point is an inflection point. We shall denote the bulk angular coordinate of the inflection point $\tilde\theta_t$.

The top panels of Fig.~\ref{concave} show that as we approach the inflection point, the centerpoints of the tangent geodesics approach $\theta_t \equiv \theta(\tilde\theta_t)$ from the same side. In other words, $\theta(\tilde\theta)$ attains an extremum at $\tilde\theta_t$. This means that, strictly speaking, we cannot speak of $\alpha(\theta)$ as a \emph{function} of $\theta$: as we trace the bulk curve, $\alpha(\theta(\tilde\theta))$ approaches the value $\alpha_t \equiv \alpha(\theta_t)$ and then reverses direction.\footnote{This was previously noted in \cite{roblast, Hubeny:2014qwa}.} This is shown in the top right panel of Fig.~\ref{concave}.

In the bulk, as the inflection point is approached from either side, the coordinate $\tilde\theta$ approaches $\tilde\theta_t$. Recall that eqs.~(\ref{invr}-\ref{invtheta}) reconstruct the bulk curve from boundary data. In particular, eq.~(\ref{invtheta}) implies that $\alpha'(\theta_t)$ is well defined and equal for both branches of $\alpha(\theta)$. We can say that the plot of $\alpha(\theta(\tilde\theta))$ develops an infinitely sharp cusp.

\paragraph{Nonconvex curves}
A closed differentiable bulk curve defines an $\alpha(\theta)$, which must return to itself after a $2\pi$ rotation in the bulk (or its multiple). This means that inflection points can only occur in pairs. After adding a second inflection point (middle panel of Fig.~\ref{concave}), we obtain a nonconvex bulk curve and $\alpha(\theta)$ shown in the bottom of Fig.~\ref{concave}. The plot in the bottom right is typical for nonconvex curves.

Interestingly, the circumference of the curve is still computed using eq.~(\ref{sdiff}), with the caveat that the segment of the boundary between the inflection points is first traversed forward, then backward, then forward again. Thus, in this case it is more correct to write:
\begin{equation}
\label{sdifft}
E = \frac{1}{2} \int_0^{2\pi} d\tilde\theta\, \frac{d\theta}{d\tilde\theta} \, \frac{dS(\alpha)}{d\alpha}\Big|_{\alpha = \alpha(\theta(\tilde\theta))} = \frac{\rm circumference}{4G} 
\end{equation}
We will comment more on the application of this formula to nonconvex curves in Sec.~\ref{intkink}.

\paragraph{Convex cover of a nonconvex curve}
The bottom panel of Fig.~\ref{concave} shows that the plot of $\alpha(\theta)$ for a nonconvex curve self-intersects. Recall that every point $(\theta, \alpha(\theta))$ on the plot defines a geodesic tangent to the curve. If the plot self-intersects, one geodesic must be tangent to the curve at two distinct points. We have marked that geodesic along with its two tangency points in the bottom panel of Fig.~\ref{concave}. This feature occurs whenever the bulk curve develops a concavity.

Using Sec.~\ref{ptmin}, we now know how to find the convex cover of a nonconvex curve. All we have to do is take the pointwise minimum of the multi-branched plot of $\alpha(\theta(\tilde\theta))$. The resulting discontinuity in $\alpha'(\theta)$ is responsible for the finite segment, along which the convex cover follows the special geodesic with two tangency points. See the bottom panels of Fig.~\ref{concave} for an illustration.

\subsection{Bulk curves with endpoints and corners}
\label{endpt}
In this and the next subsection we gather some results, which clarify our hole-ographic construction of bulk points in Sec.~\ref{secpoints}. 

We start with eq.~(\ref{open}), which gives a differential entropy formula for the length of a bulk curve with endpoints. The function $f(\alpha, \theta, \tilde\theta)$ featured in eq.~(\ref{open}) is dissatisfying in a hole-ographic context, because it depends on the bulk coordinate $\tilde\theta$. We would like to relate $f$ to the function $\alpha_A(\theta)$, which defines the endpoint of the bulk curve according to Sec.~\ref{secpoints}.

\begin{figure}[t]
\centering
\includegraphics[width=.6 \textwidth]{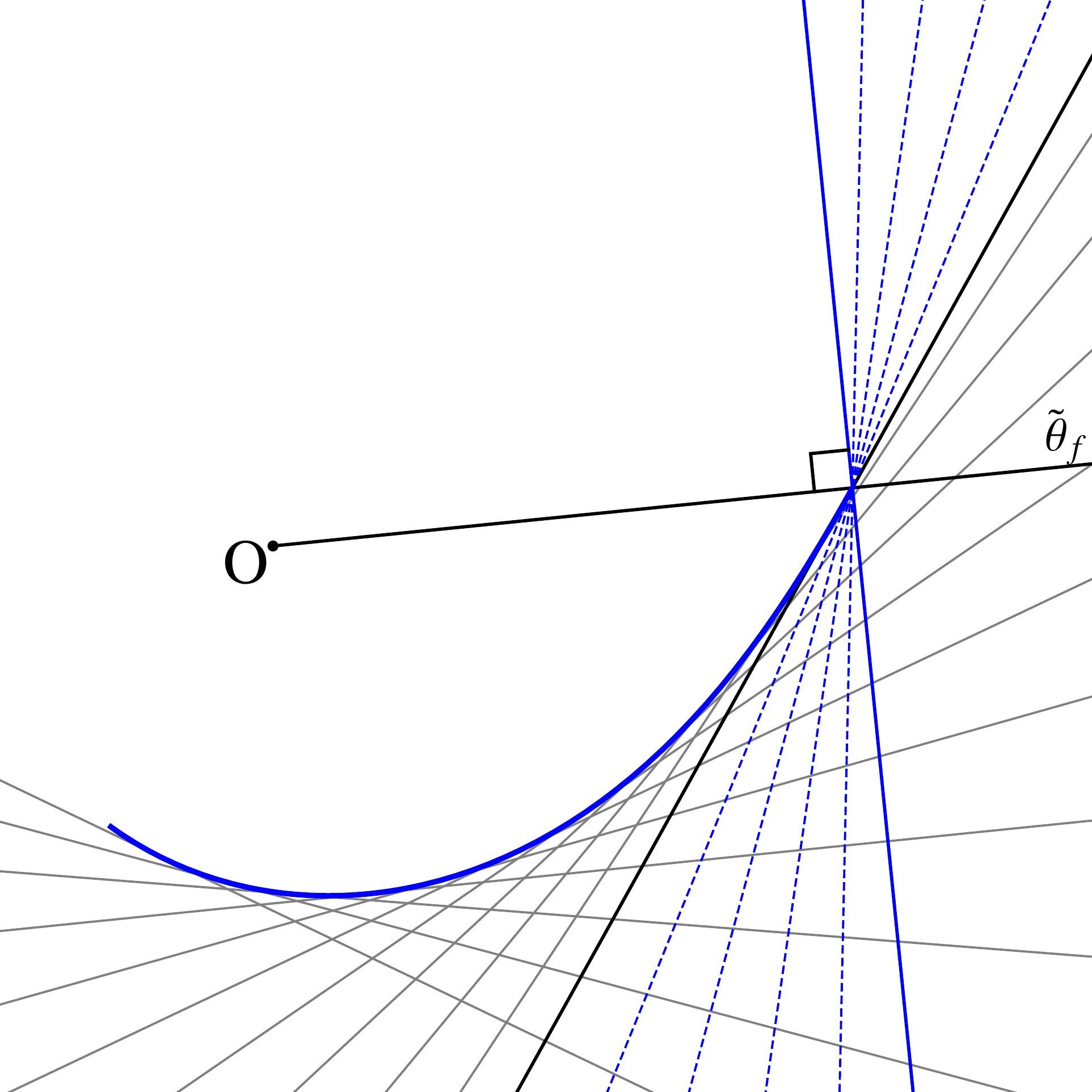}
\caption{An open bulk curve $a$ with an endpoint $A$ at $\tilde\theta = \tilde\theta_f$. We trade the inhomogeneous term $f(\tilde\theta_f)$ in eq.~(\ref{open}) for an extension of $\alpha_a(\theta)$ given by $\alpha_A(\theta)$ in eq.~(\ref{defA0}). The extension describes all geodesics (shown in dashed blue) that pass through the endpoint of the bulk curve until $\theta = \tilde\theta_f$, where the inhomogeneous term is set to zero.}
\label{endpoint}
\end{figure}

Suppose that the endpoint $A$ of an open bulk curve $a$ falls at $\tilde\theta_f$.  The boundary function $\alpha_a(\theta)$ of the curve (not $\alpha_A(\theta)$ of the endpoint) terminates at some $\theta_f$. Generically $\theta_f \neq \tilde\theta_f$ and $f$ is nonvanishing; see eq.~(\ref{deff}). Now take $\alpha_A(\theta)$ over the range $\theta_f < \theta < \tilde\theta_f$ and concatenate it with $\alpha_a(\theta)$ of the curve. The bulk meaning of this operation is shown in Fig.~\ref{endpoint}: it adds to the set of geodesics tangent to curve $a$ a further set of geodesics, which pass through endpoint $A$. Extending the boundary function in this way does not affect the bulk curve at all. Using Sec.~\ref{kinkina}, this directly implies that
\begin{equation}
\alpha_a(\theta_f) = \alpha_A(\theta_f) \qquad {\rm and} \qquad
\alpha_a'(\theta_f) = \alpha_A'(\theta_f)\,,
\label{extdiff}
\end{equation}
for any discontinuity in $\alpha'(\theta)$ would add to the bulk curve a finite piece of a geodesic.
Eqs.~(\ref{extdiff}) express the fact that point $A$ lives on the bulk curve $a$; the tangent to $a$ at $A$ is centered at $\theta_f$.

We now compute the differential entropy of the concatenated range of $\alpha_A(\theta)$. Referring to eq.~(\ref{adsgeodesic}), the geodesic with opening angle $\alpha_f$ centered at $\theta_f$  can be recast in the form:
\begin{equation}
\frac{\cos\alpha_f}{\cos(\theta_f - \tilde\theta_f)} = \frac{R}{\sqrt{L^2 + R^2}}\,.
\end{equation}
Because endpoint $A$ lives on that geodesic, we can express $\alpha_A(\theta)$ of eq.~(\ref{point}) as:
\begin{equation}
\alpha_A(\theta) = 
\cos^{-1}  \frac{\cos\alpha_f \cos(\theta-\tilde\theta_f)}{\cos(\theta_f - \tilde\theta_f)}\,.
\label{defA0}
\end{equation}
It is now straightforward to confirm that:
\begin{equation}
\frac{1}{2} \int_{\theta_f}^{\tilde\theta_f} d\theta\, \frac{dS(\alpha)}{d\alpha}\Big|_{\alpha =\alpha_A(\theta)} = f(\alpha_f, \theta_f, \tilde\theta_f)\,.
\label{newf}
\end{equation}
This gives a new perspective on the function $f$: it is the differential entropy of the zero length extension of the open bulk curve, which resets $\theta_f$ to $\tilde\theta_f$. A consequence of the resetting is that $f$ is set to zero.  Note that when $\theta_f > \tilde\theta_f$ the integral in eq.~(\ref{newf}) ``runs backwards'' and $f$ is negative.

\paragraph{A corner in the bulk} 
Putting together two curves with coincident endpoints we can construct a bulk curve with a ``corner.'' Two open curves $a$ and $b$ have a common endpoint $A$ if:
\begin{eqnarray}
\alpha_a(\theta_L) = \alpha_A(\theta_L) 
& \qquad {\rm and} \qquad &
\alpha_a'(\theta_L) = \alpha_A'(\theta_L) 
\label{corner1} \\
\alpha_b(\theta_R) = \alpha_A(\theta_R) 
& \qquad {\rm and} \qquad &
\alpha_b'(\theta_R) = \alpha_A'(\theta_R) 
\label{corner2}
\end{eqnarray}
The values $\theta_{L,R}$ are centerpoints of the two geodesics, which are tangent to $a$ and to $b$ at $A$. The location of $A$, the corner of $a \cup b$, is set by the choice of $\alpha_A(\theta)$, which satisfies conditions (\ref{corner1}-\ref{corner2}). To obtain the circumference of such a curve, we substitute in the differential entropy formula (\ref{sdiff}) the boundary function, which is obtained by concatenating the boundary functions of curve $a$, corner point $A$, and curve $b$.

In summary, a piecewise differentiable bulk curve can be represented on the boundary by a continuous and differentiable $\alpha(\theta)$. A corner -- that is, a discontinuity in $dR/d\tilde\theta$ -- occurs when $\alpha(\theta)$ follows over a finite range of $\theta$ 
one of our point functions $\alpha_A(\theta)$ defined in eq.~(\ref{point}).

\subsection{Orientation and signed lengths}
\label{intkink}
\begin{figure}[t!]
\centering
\begin{tabular}{cc}
\includegraphics[width=.48\textwidth]{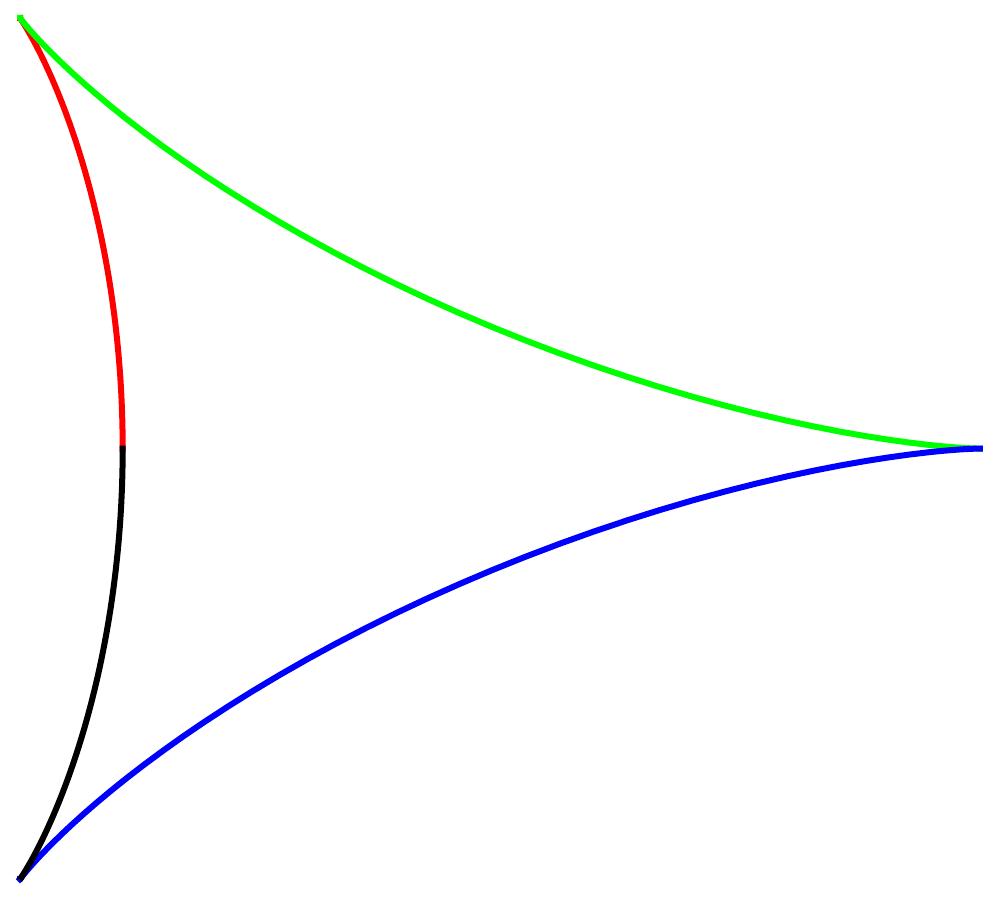} & 
\includegraphics[width=.48\textwidth]{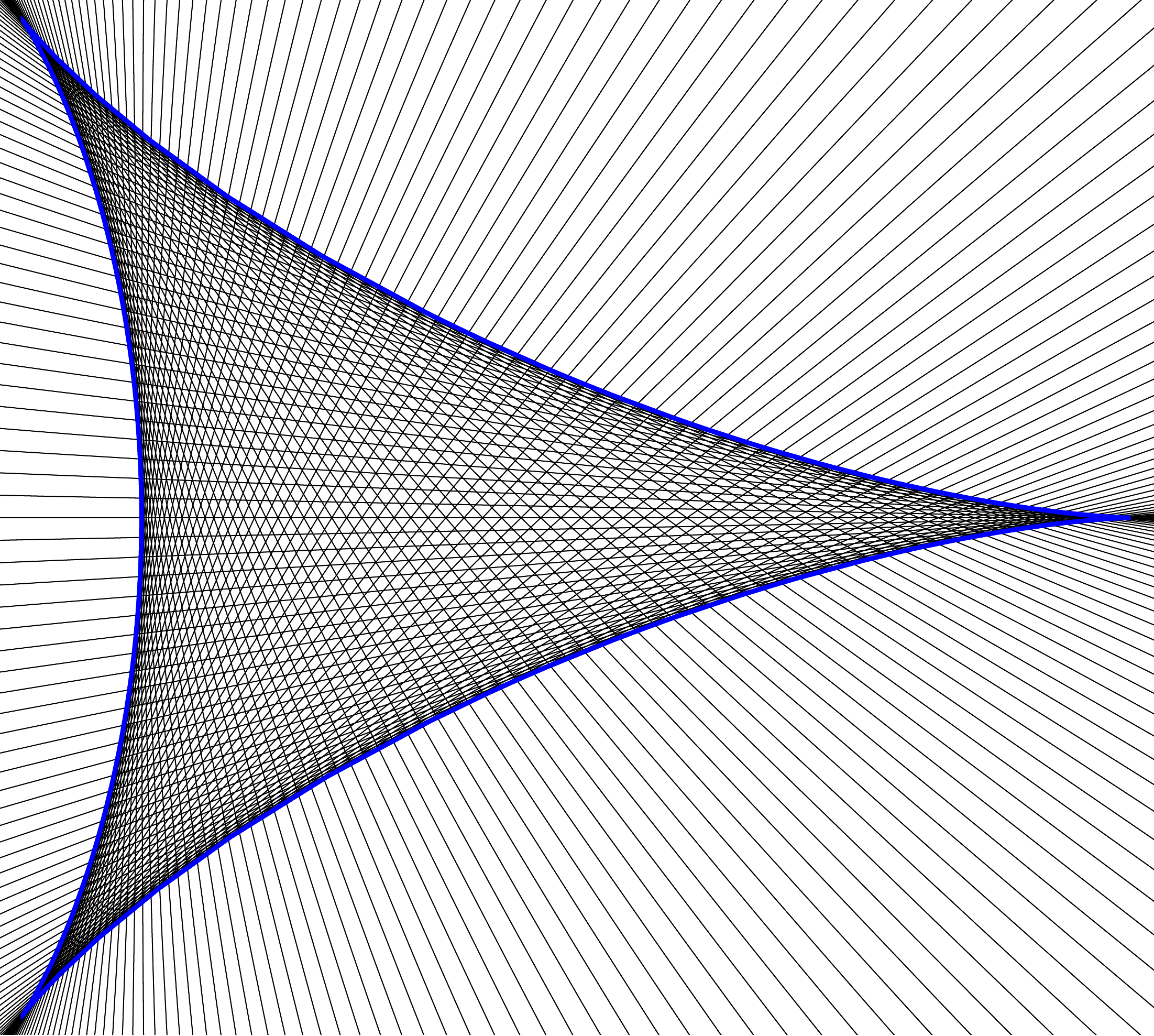} 
\end{tabular}
\caption{The bulk curve defined by eq.~(\ref{pseudopt}). Left: The color coding is with reference to eq.~(\ref{evanish}). Right: We display $N=80$ geodesics tangent to the curve.\\}
\label{3kinks}
\end{figure}

In addition to the corners discussed in Sec.~\ref{endpt}, bulk curves may develop another, qualitatively different type of singularity. 
As an example, consider:
\begin{equation}
\alpha(\theta) = \cos^{-1} \frac{\cos\theta}{5+\cos2\theta}\,.
\label{pseudopt}
\end{equation}
Using eqs.~(\ref{invr}-\ref{invtheta}) we find the bulk curve shown in Fig.~\ref{3kinks}. It has several surprising features. First, it contains three cusps, even though it does not agree with eq.~(\ref{point}) on any finite size interval. Next, as $\theta$ varies from 0 to $2\pi$, the curve is traversed twice. Finally, plugging eq.~(\ref{pseudopt}) into the differential entropy formula (\ref{sdiff}) gives $E = 0$, even though the curve has nonvanishing length.

\begin{figure}[t!]
\centering
\begin{tabular}{lr}
\includegraphics[width=.45\textwidth]{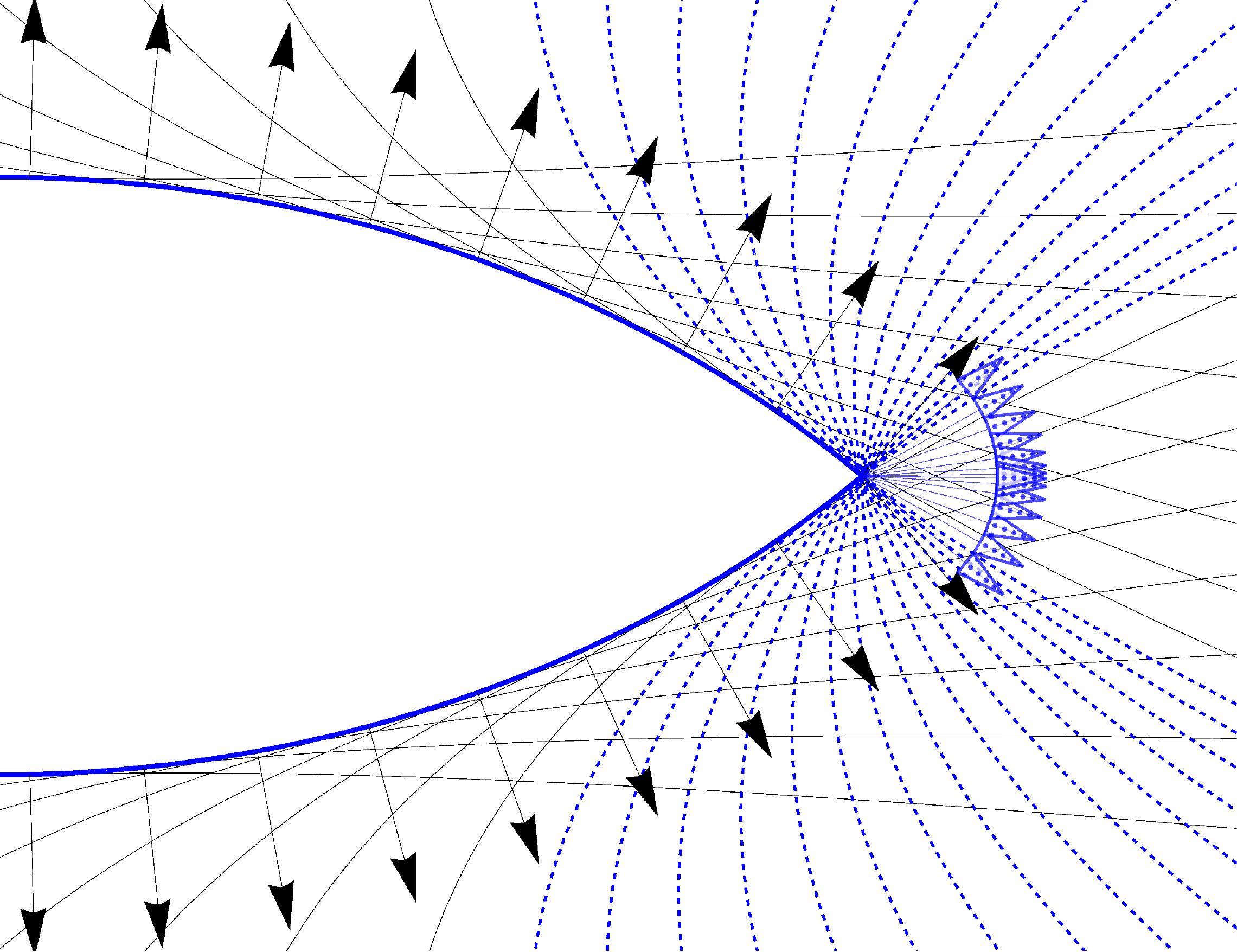} & 
\includegraphics[width=.45\textwidth]{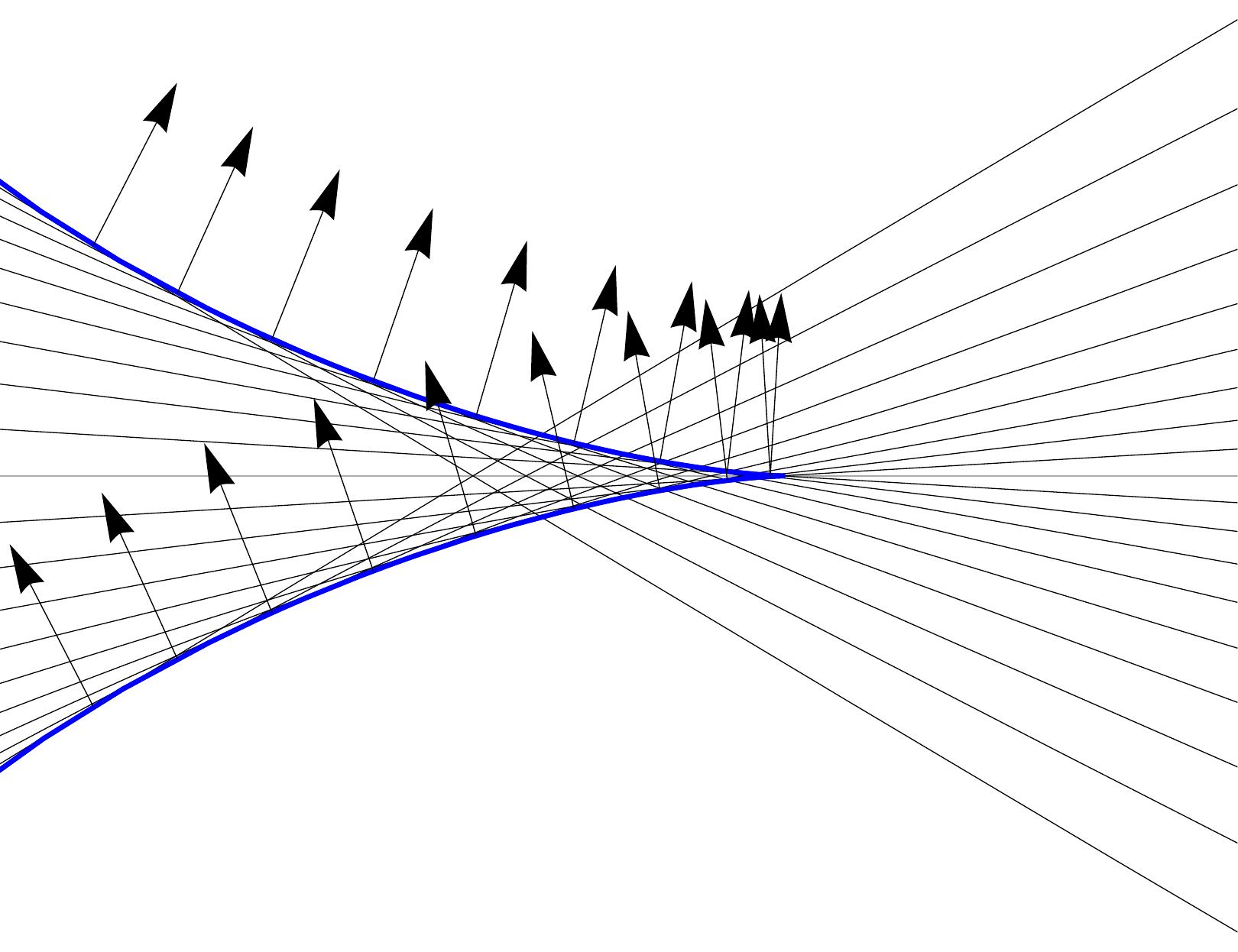} 
\end{tabular}
\caption{
An example corner (discussed in Sec.~\ref{endpt}, left panel) and cusp (Sec.~\ref{intkink}, right). The normal vectors, which define an orientation on the curve, point toward $\theta(\tilde\theta)$ -- the centerpoints of the boundary intervals selected by $\alpha(\theta)$ in eq.~(\ref{tantheta}). \vspace{.3cm} \\
Left: $dR/d\tilde\theta$ is discontinuous at a corner. In Sec.~\ref{endpt} we explained how to define a continuous boundary function $\alpha(\theta)$ around a corner. The discontinuity in $dR/d\tilde\theta$ is filled with $\alpha_A(\theta)$, the point function of the corner (eq.~\ref{point}). The ``normal vectors'' in this range are shown in blue; they preserve the orientation on the curve. \vspace{.3cm}\\
Right: $dR/d\tilde\theta$ is continuous at a cusp. The function $\alpha(\theta)$ is continuous and differentiable, because the tangent geodesics vary continuously along the curve. However, the orientation on the curve (relative to $\theta$) is reversed.}
\label{contrast}
\end{figure}

To understand this, we read off from eq.~(\ref{pseudopt}) and draw the set of geodesics tangent to the curve. In contrast to the cases discussed in Sec.~\ref{endpt}, the geodesics tangent to the curve vary smoothly in the neighborhood of each cusp. This is equivalent to saying that the cusp is infinitely sharp: both segments incident on the cusp tend to the same limiting gradient. Yet another way to phrase this is that $dR/d\theta$ and $d\tilde\theta / d\theta$ simultaneously vanish, so $dR/d\tilde\theta$ is well defined and not equal to zero. A corner and a cusp are contrasted in Fig.~\ref{contrast}. 

\paragraph{Signed length of oriented curves}
For each $\theta$, draw at the corresponding tangency point an arrow aimed toward $\theta$ on the boundary. This procedure endows the bulk curve with an orientation. Fig.~\ref{contrast} shows that cusps reverse the orientation while corners, treated with our prescription of Sec.~\ref{endpt}, preserve it. One easily confirms that after passing a cusp, the differential entropy formula computes the length of the curve with a relative minus sign. In other words, formula~(\ref{sdiff}) computes the \emph{signed} length of an \emph{oriented} curve.\footnote{This was first observed in \cite{robproof}.} For example, for the bulk curve of eq.~(\ref{pseudopt}) we have:
\begin{equation}
E = 0 = 
\raisebox{-.25cm}{\includegraphics[scale=.07]{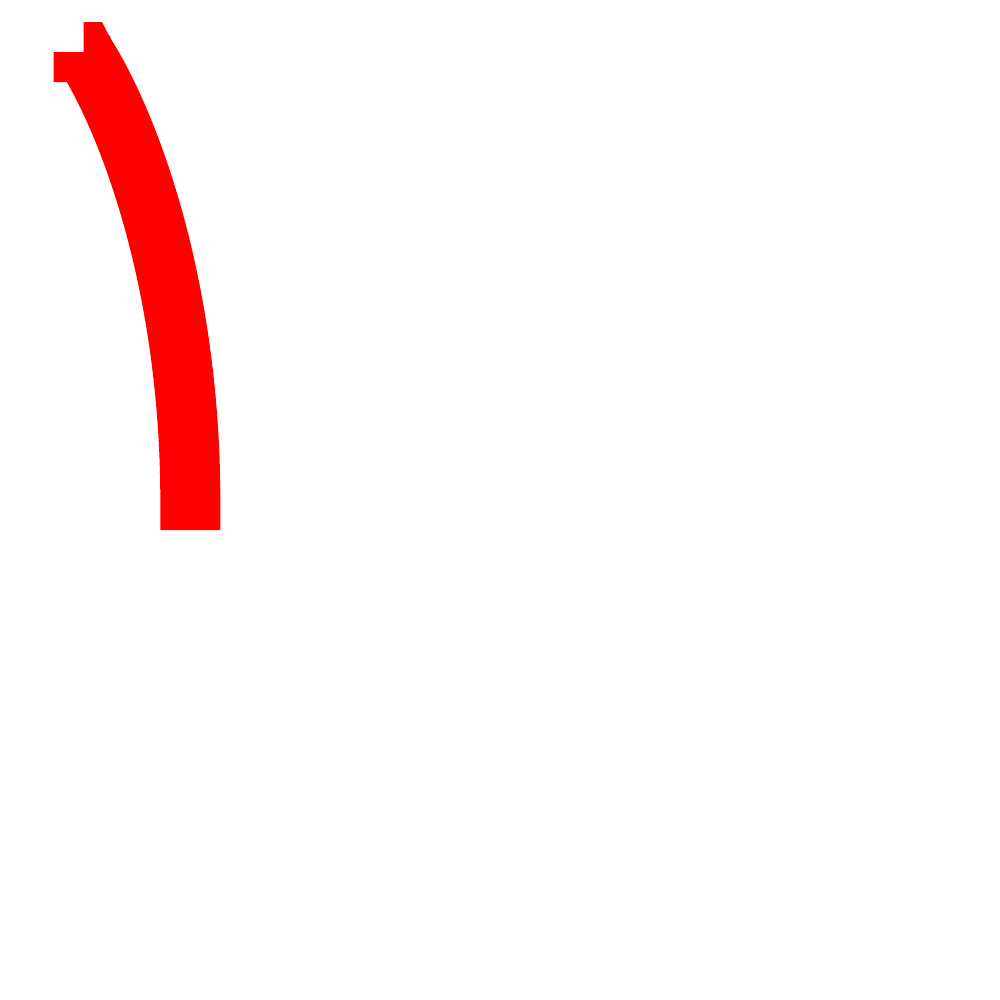}} \!\!\!\!\!\!
- \raisebox{-.25cm}{\includegraphics[scale=.07]{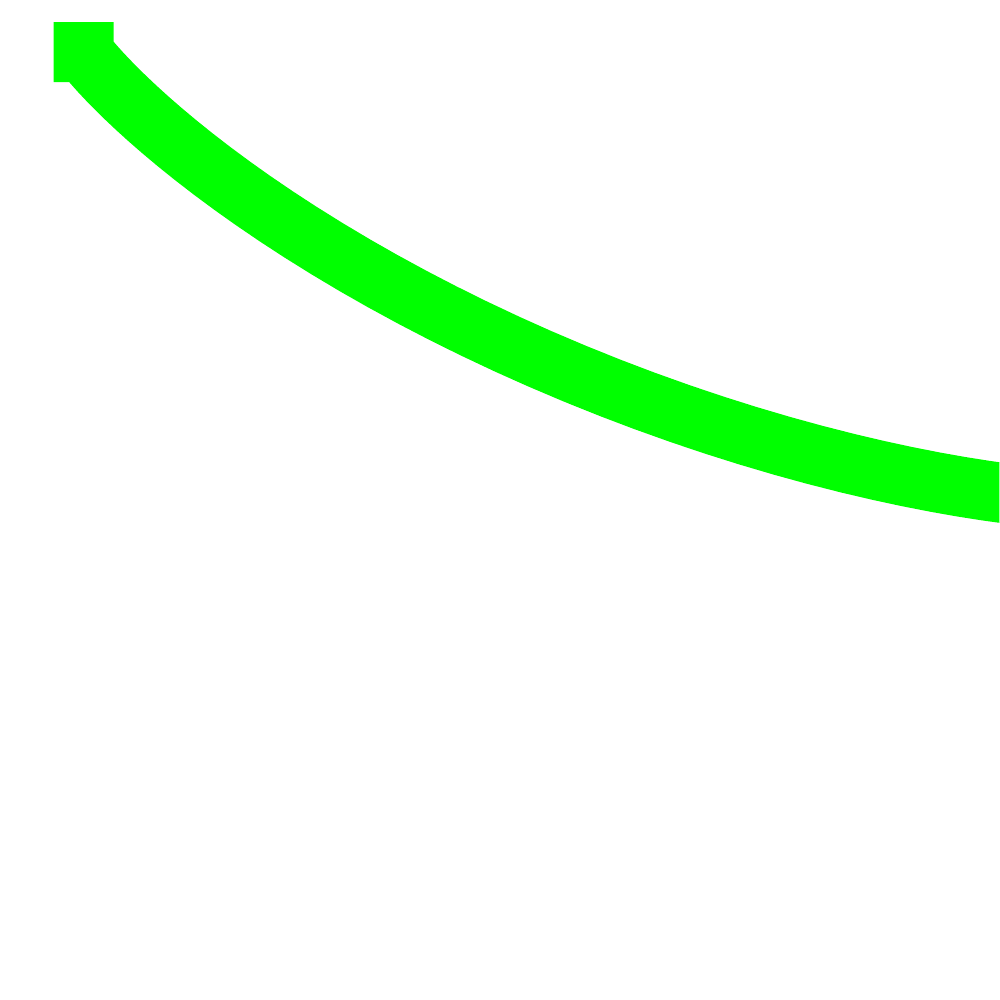}}\,
+ \raisebox{-.25cm}{\includegraphics[scale=.07]{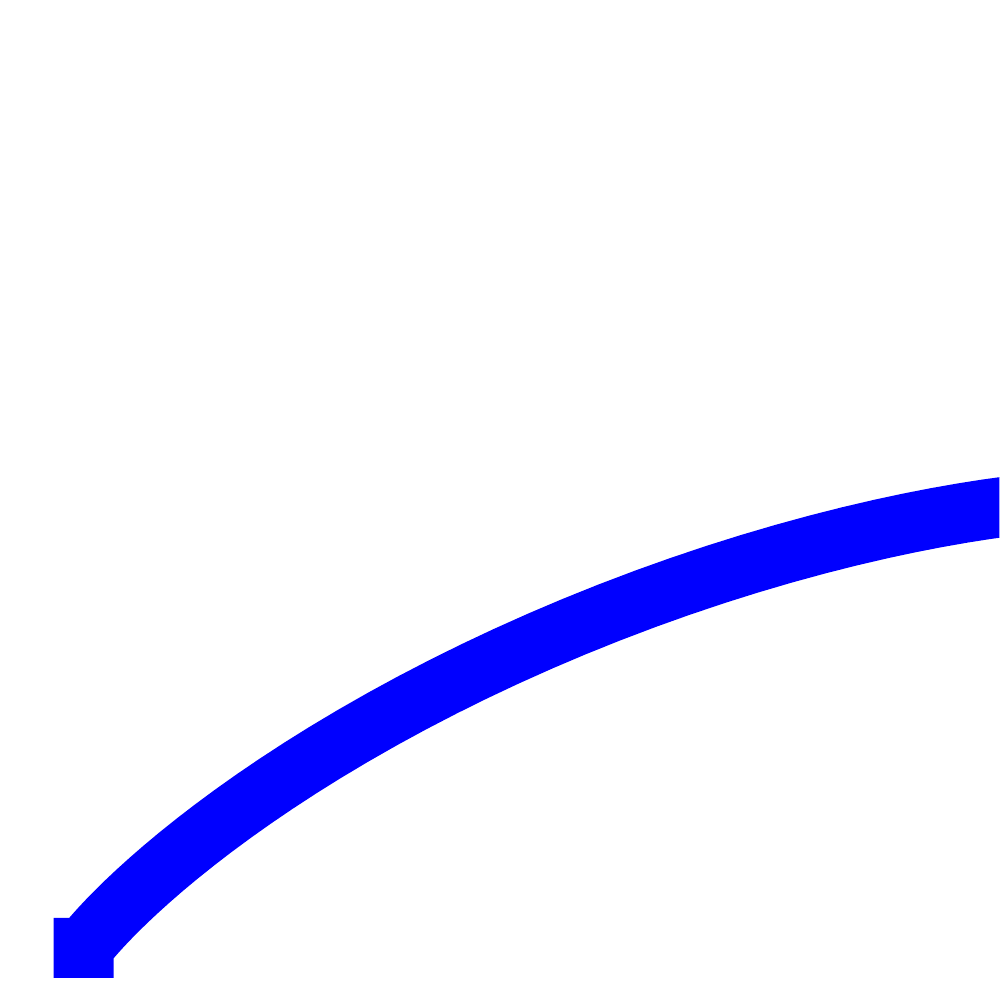}}\,
-\, \raisebox{-.25cm}{\includegraphics[scale=.07]{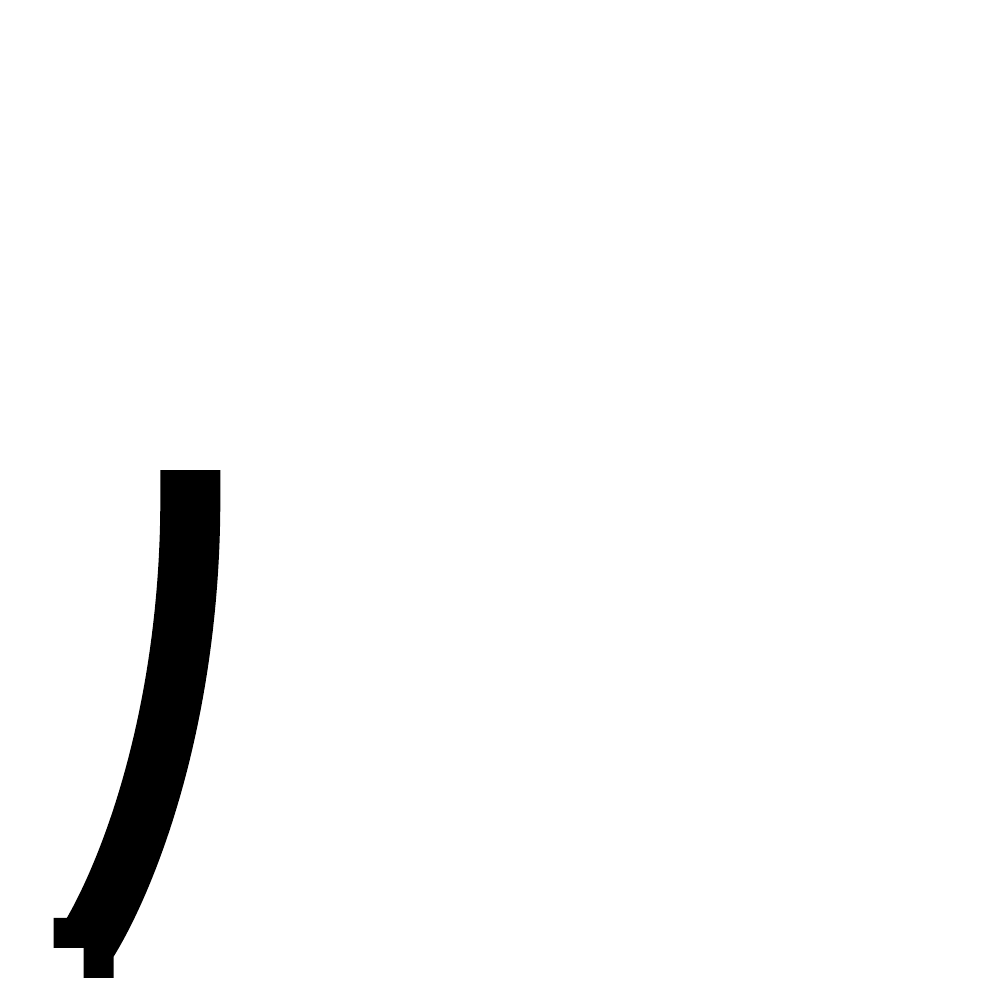}}\!\!\!\!\!\!
- \raisebox{-.25cm}{\includegraphics[scale=.07]{piecered.pdf}} \!\!\!\!\!\!
+ \raisebox{-.25cm}{\includegraphics[scale=.07]{piecegreen.pdf}}\,
- \raisebox{-.25cm}{\includegraphics[scale=.07]{pieceblue.pdf}}\,
+\, \raisebox{-.25cm}{\includegraphics[scale=.07]{pieceblack.pdf}}
\label{evanish}
\end{equation}
The color-coding follows the left panel of Fig.~\ref{3kinks}. Each term appears twice (albeit with a different sign), because as $\theta$ ranges from 0 to $2\pi$, the bulk curve is traversed twice. Example (\ref{pseudopt}) illustrates why $E[\alpha(\theta)] = 0$ does not work as a boundary definition of a bulk point. Instead, identifying points involves the extrinsic curvature of the curve.

\paragraph{Orientation}
How can we diagnose a reversal of orientation? It occurs at common zeroes of $dR/d\theta$ and $d\tilde\theta / d\theta$. Differentiating eqs.~(\ref{invr}-\ref{invtheta}), we identify the common factor:
\begin{equation}
\cos\alpha(\theta) \big(1-\alpha'(\theta)^2\big) - \sin\alpha(\theta) \,\alpha''(\theta)\gtrless 0. \label{orient}
\end{equation}
In this diagnostic, the upper sign selects the outward orientation for a circle $R = const. > 0$. 
Comparing (\ref{orient}) with eq.~(\ref{pointdiffeq}), we see that points are precisely those ``curves'' whose orientation is everywhere undefined. Recall that we derived eq.~(\ref{pointdiffeq}) in Sec.~\ref{intersectpoint} by demanding that the intersection point of two infinitesimally separated geodesics remain at constant $\tilde\theta$ in the bulk. Condition~(\ref{orient}) distinguishes when the said point moves in the direction of increasing or decreasing $\theta$. Because intersections of neighboring geodesics are what defines the bulk curve, this is equivalent to deciding in which direction we scan the bulk curve as $\theta$ on the boundary increases.

\paragraph{Discretized differential entropy} 
When the orientation of the bulk curve changes, the discrete approximation to differential entropy given in eq.~(\ref{differentials}) must be supplemented by a second case:
\begin{equation}
E = \lim_{N \to \infty} \sum_{k=1}^N 
S(I_k) - \left\{ \begin{array}{ll} 
S(I_k \cap I_{k+1}) \quad & \textrm{if (\ref{orient}) is positive} \\
S(I_k \cup I_{k+1}) \quad & \textrm{if (\ref{orient}) is negative}
\end{array} \right.
\label{gendifferential}
\end{equation}
This equation was first observed in \cite{roblast} and explained formally in \cite{robproof}. To check that the second case correctly captures the extra minus sign occasioned by an orientation flip, define $J_k = I_k \cap I_{k+1}$ so that $S(J_k) - S(J_{k-1} \cup J_{k}) = - \big( S(I_k) - S(I_k \cap I_{k+1})\big)$. We now note that for infinitesimally separated intervals the differences between $I_k$ and $J_k$ and between $I_k \cup I_{k+1}$ and $I_k \cup I_{k-1}$ can be ignored. 

A reversal of orientation could be seen as replacing a boundary interval $I_k$ with its complement. This is because the normal to the curve, which used to point toward the center of $I_k$, gets flipped toward the center of $\bar{I}_k$. In a pure state such as the vacuum dual to AdS$_3$, this allows us to rewrite eq.~(\ref{gendifferential}) in an arguably more symmetric fashion:
\begin{equation}
E = \lim_{N \to \infty} \sum_{k=1}^N 
\,\left\{ \begin{array}{ll} 
S(I_k) - S(I_k \cap I_{k+1}) \quad & \textrm{if (\ref{orient}) is positive} \\
S(\bar{I}_k) - S(\bar{I}_k \cap \bar{I}_{k+1}) \quad & \textrm{if (\ref{orient}) is negative}
\end{array} \right.
\label{gendifferential2}
\end{equation}

\paragraph{Nonconvex curves} Eq.~(\ref{gendifferential}) was first observed in \cite{roblast} to apply to nonconvex curves. Specifically, the second case applies in the region between two inflection points. We first verify this statement using eq.~(\ref{orient}). We then explain why the differential entropy formula correctly computes circumferences of nonconvex curves, despite a change of sign in (\ref{orient}).

First, look at condition (\ref{orient}). We observed in Sec.~\ref{nonconvex} that in a neighborhood of an inflection point both $\alpha(\theta)$ and $\alpha'(\theta)$ are continuous. However,
\begin{equation}
\alpha''(\theta) = \frac{d\tilde\theta}{d\theta} \,\frac{d\alpha'(\theta)}{d\tilde\theta}
\label{2ndderchain}
\end{equation}
blows up at inflection points, because they are characterized by a reversal of direction of $\theta(\tilde\theta)$ so $d\theta/d\tilde\theta = 0$. The blow-up means that the sign of $\alpha''(\theta)$ in eq.~(\ref{orient}) alone sets the orientation of the curve. Because $\theta(\tilde\theta)$ attains a maximum or minimum at an inflection point, $d\theta/d\tilde\theta$ necessarily changes sign, so the orientation of the curve as defined by (\ref{orient}) is reversed between inflection points. This confirms that a concave stretch of a bulk curve falls under the second case of eq.~(\ref{gendifferential}).

\begin{figure}[t!]
\centering
\begin{tabular}{ccccc}
\includegraphics[width=.28\textwidth]{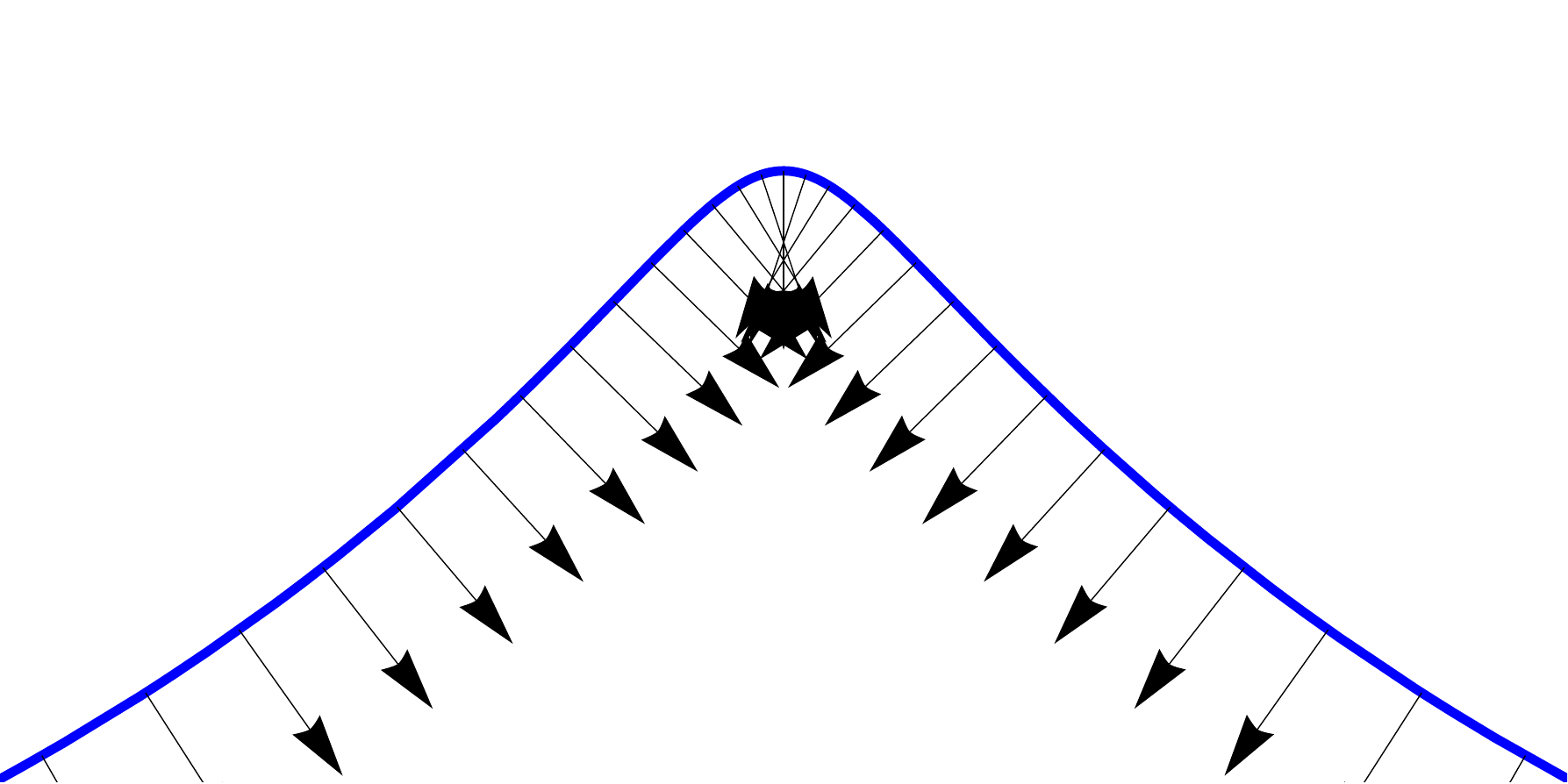} & 
\raisebox{1.2cm}{=} &
\includegraphics[width=.28\textwidth]{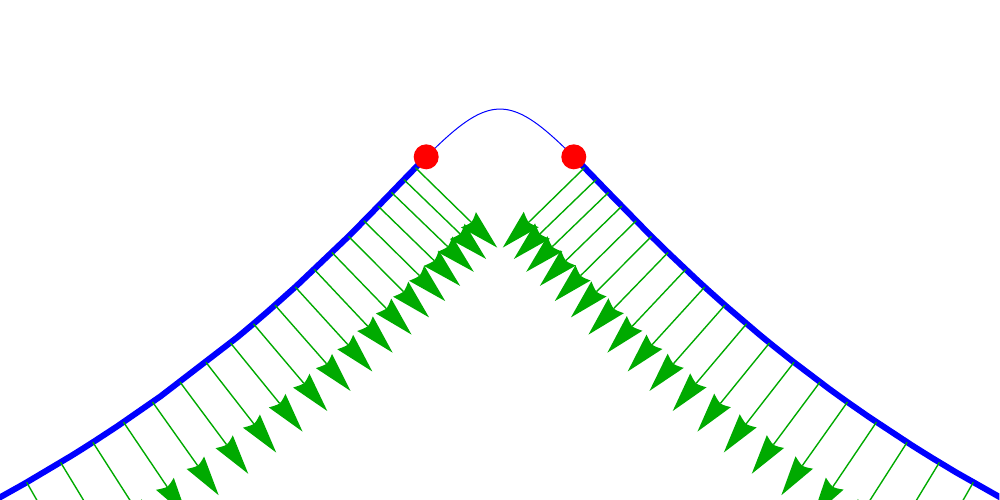} &
\raisebox{1.2cm}{-} &
\includegraphics[width=.28\textwidth]{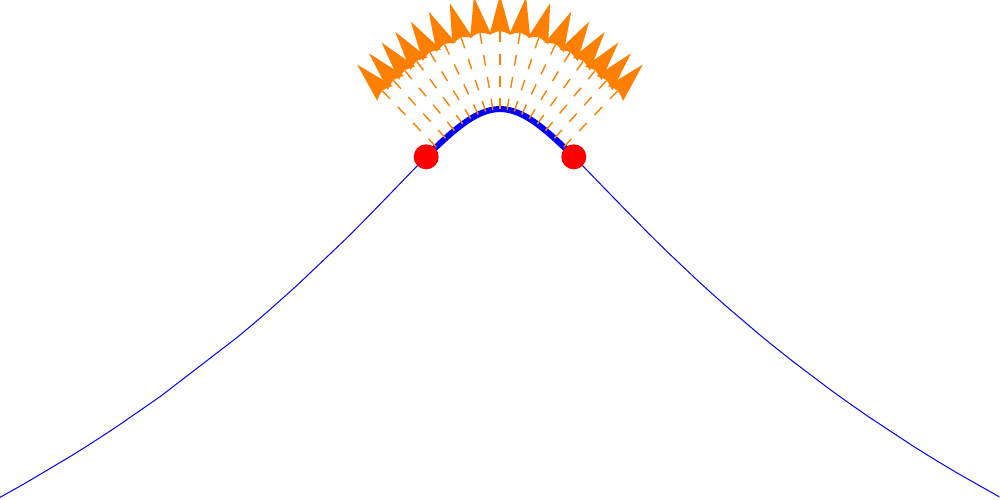}
\end{tabular}
\caption{The differential entropy formula computes the total length of a nonconvex curve thanks to two sign changes, one from a flip in orientation and one from reversing the direction of integration.}
\label{doubleflip}
\end{figure}

And yet, as we remarked in eq.~(\ref{sdifft}), the differential entropy of a nonconvex curve reproduces exactly the circumference of the curve, without an extra minus sign. To understand why the length of the curve between inflection points does not contribute negatively, return to condition~(\ref{orient}). As already remarked, it specifies in which direction the bulk curve is scanned as $\theta$ increases. However, in Sec.~\ref{nonconvex} we found that between inflection points $\theta$ is traversed from right to left. This means that after orientation reversal the bulk curve is again scanned in the correct direction -- from left to right. In this way, eq.~(\ref{sdifft}) is a consequence of two compensating sign changes, which are illustrated in Fig.~\ref{doubleflip}.

\section{Conical defect geometry}
\label{condef}
From the bulk point of view, Sec.~\ref{intersectpoint} constructs a point in AdS$_3$ by selecting the complete set of geodesics that pass through it. The input to our construction must therefore encompass (boundary avatars of) all spatial geodesics in the bulk. When the spacetime contains geodesics, which cannot be interpreted as entanglement entropies, it is necessary to supplement the set of all entanglement entropies with additional boundary data. 

This section presents the simplest example, where data beyond entanglement entropy first becomes necessary. The example is the conical defect geometry and the new data was named \emph{entwinement} in \cite{entwinement}. In the next section we will discuss the BTZ geometry, where the input to the hole-ographic construction will include yet another novel ingredient.

\subsection{Review of the conical defect geometry and entwinement} 
\begin{figure}[t!]
\centering
\begin{tabular}{ccc}
\includegraphics[width=.34\textwidth]{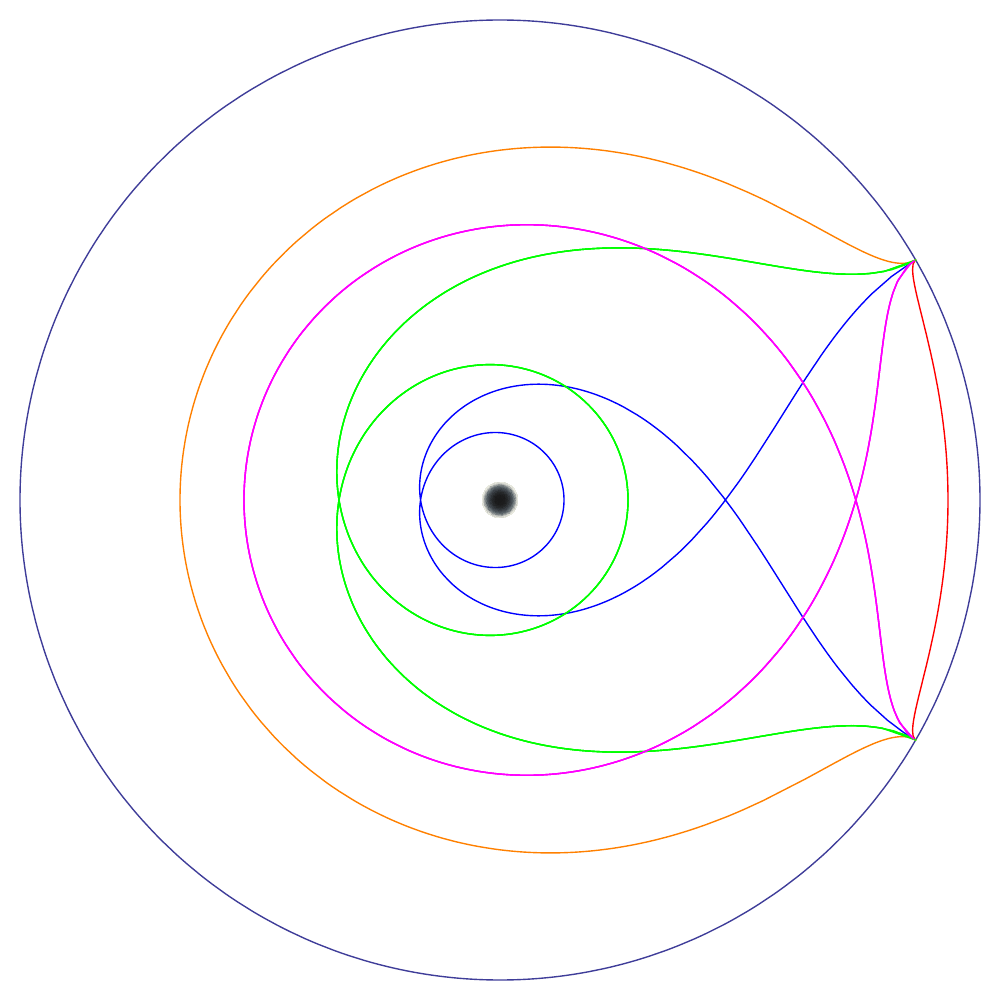} & \qquad\qquad\qquad &
\includegraphics[width=.34\textwidth]{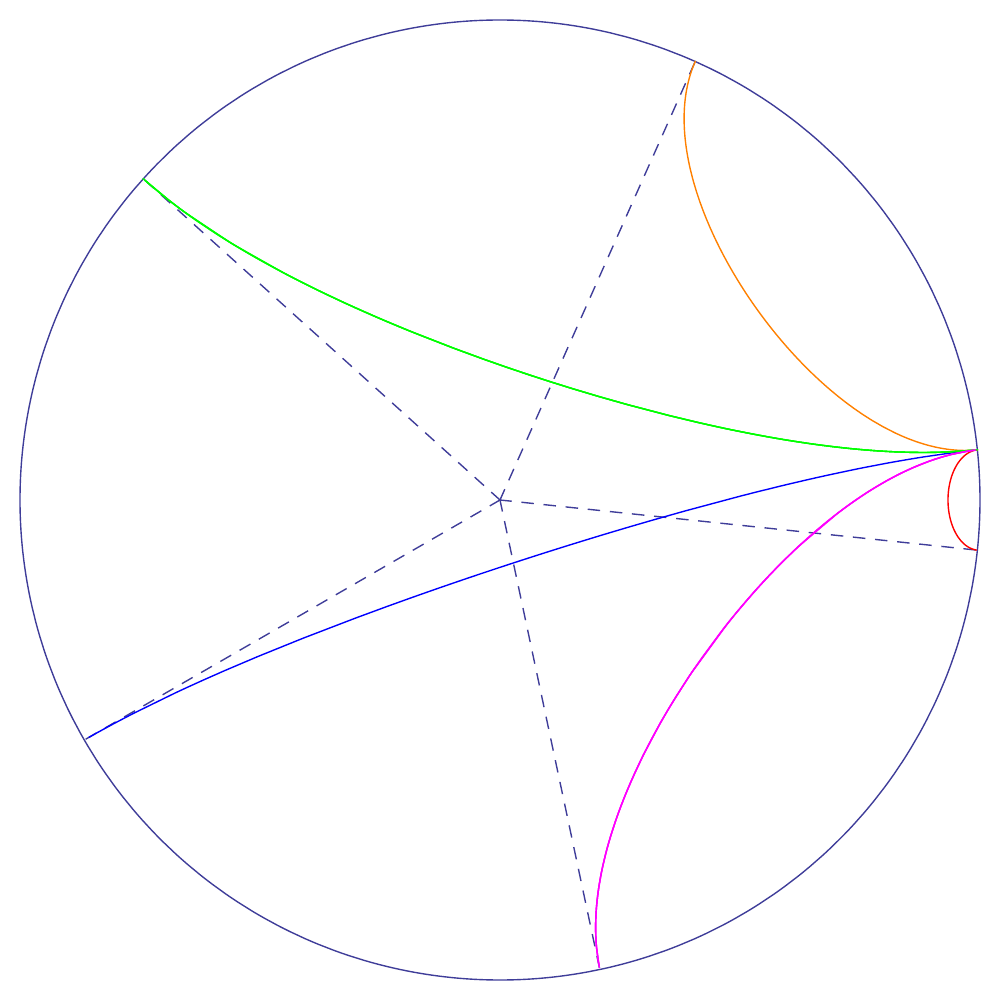}
\end{tabular}
\caption{A spatial slice of the conical defect geometry (left) and of its covering space (right); here $n=5$. Spatial geodesics in the defect geometry descend from geodesics in the covering space, where one endpoint ranges over a sequence of $(2 \pi / n)$-translates in metric~(\ref{ads3metric}).}
\label{geodesicspic}
\end{figure}

The conical defect metric (see e.g. \cite{matschull, luninmathur, cdvijay}) is:
\begin{equation}
ds^2 = - \left( \frac{1}{n^2} + \frac{r^2}{L^2}\right) dt^2 + \left( \frac{1}{n^2} + \frac{r^2}{L^2}\right)^{-1} dr^2 + r^2 d\tilde\vartheta^2 \,,
\label{defmetric}
\end{equation}
with $\tilde\vartheta$ ranging from $0$ to $2\pi$. To see the conical defect singularity at $r=0$, change coordinates according to
\begin{equation}
\tilde\vartheta = n \tilde\theta \qquad {\rm and} \qquad r = R / n \qquad {\rm and} \qquad t = n T
\label{rescalings}
\end{equation}
to obtain metric (\ref{ads3metric}), but with $\tilde\theta$ in the range from 0 to $2\pi/n$. When $n$ is an integer, this is AdS$_3/\mathbb{Z}_n$. Applying the inverse of this coordinate change to the spatial geodesic (\ref{adsgeodesic}) in AdS$_3$, we obtain the geodesics in the conical defect:
\begin{equation}
\tan^2\tilde\theta = \frac{R^2 \tan^2\tilde\alpha - L^2}{R^2 + L^2} 
\qquad \Leftrightarrow \qquad
\tan^2(\tilde\vartheta/n) = \frac{n^2 r^2 \tan^2(\alpha/n) - L^2}{n^2 r^2 + L^2} 
\label{defgeodesic}
\end{equation}
Here $\tilde\alpha$ is the opening angle (measured in $\tilde\theta$) of a geodesic in pure AdS$_3$, which becomes the geodesic of opening angle $\alpha$ (measured in $\tilde\vartheta$) via eq.~(\ref{rescalings}). The relation between the two is $\alpha = n \tilde\alpha$. The length of geodesic (\ref{defgeodesic}) in Planck units can likewise be read off from eq.~(\ref{entinterval}):
\begin{equation}
S(\alpha) = 
\frac{L}{2G} \log \left( \frac{2L}{\mu} \sin\tilde\alpha \right) = 
\frac{L}{2G} \log \left( \frac{2L}{\mu} \sin(\alpha/n) \right).
\label{deflength}
\end{equation}
Note that as $\tilde\alpha \to \tilde\alpha + 2 \pi/n$, $\alpha$ changes by $2\pi$, so an $\alpha$-arc connects the same pair of boundary points. As a consequence, in the conical defect geometry each pair of boundary points is connected by $n$ or $n-1$ distinct geodesics (if $n$ is an integer, there are $n$ distinct geodesics for every pair of points). Graphically, these geodesics can be obtained by fixing one endpoint in the covering space of the conical defect in coordinates (\ref{ads3metric}) and varying the other by $\Delta \tilde\theta = 2 \pi / n$; see Fig.~\ref{geodesicspic}.

\paragraph{Long geodesics and entwinement}
When $\alpha \leq \pi/2$, the geodesic (\ref{defgeodesic}) is the shortest path between its endpoints, so according to the Ryu-Takayanagi proposal it computes the entanglement entropy of an interval in the dual field theory. The geodesics with $\alpha > \pi/2$ connect the same pairs of points, but they are strictly longer. Because they fail the minimality condition of \cite{rt1, rt2}, they do not compute entanglement entropies of any boundary interval. We call such geodesics long geodesics.

Ref.~\cite{entwinement} proposed a boundary quantity, which reproduces lengths of long geodesics. This object, called entwinement, was designed to capture an analogue of entanglement appropriate for internal degrees of freedom, which may not be spatially organized. In a general system, one may diagnose the relevance of such degrees of freedom by inspecting the energy gaps: when they are smaller than the inverse size of the system, we may infer that the internal architecture of the theory, such as its internal symmetries or division into sectors, is as important as its spatial organization \cite{longstring}. The importance of entwinement in a holographic reconstruction of spacetime is not surprising. Indeed, bulk locality on sub-AdS scales is generally thought to be related to the internal structure of the dual field theory, such as its matrix degrees of freedom \cite{lenny99, Bousso:2005ie, Berenstein:2008eg, Asplund:2008xd, idseetal}.

On a technical level, a key challenge in defining entwinement is related to the presence of gauge symmetries. A discussion of similar difficulties in defining conventional entanglement entropy in gauge theories can be found in \cite{donnelly, entmaxwell, casinihuertaongaugeent, yuji, donnelly2}. For a technical definition and a thorough discussion of entwinement, we refer the reader to \cite{entwinement}. Here we content ourselves with stating that eq.~(\ref{deflength}) can be meaningfully extended to the range $\alpha > \pi/2$ and that the complete set of $S(\alpha)$s is in principle available in the dual field theory -- in the form of entanglement entropies and entwinements. Just like entanglement entropy, entwinement is a concave function of $\alpha$:
\begin{equation}
\frac{d^2S}{d\alpha^2} < 0.
\end{equation}

\subsection{Conical defect hole-ography}
\label{cdholes}
In Sec.~\ref{rev} we collected some hole-ographic formulas, which pertain to pure AdS$_3$. Here we briefly list their conical defect counterparts. As before, $\vartheta$ is the angular coordinate on the boundary while $\tilde\vartheta$ is the coordinate in the bulk.

The bulk-to-boundary map can be obtained from eqs.~(\ref{tanalpha}-\ref{tantheta}) and relations (\ref{rescalings}):
\begin{eqnarray}
\tan \big(\alpha(\tilde\vartheta)/n\big) & = & \frac{L}{nr}\, \sqrt{1 + \frac{n^2 L^2}{n^2 r^2 + L^2} \left( \frac{d \log r}{d\tilde\vartheta}\right)^2} \label{cdtanalpha} \\
\tan\,\frac{\tilde\vartheta - \vartheta(\tilde\vartheta)}{n} & = & \frac{n L^2}{n^2 r^2 + L^2}  \frac{d \log r}{d\tilde\vartheta} \label{cdtantheta}
\end{eqnarray}
Similarly, the boundary-to-bulk map becomes:
\begin{eqnarray}
r(\vartheta) & = & \frac{L}{n} \cot \big(\alpha(\vartheta)/n\big) \, \sqrt{\frac{1 + \alpha'(\vartheta)^2 \tan^2\! \big(\alpha(\vartheta)/n\big)}{1-\alpha'(\theta)^2}} \label{invrcd} \\
\tan\,\frac{\vartheta - \tilde\vartheta(\vartheta)}{n} & = & \alpha'(\vartheta) \tan\big(\alpha(\vartheta)/n\big)  
\label{invthetacd}
\end{eqnarray}
The differential entropy formula continues to hold in its most general form:
\begin{equation}
\pm \frac{\rm length}{4G} = \frac{\pm 1}{4G} \int_{\tilde\vartheta_i}^{\tilde\vartheta_f} d\tilde\vartheta\, 
\sqrt{\left(\frac{1}{n^2} + \frac{r^2}{L^2}\right)^{-1} \!\! \left(\frac{dr}{d\tilde\theta}\right)^2 + r^2} =
\frac{1}{2} \int_{\vartheta_i}^{\vartheta_f} d\vartheta\, \frac{dS(\alpha)}{d\alpha}\Big|_{\alpha = \alpha(\vartheta)} + f(\tilde\vartheta_f) - f(\tilde\vartheta_i)
\label{opencd}
\end{equation}
The sign ambiguity is related to the issue of orientation, which we discussed in Sec.~\ref{intkink}.
The function $f$ is related to its AdS$_3$ precursor in the obvious way:
\begin{equation}
f(\tilde\vartheta) = \frac{L}{8G} \log\frac{\sin \big( \alpha(\tilde\vartheta) + \tilde\vartheta - \vartheta(\tilde\vartheta)\big)/n}{\sin \big( \alpha(\tilde\vartheta) - \tilde\vartheta + \vartheta(\tilde\vartheta)\big)/n}
\label{deffcd}
\end{equation}

\paragraph{Relevance of entwinement} It is interesting to observe which curves can be described in terms of entanglement alone and which require the introduction of entwinement. Entwinement makes an appearance whenever
\begin{equation}
\frac{L}{nr}\, \sqrt{1 + \frac{n^2 L^2}{n^2 r^2 + L^2} \left( \frac{d \log r}{d\tilde\vartheta}\right)^2} > \tan(\pi/2n),
\label{entcritical}
\end{equation}
that is when eq.~(\ref{cdtanalpha}) returns $\alpha (\tilde\vartheta) > \pi/2$. Specifically, curves that are nearly radial tend to involve entwinement; the critical ``slope'' $dr/d\tilde\vartheta$ as a function of $r$ can be read off from eq.~(\ref{entcritical}). However, within a coordinate distance 
\begin{equation}
r < \frac{L}{n} \cot(\pi/2n)
\label{rcrit}
\end{equation}
of the singularity every curve is described entirely in terms of entwinement. 

\paragraph{Domain of $\alpha(\vartheta)$.} When the bulk curve encircles the singularity, the boundary function $\alpha(\vartheta)$ is defined over a range of $\vartheta$ of width $2\pi$. For example, a circle $r = r_0$ has:
\begin{equation}
\alpha(\vartheta) = n \cot^{-1} \frac{n r_0}{L} \qquad {\rm with} \quad -\pi \leq \vartheta < \pi.
\end{equation}
In contrast, the boundary function of a closed bulk curve that does not surround the singularity is defined over a range of width $2\pi n$. One example is the curve obtained with eqs.~(\ref{invrcd}-\ref{invthetacd}) from the function:
\begin{equation}
\alpha(\vartheta) = n \cos^{-1} \frac{nr \cos(\vartheta/n)}{\sqrt{n^2 r^2 + L^2}} - p
\qquad {\rm with} \quad -n \pi \leq \vartheta < n \pi\,.
\label{outexample}
\end{equation}
The ranges of $\vartheta$ (outside $-\pi \leq \vartheta \leq \pi$) and of $\alpha$ (outside $-\pi/2 \leq \alpha \leq \pi/2$) reflect the fact that any closed curve which does not encircle the singularity necessarily violates condition~(\ref{entcritical}) and involves entwinement. 

\begin{figure}[t]
\centering
\includegraphics[width=.47\textwidth]{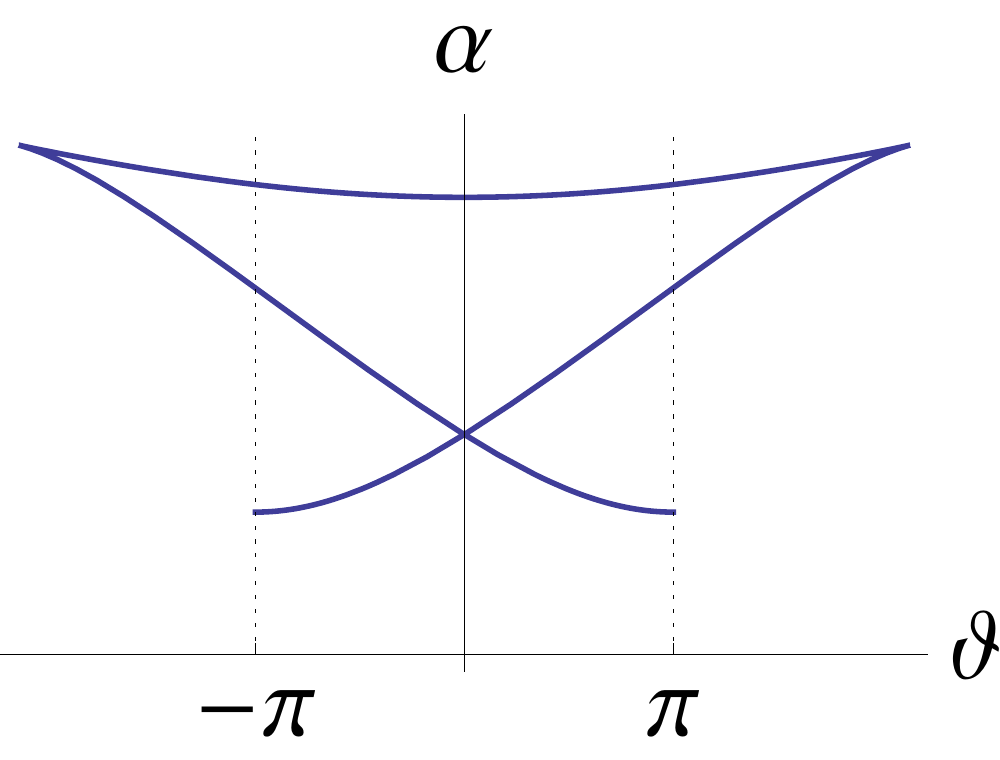}
\caption{The boundary function $\alpha(\vartheta(\tilde\vartheta))$ of the curve defined in eq.~(\ref{funnyexample}); here $n = 5$. Because the curve surrounds the singularity, going around the curve once increases $\vartheta(\tilde\vartheta)$ by $2\pi$. However, $\vartheta(\tilde\vartheta)$ leaves the range $(-\pi, \pi)$ over the concave segment of the curve.}
\label{funnygraph}
\end{figure}

In Sec.~\ref{nonconvex} we saw that nonconvex bulk curves produce boundary functions, which are not single-valued: on concave segments the plot of $\alpha(\vartheta(\tilde\vartheta))$ backtracks toward decreasing $\vartheta$. This means that even if a bulk curve goes around the singularity, $\vartheta$ can intermittently go outside the range $(-\pi, \pi)$. This occurs when the curve has a sufficiently concave segment. After sweeping the full curve, however, we still have $\vartheta_{\rm final} - \vartheta_{\rm initial} = 2\pi$. An example of this is the curve
\begin{equation}
r(\tilde\vartheta) =  \frac{L}{n}\, (1.2 - \cos \tilde\vartheta)
\qquad ({\rm for} \quad n \geq 3)
\label{funnyexample}
\end{equation}
whose $\alpha(\vartheta)$ is shown in Fig.~\ref{funnygraph}. In all these cases, the circumference of the curve is correctly reproduced by the differential entropy formula:
\begin{equation}
\label{sdifftt}
E[\alpha(\vartheta(\tau))] = \frac{1}{2} \oint d\tau\, \frac{d\vartheta}{d\tau} \, \frac{dS(\alpha)}{d\alpha}\Big|_{\alpha = \alpha(\vartheta(t))} = \frac{\rm circumference}{4G} 
\end{equation}
Here $\tau$ is a parameter along the bulk curve, which allows us to account for backtracking in $\vartheta$ occasioned by concave regions.

\subsection{Points and distances}
\label{ptscd}
\paragraph{Points} We define them as discussed in Sec.~\ref{pointsent}, that is by extremizing action:
\begin{equation}
I = \oint d\vartheta\, \sqrt{-\frac{d^2S}{d\alpha^2}\big(1 - \alpha'(\vartheta)^2\big)}
= \frac{L}{2nG} \int_0^{2\pi n} \frac{d\vartheta \, \sqrt{1 - \alpha'(\vartheta)^2}}{\sin\big(\alpha(\vartheta)/n\big)}
\label{cdaction}
\end{equation} 
A computation similar to Appendix~\ref{extrinsic} and eqs.~(\ref{cdtanalpha}-\ref{cdtantheta}) confirm that the Lagrangian is equal to $(d\tau/d\vartheta) \sqrt{h}\, K$ of the corresponding bulk curve. The equation of motion (\ref{eom}) becomes:
\begin{equation}
n \alpha_A''
+ \big(1-\alpha_A'^2\big) \cot(\alpha_A/n) = 0
\label{cdeom}
\end{equation}
The solutions of (\ref{cdeom}) are the $p \to 0$ limits of eq.~(\ref{outexample}):
\begin{equation}
\alpha_A(\vartheta) = n \cos^{-1} \frac{nr_A \cos\big((\vartheta - \tilde\vartheta_A)/n\big)}{\sqrt{n^2 r_A^2 + L^2}}
\qquad {\rm with} \quad -n \pi \leq \vartheta < n \pi\,.
\label{cdpoint}
\end{equation}
Note that these are very different from points in pure AdS$_3$. For example, consider $\alpha_O(\theta) = \pi/2$, which describes the point at the origin in AdS$_3$. If we substitute it into eqs.~(\ref{invrcd}-\ref{invthetacd}), the resulting bulk curve is a circle at the critical radius (\ref{rcrit}). 

\paragraph{Distances} We employ the definition from Sec.~\ref{secdistance}:
\begin{equation}
d(A,B) = \frac{1}{4} \int_0^{2\pi n} d\vartheta\, \frac{dS(\alpha)}{d\alpha}\Big|_{\alpha = \min\{\alpha_A(\vartheta), \alpha_B(\vartheta)\}}
\label{cddistfull}
\end{equation}
An astute reader may remark that in the conical defect geometry two points $A$ and $B$ are generically connected by more than one geodesic. Which geodesic distance is recovered by eq.~(\ref{cddistfull})? Observe that every solution $\alpha_A(\vartheta)$ of (\ref{cdeom}) has a minimum. From the bulk point of view this minimum falls at $\vartheta = \tilde\vartheta_A$, the angular coordinate of the bulk point $A$. In consequence, redefining
\begin{equation}
\alpha_A(\vartheta) \to \alpha_A(\vartheta + 2\pi k)
\label{extrawind}
\end{equation}
leaves the bulk point unchanged. If we do this while keeping $\alpha_B(\vartheta)$ fixed, however, we affect the integral (\ref{cddistfull}), because the ranges where $\alpha_{A,B}(\vartheta)$ are minimal get altered. Now recall from Sec.~\ref{endpt} that a discontinuity in the derivative of the boundary function means that the bulk curve follows a geodesic for a finite distance. In the case of $\gamma(\vartheta) = \min\{\alpha_A(\vartheta), \alpha_B(\vartheta)\}$, 
such a discontinuity occurs at the location $\vartheta_{\rm int}$ where $\alpha_A(\vartheta)$ and $\alpha_B(\vartheta)$ intersect. This location selects the geodesic along which $A$ connects to $B$. Translations (\ref{extrawind}) lead to different $\vartheta_{\rm int}$, so they yield different geodesic distances between $A$ and $B$. The minimal distance is recovered when the minima of the two boundary functions are no more than $\pi$ apart.

The distance can be equivalently computed in a slightly different way, which we will find useful in Sec.~\ref{btzthermal}. 
Consider the integral (\ref{cddistfull}) extended not across the full $2\pi n$-range of $\vartheta$, but only from
\begin{equation}
\tilde\vartheta_A = \textrm{minimum of }\alpha_A(\vartheta) 
\qquad {\rm to} \qquad
\tilde\vartheta_B = \textrm{minimum of }\alpha_B(\vartheta).
\label{trickrange}
\end{equation}
As discussed in Sec.~\ref{endpt}, this segment of the boundary function describes an open bulk curve with endpoints at $A$ and $B$. By construction, the curve traverses from $A$ to $B$ along a geodesic selected by $\vartheta_{\rm int}$, so its length is exactly $d(A,B)$. The endpoint conditions (\ref{trickrange}) ensure that the inhomogeneous terms $f$ in eq.~(\ref{opencd}) do not appear, so we have:
\begin{equation}
d(A,B) = 
\frac{1}{2} \left| 
\int_{\tilde\vartheta_A}^{\vartheta_{\rm int}} d\vartheta\, \frac{dS(\alpha)}{d\alpha}\Big|_{\alpha = \alpha_A(\vartheta)} 
+ 
\int_{\vartheta_{\rm int}}^{\tilde\vartheta_B} d\vartheta\, \frac{dS(\alpha)}{d\alpha}\Big|_{\alpha = \alpha_B(\vartheta)} 
\right|
\label{cddisttrick}
\end{equation}
We write formula (\ref{cddisttrick}) explicitly as a sum of two integrals in order to account for special cases, where $\vartheta_{\rm int}$ is greater than (or less than) both $\tilde\vartheta_A$ and $\tilde\vartheta_B$.

\section{The static BTZ spacetime}
\label{btzthermal}
\begin{figure}[t!]
\centering
\begin{tabular}{cc}
\includegraphics[width=.48\textwidth]{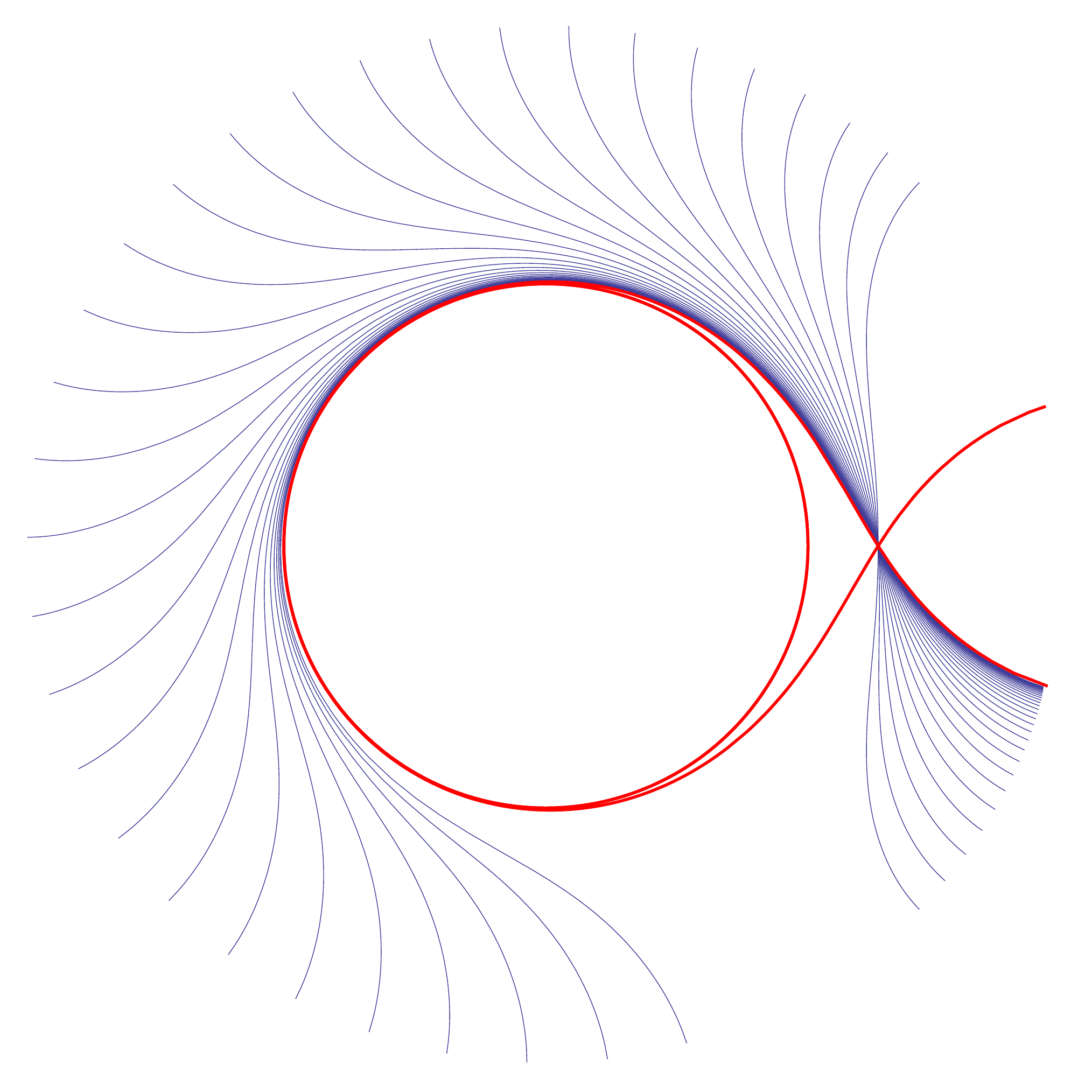} & 
\includegraphics[width=.48\textwidth]{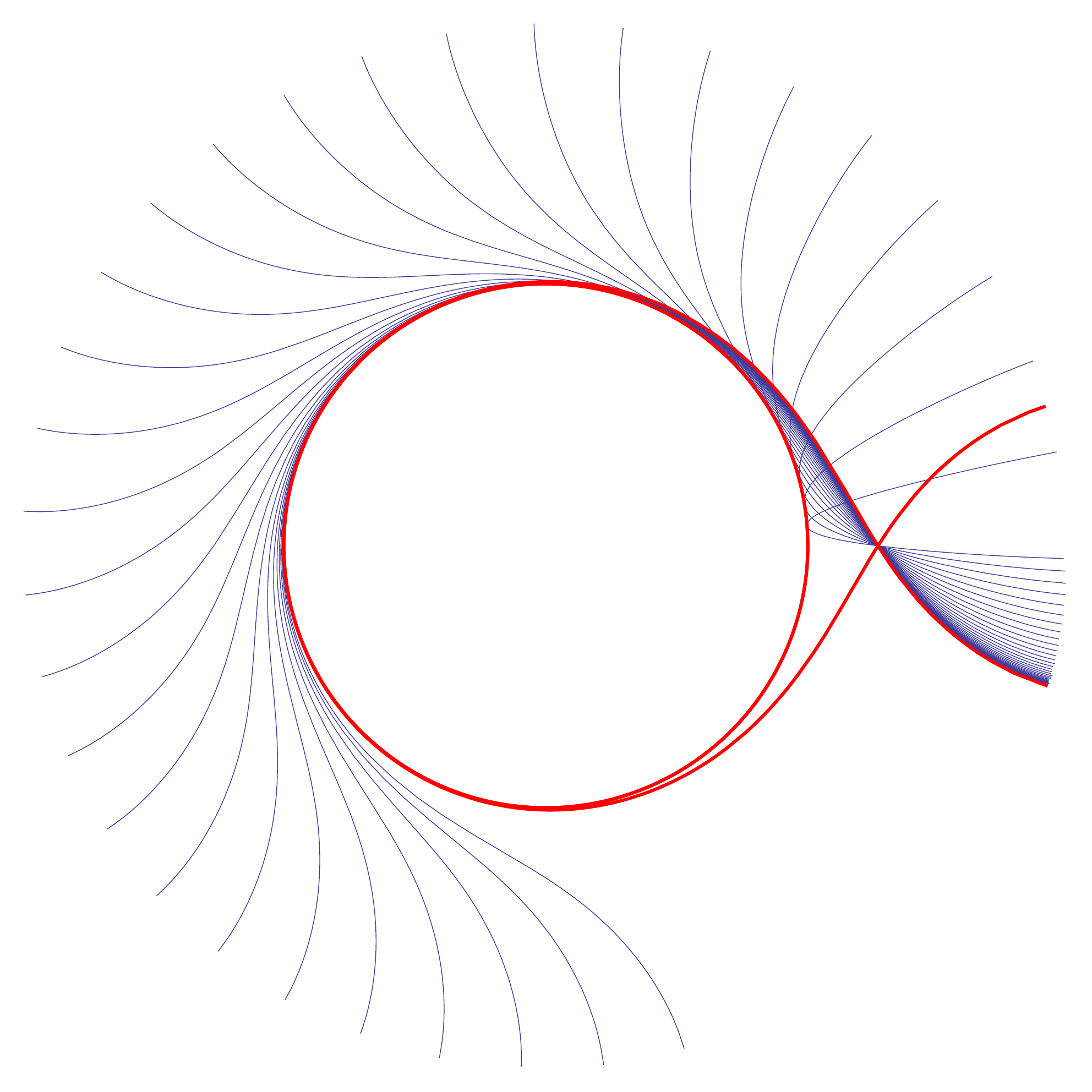}
\end{tabular}
\caption{Geodesics of type (\ref{btzgeod1}) (left panel) and of type (\ref{btzgeod2}) (right panel), which pass through the same bulk point $A$. The two families span complementary directions, along which $A$ may be passed. We only display geodesics with opening angles $\alpha, \beta < 3 \pi/4$, but $\alpha, \beta$ can be arbitrarily large, which produces winding around the black hole horizon. In contrast to geodesics (\ref{btzgeod2}), geodesics (\ref{btzgeod1}) do not touch the horizon, though this is obscured on the graph. The critical geodesic (\ref{btzgeodinf}), which is obtained by sending $\alpha, \beta \to \infty$, is shown in red. It circles around the black hole infinitely many times.}
\label{BTZgeodesics}
\end{figure}

The BTZ geometry \cite{btzoriginal, btzreview} contains a qualitatively novel type of spatial geodesics: ones that connect the asymptotic boundary with the horizon and the second asymptotic region beyond it (Fig.~\ref{BTZgeodesics}). These geodesics have been used before to probe the region behind the horizon \cite{excursions}. The hole-ographic construction of a static slice of the BTZ geometry uses all spatial geodesics: the ones dual to entanglement entropies, the long geodesics analogous to entwinement, and the 2-sided geodesics. In this section we assume that all this data is in principle available in the boundary field theory and in its purification. Including the purification is dual to extending the BTZ geometry to a second asymptotic region \cite{Horowitz:1998xk, vijayprobes, juanbtz}. 

\subsection{The BTZ geometry}
The metric of a static BTZ black hole with temperature $R_+/2\pi$ is:
\begin{equation}
ds^2 = - \frac{R^2 - R_+^2}{L^2}\, dT^2 + \frac{L^2}{R^2 - R_+^2}\, dR^2 + R^2 d\tilde\theta^2\,.
\label{btzmetric}
\end{equation}
From here on we will set $L \equiv 1$ to avoid cluttering the formulae. Geodesics on a static slice fall into two classes (see Fig.~\ref{BTZgeodesics}):
\begin{eqnarray}
\!\!\!\!\!\!\!\!\!\!
R = R_+ \frac{\cosh (R_+ \alpha)}{\sqrt{\sinh^2 (R_+ \alpha) - \sinh^2 (R_+ (\tilde\theta\!-\!\theta))}} 
& \!\Leftrightarrow\! &
\tanh^2(R_+ (\tilde\theta\!-\!\theta)) = \frac{R^2 \tanh^2(R_+ \alpha) - R_+^2}{R^2-R_+^2}\,
\label{btzgeod1} \\
\!\!\!\!\!\!\!\!\!\!
R = R_+ \frac{\sinh (R_+ \beta)}{\sqrt{\cosh^2 (R_+ \beta) - \cosh^2 (R_+ (\tilde\theta\!-\!\theta)}}
& \!\Leftrightarrow\! &
\coth^2(R_+ (\tilde\theta\!-\!\theta)) = \frac{R^2 \coth^2(R_+ \beta) - R_+^2}{R^2-R_+^2}\,
\label{btzgeod2}
\end{eqnarray}
They are unified under the identification
\begin{equation}
\beta \to \alpha + \frac{i \pi}{2 R_+} \qquad {\rm and} \qquad \theta \to \theta + \frac{i \pi}{2 R_+}\,,
\label{imshift}
\end{equation}
which keeps the left endpoint of a boundary interval, $\theta - \alpha$, real. Geodesics (\ref{btzgeod1}) are similar to the geodesics we encountered in pure AdS$_3$ and the conical defect. They subtend the angular wedge $-\alpha < \tilde\theta < \alpha$ and penetrate the bulk down to a depth $R_{\rm min} = R_+ \coth(R_+\alpha)$. Geodesics (\ref{btzgeod2}) likewise subtend the angular wedge $-\beta < \tilde\theta < \beta$, but they reach down to the horizon for all values of $\beta$. Though eq.~(\ref{btzgeod2}) is differentiable everywhere, this is a coordinate artifact. If we parameterize the radial coordinate by proper length, $\rho = \cosh^{-1} (R/R_+)$, then
\begin{equation}
\frac{d\rho}{d\tilde\theta}\Big|_{\theta} = \lim_{R \to R_+} \frac{d\rho}{dR}\, \frac{dR}{d\tilde\theta} = \lim_{R \to R_+} \frac{1}{\sqrt{R^2 - R_+^2}} \frac{R_+ \sqrt{R^2 - R_+^2}}{\sinh^2(R_+\beta)} = \frac{R_+}{\sinh^2 (R_+ \beta)}\,.
\label{crosshorizon}
\end{equation}
We see that the geodesics intersect the horizon rather than skirting it tangentially. The special case $\beta = 0$ is the radial geodesic. In this way, eq.~(\ref{btzgeod2}) actually describes two halves of mutually intersecting geodesics rather than a single geodesics.

The regulated length (in units of $4G$) of geodesic (\ref{btzgeod1}) is:
\begin{equation}
S(\alpha) = \frac{1}{2G} \log \frac{2R_+ \sinh(R_+ \alpha)}{\mu} \,.
\label{btzlength}
\end{equation}
We will need a similar expression for geodesics (\ref{btzgeod2}), but we have to be careful about regularizing the length. When we impose a radial cutoff at $R_* = R^2_+/\mu$, the length behaves analytically. Thus, we can analytically continue eq.~(\ref{btzlength}) according to (\ref{imshift}) to obtain:
\begin{equation}
S(\beta) = \frac{1}{2G} \log \frac{2R_+ \cosh(R_+ \beta)}{\mu} + \frac{1}{2G} \cdot \frac{i\pi}{2}
\label{btzlength2}
\end{equation}
As explained in \cite{excursions}, the imaginary piece is associated with the geodesic crossing the horizon and continuing into the second asymptotic region. If we ask strictly for the length of curve (\ref{btzgeod2}), which stays in the same asymptotic region before and after reaching the horizon, we drop the imaginary piece. There is yet another sense in which we could compute the length of a $\beta$-type geodesic: we can start from $\tilde\theta = \theta$ and ask for the segment of the geodesic contained in a fixed angular wedge. This is what we used to define the functions $f$ in eqs.~(\ref{deff}) and (\ref{deffcd}), which allowed us to compute lengths of curves with endpoints. For this question, we discover that:
\begin{equation}
\left(\frac{d\,\rm{length}}{d\tilde\theta}\right)^2 = 
\frac{(dR_\alpha / d\tilde\theta)^2}{R_\alpha^2 - R_+^2} + R_\alpha^2 = \frac{(dR_\beta / d\tilde\theta)^2}{R_\beta^2 - R_+^2} + R_\beta^2
\qquad {\rm for}~\alpha = \beta.
\label{samef}
\end{equation}
Consequently, we will encounter only one function $f$, which will apply to all differentiable open curves, regardless of whether the tangent geodesic at endpoint is of type (\ref{btzgeod1}) or (\ref{btzgeod2}).

There is a critical geodesic, which belongs to both classes as a limit $\alpha \to \infty$ and $\beta \to \infty$. It circles around the black hole infinitely many times before reaching the horizon tangentially. Its explicit form is:
\begin{equation}
\label{btzgeodinf}
R = R_+ \left(1 - e^{-2 R_+ \tilde\theta}\right)^{-1/2}.
\end{equation}

\paragraph{Interpretation} For small $\alpha$, $S(\alpha)$ computes the entanglement entropy in the thermal state of a boundary interval of angular size $2\alpha$.
This interpretation persists up to the critical scale
\begin{equation}
\alpha_* = \frac{1}{R_+} \coth^{-1} \big(2 \coth (R_+ \pi) - 1\big) ,
\end{equation}
where the Araki-Lieb inequality $|S(\alpha) - S(\pi - \alpha)| \leq 2\pi R_+/4G$ is saturated \cite{plateau}.
For $\alpha > \alpha_*$ the minimal surface in the Ryu-Takayanagi prescription is not the geodesic (\ref{btzgeod1}), but the union of the black hole horizon and the geodesic spanning the complementary interval. In this regime it is not clear how to extract $S(\alpha)$ from the boundary theory most cleanly. One option is to associate it with subleading saddle points in the semiclassical computation of a boundary two-point function. Another possibility is to extend the definition of entwinement \cite{entwinement} to the thermal state. In this paper we simply assume that $S(\alpha)$ can in principle be extracted from the boundary theory, regardless of the precise algorithm to do so.

The geodesics (\ref{btzgeod2}) are best interpreted in the two-sided extension of the BTZ black hole, which is dual to the thermofield double state \cite{Horowitz:1998xk, vijayprobes, juanbtz}. In the semiclassical approximation, computing a two-sided correlator in the thermofield double state reduces to finding the shortest geodesic connecting the insertion points, a fact that was previously used to study black hole interiors in \cite{excursions}. This interpretation applies directly to geodesics with $\beta \leq \pi/2$, which are the shortest routes from one boundary to the other. For $\beta > \pi/2$, we may think of $S(\beta)$ as determining the subleading saddle points in the same semiclassical computation.

\subsection{BTZ hole-ography}
We collect here BTZ analogues of formulas given in Secs.~\ref{rev} and \ref{cdholes}. The bulk-to-boundary map can be obtained in a variety of ways, one of which is the continuation $n^{-1} \to i R_+$:
\begin{eqnarray}
\tanh \big(R_+ \alpha(\tilde\theta)\big) & = & \frac{R_+}{R}\, \sqrt{1 + \frac{1}{R^2 - R_+^2} \left( \frac{d \log R}{d\tilde\theta}\right)^2} \label{btztanalpha} \\
\tanh \big(R_+ (\tilde\theta - \theta(\tilde\theta))\big) & = & \frac{R_+}{R^2 - R_+^2}  \frac{d \log R}{d\tilde\theta} \label{btztantheta}
\end{eqnarray}
These equations pertain to those stretches of bulk curves, which are tangent to geodesics~(\ref{btzgeod1}). In this regime the right hand side of eq.~(\ref{btztanalpha}) is smaller than 1. In the borderline case, where it equals 1, the curve becomes tangent to the critical geodesic~(\ref{btzgeodinf}). In fact, the simplest way of finding (\ref{btzgeodinf}) is to solve:
\begin{equation}
\frac{R_+}{R}\, \sqrt{1 + \frac{1}{R^2 - R_+^2} \left( \frac{d \log R}{d\tilde\theta}\right)^2} = 1.
\label{steepness}
\end{equation}
When this expression is greater than 1, the bulk curve becomes tangent to geodesics from family (\ref{btzgeod2}). We shall refer to this regime as ``steep.'' Because geodesics (\ref{btzgeod1}) and (\ref{btzgeod2}) are related to one another by the simple exchange $\tanh \leftrightarrow \coth$, in the steep regime the boundary-to-bulk map becomes:
\begin{eqnarray}
\coth \big(R_+ \beta(\tilde\theta)\big) & = & \frac{R_+}{R}\, \sqrt{1 + \frac{1}{R^2 - R_+^2} \left( \frac{d \log R}{d\tilde\theta}\right)^2} \label{btztanalpha2} \\
\coth \big(R_+ (\tilde\theta - \theta(\tilde\theta))\big) & = & \frac{R_+}{R^2 - R_+^2}  \frac{d \log R}{d\tilde\theta} \label{btztantheta2}
\end{eqnarray}

The inverse relations to eqs.~(\ref{btztanalpha}-\ref{btztantheta}) are:
\begin{eqnarray}
R(\theta) & = & R_+ \sqrt{\frac{\coth^2 \big(R_+ \alpha(\theta)\big) - \alpha'(\theta)^2}{1-\alpha'(\theta)^2}} \label{invrbtz} \\
\tanh\,\left(R_+(\theta - \tilde\theta(\theta))\right) & = & \alpha'(\theta) \tanh\big(R_+ \alpha(\theta)\big)  
\label{invthetabtz}
\end{eqnarray}
For example, the function $\alpha(\theta) = \pi/2$, which describes the point at the origin in pure AdS$_3$, now corresponds to a circle $R = R_+ \coth (R_+ \pi/2)$. For $R_+ = 1 \, (= L)$, this is a geodesic distance of $0.42L$ from the horizon. We referred to this fact in the introduction. Note, however, that this distance tends to 0 in the limit of large horizon size.

Eqs.~(\ref{btztanalpha2}-\ref{btztantheta2}) have their own inverses:
\begin{eqnarray}
R(\theta) & = & R_+ \sqrt{\frac{\beta'(\theta)^2 - \tanh^2 \big(R_+ \beta(\theta)\big)}{\beta'(\theta)^2-1}} \label{invrbtz2} \\
\coth\,\left(R_+(\theta - \tilde\theta(\theta))\right) & = & \beta'(\theta) \coth\big(R_+ \beta(\theta)\big)  
\label{invthetabtz2}
\end{eqnarray}
While in the non-steep regime we continue to impose condition (\ref{consist}), in the steep regime we have $|\beta'(\theta)| > 1$.

A bulk curve whose radial derivative never exceeds (\ref{steepness}) can be represented on the boundary as a function $\alpha(\theta)$ or, in case of concavities, as a multi-valued $\alpha(\theta(\tilde\theta))$ as discussed in Sec.~\ref{nonconvex}. For such a curve, the differential entropy formula reads:
\begin{equation}
\pm \frac{\rm length}{4G} = \frac{\pm 1}{4G} \int_{\tilde\theta_i}^{\tilde\theta_f} d\tilde\theta\, 
\sqrt{\left(R^2 - R_+^2\right)^{-1} \!\! \left(\frac{dR}{d\tilde\theta}\right)^2 + R^2} =
\frac{1}{2} \int_{\theta_i}^{\theta_f} d\theta\, \frac{dS(\alpha)}{d\alpha}\Big|_{\alpha = \alpha(\theta)} + f(\tilde\theta_f) - f(\tilde\theta_i)
\label{openbtz}
\end{equation}
The sign ambiguity, which is related to the orientation of the curve, was discussed in Sec.~\ref{intkink}.
The nonhomogeneous term is again the length of the geodesic tangent at the endpoint, which is contained in the angular wedge $\theta_f < \tilde\theta < \tilde\theta_f$ (and likewise for the initial point):
\begin{equation}
f(\tilde\theta) = \frac{1}{8G} \log\frac{\sinh R_+ \big( \alpha(\tilde\theta) + \tilde\theta - \theta(\tilde\theta)\big)}{\sinh R_+ \big( \alpha(\tilde\theta) - \tilde\theta + \theta(\tilde\theta)\big)}
\label{deffbtz}
\end{equation}
Similarly, a curve that is everywhere steep can be represented on the boundary with  $\beta(\theta)$. Of course such a curve can never be closed; this can be seen on the boundary from $|\beta'(\theta)| > 1$. In this case the length of the curve is computed by a direct analogue of eq.~(\ref{openbtz}):
\begin{equation}
\pm \frac{\rm length}{4G} = 
\frac{1}{2} \int_{\theta_i}^{\theta_f} d\theta\, \frac{dS(\beta)}{d\beta}\Big|_{\beta = \beta(\theta)} + f(\tilde\theta_f) - f(\tilde\theta_i)
\label{openbtz2}
\end{equation}
Because of eq.~(\ref{samef}), we do not analytically continue (\ref{deffbtz}), but simply substitute $\alpha(\tilde\theta) \to \beta(\tilde\theta)$ into the hyperbolic sines. However, the different dependence of $\theta$ on $\tilde\theta$ in eqs.~(\ref{btztantheta}) and (\ref{btztantheta2}) means that $f(\tilde\theta)$ behaves differently in the $\alpha$- and $\beta$-branch. This will be important in our discussion of points below.

The most general bulk curve contains both types of segments -- steep and non-steep. If the curve is differentiable, the transition can only occur at $\alpha = \beta = \theta = \infty$: this is the only place where we can have $\alpha'(\theta) \to 1$ at finite $R$. Thus, in the vicinity of a steep / non-steep transition the differential entropy formula becomes:
\begin{equation}
\pm \frac{\rm length}{4G} = - f(\tilde\theta_i) +
\frac{1}{2} \int_{\theta_i}^\infty d\theta\, \frac{dS(\alpha)}{d\alpha}\Big|_{\alpha = \alpha(\theta)} +
\frac{1}{2} \int_{\infty}^{\theta_f} d\theta\, \frac{dS(\beta)}{d\beta}\Big|_{\beta = \beta(\theta)} + f(\tilde\theta_f)
\label{openbtzgen}
\end{equation}
Both integrals diverge in the same way at their infinite limits, but the divergences come with opposite signs.

\subsection{Points and distances}
\label{ptsbtz}

\paragraph{Points} We follow the same steps as in Sec.~\ref{ptscd}. To define points, we extremize action (\ref{action}):
\begin{equation}
I = \int d\theta\, \sqrt{-\frac{d^2S}{d\alpha^2}\big(1 - \alpha'(\theta)^2\big)}
= \frac{R_+}{2G} \int \frac{d\theta \, \sqrt{1 - \alpha'(\theta)^2}}{\sinh\big(R_+ \alpha(\theta)\big)}
\label{btzaction}
\end{equation} 
As in the conical defect case, eqs.~(\ref{btztanalpha}-\ref{btztantheta}) and a computation analogous to Appendix~\ref{extrinsic} establish that the Lagrangian equals $(d\tau/d\theta) \sqrt{h}\, K$ of the corresponding bulk curve. The equation of motion becomes:
\begin{equation}
\alpha_A''
- \big(1-\alpha_A'^2\big) R_+ \coth(R_+ \alpha_A) = 0
\label{btzeom}
\end{equation}
The solutions are:
\begin{equation}
\alpha_A(\theta) = \frac{1}{R_+} \cosh^{-1} \frac{R_A \cosh\big(R_+ (\theta - \tilde\theta_A)\big)}{\sqrt{R_A^2 - R_+^2}}
\qquad {\rm with} \quad -\infty < \theta < \infty
\label{btzpoint}
\end{equation}
But eq.~(\ref{btzpoint}) only contains information about geodesics of class (\ref{btzgeod1}). To learn about geodesics of type (\ref{btzgeod2}) that pass through a bulk point, we need to repeat the exercise for the function $\beta(\theta)$. However, we found that such functions satisfy $\beta'(\theta)^2>1$, so we must be careful in applying action (\ref{action}). Specifically, if the Lagrangian is to reproduce $(d\tau/d\theta) \sqrt{h}\, K$ of bulk curves then the other factor under the square root must change sign. How to interpret $d^2 S / d\alpha^2$ in the context of $\beta$-geodesics?

The right answer is the continuation (\ref{imshift}). Treating $S(\alpha)$ as an analytic function of a complex variable and using $\beta = \alpha + i \pi / 2 R_+$, action (\ref{action}) becomes:
\begin{equation}
I = \frac{R_+}{2G} \int \frac{d\theta \, \sqrt{\beta'(\theta)^2-1}}{\cosh\big(R_+ \beta(\theta)\big)}
\label{btzaction2}
\end{equation} 
Using eqs.~(\ref{btztanalpha2}-\ref{btztantheta2}) one can verify that the Lagrangian is the extrinsic curvature of the bulk curve. The equation of motion becomes
\begin{equation}
\beta_A''
+ \big(\beta_A'^2-1\big) R_+ \tanh(R_+ \beta_A) = 0
\label{btzeom2}
\end{equation}
and the solutions read:
\begin{equation}
\beta_A(\theta) = \frac{1}{R_+} \sinh^{-1} \frac{R_A \sinh \big(R_+ (\theta - \tilde\theta_A)\big)}{\sqrt{R_A^2 - R_+^2}}
\label{btzpoint2}
\end{equation}

In order to identify solutions (\ref{btzpoint}) and (\ref{btzpoint2}) as describing the same bulk point, we need two conditions, one for each integration variable of eqs.~(\ref{btzeom}) and (\ref{btzeom2}). The first of these fixes the relative shift in $\theta$ between the $\alpha$- and $\beta$-branches. The second condition is that $\alpha_A(\theta)$ and $\beta_A(\theta)$ asymptote to the same straight line, which is:
\begin{equation}
\alpha_A(\theta),\, \beta_A(\theta)\quad \stackrel{\theta \to \infty}{\longrightarrow} \quad \theta - \tilde\theta_A + \frac{1}{R_+} \log \frac{R_A}{\sqrt{R_A^2 - R_+^2}}
\label{samelimit}
\end{equation}
In the bulk, this means that the points described by $\alpha_A(\theta)$ and $\beta_A(\theta)$ lie on the same critical geodesic (\ref{btzgeodinf}). In Appendix \ref{btzpointlength} we show how our construction recovers the fact that the circumference of a point ``curve'' vanishes. 

\begin{figure}[t!]
\centering
\begin{tabular}{ccc}
\includegraphics[width=.31\textwidth]{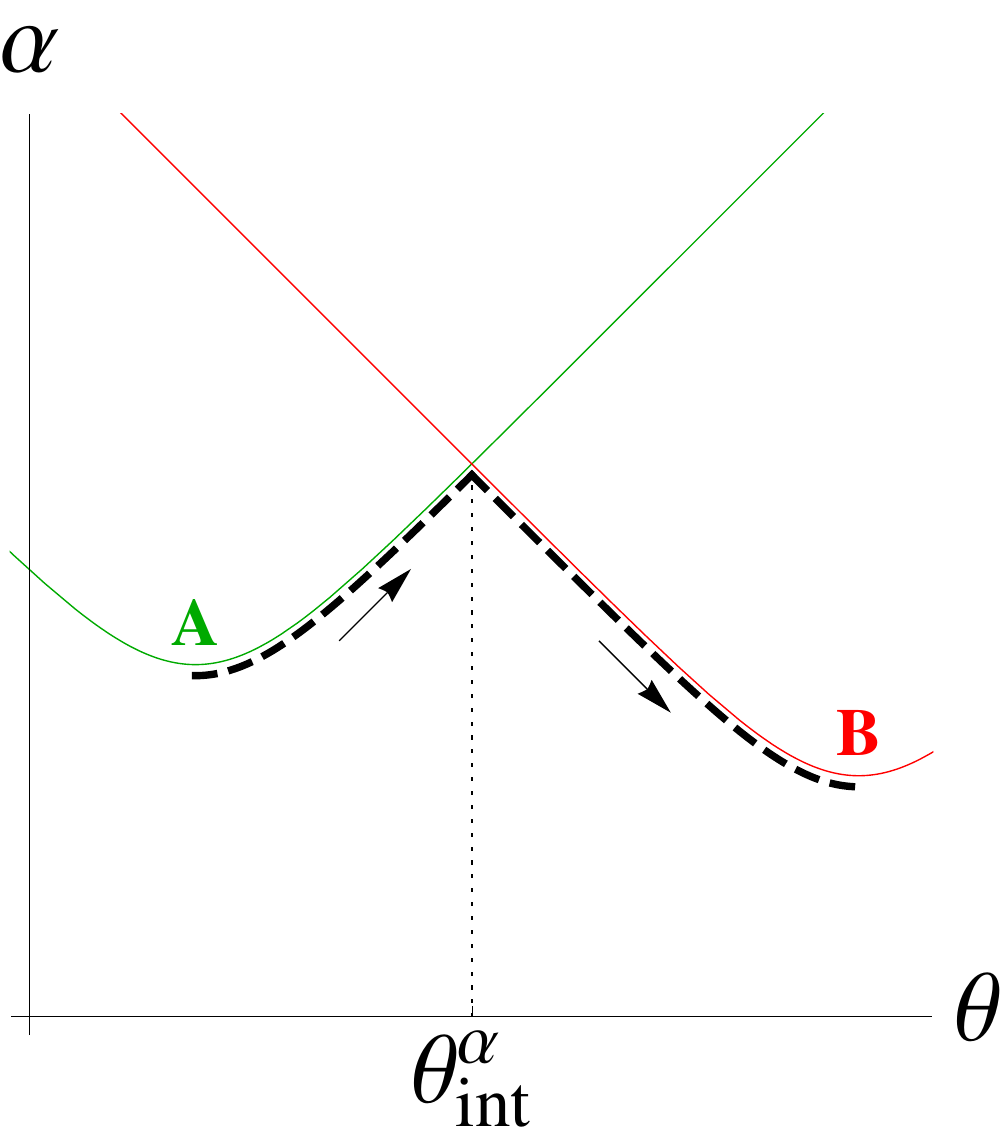} & 
\includegraphics[width=.31\textwidth]{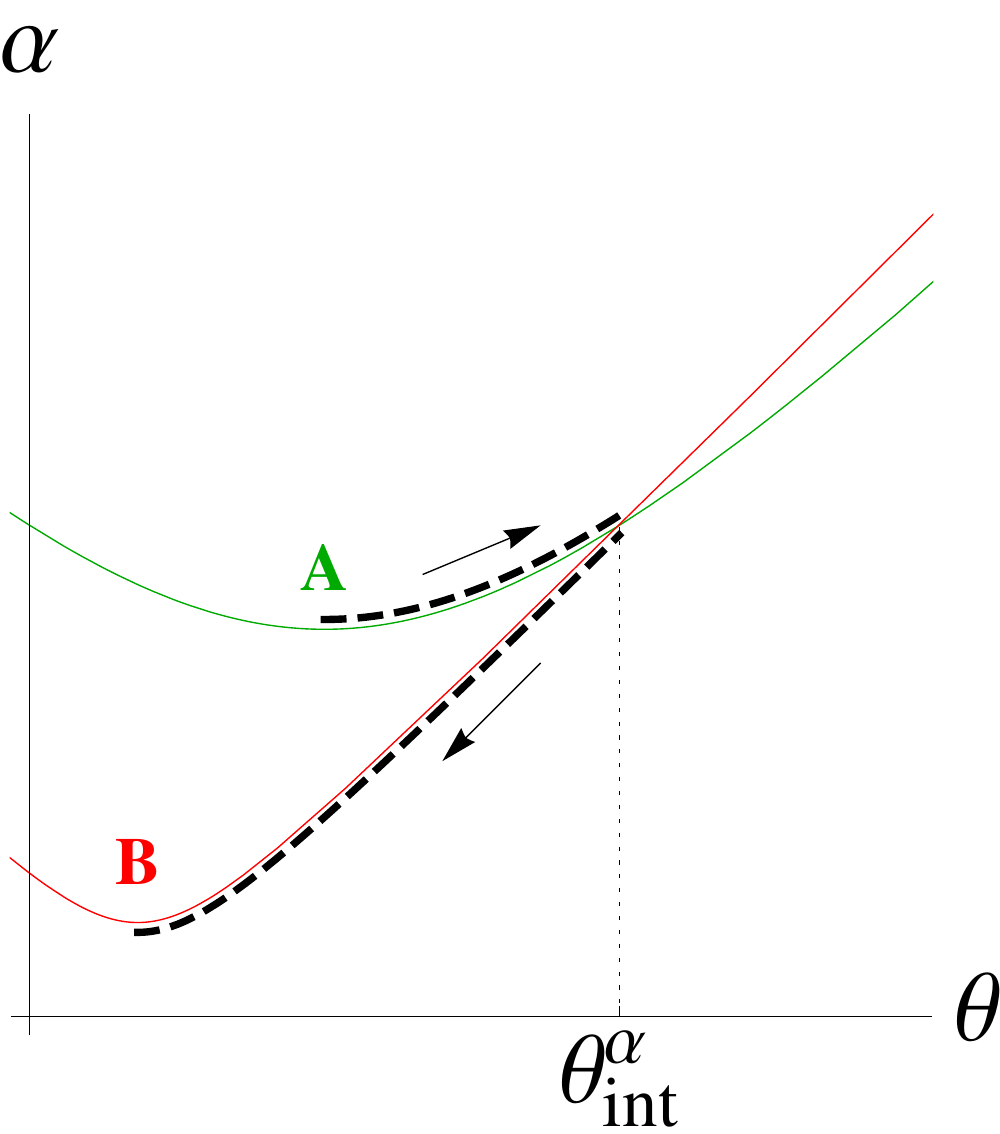} & 
\includegraphics[width=.31\textwidth]{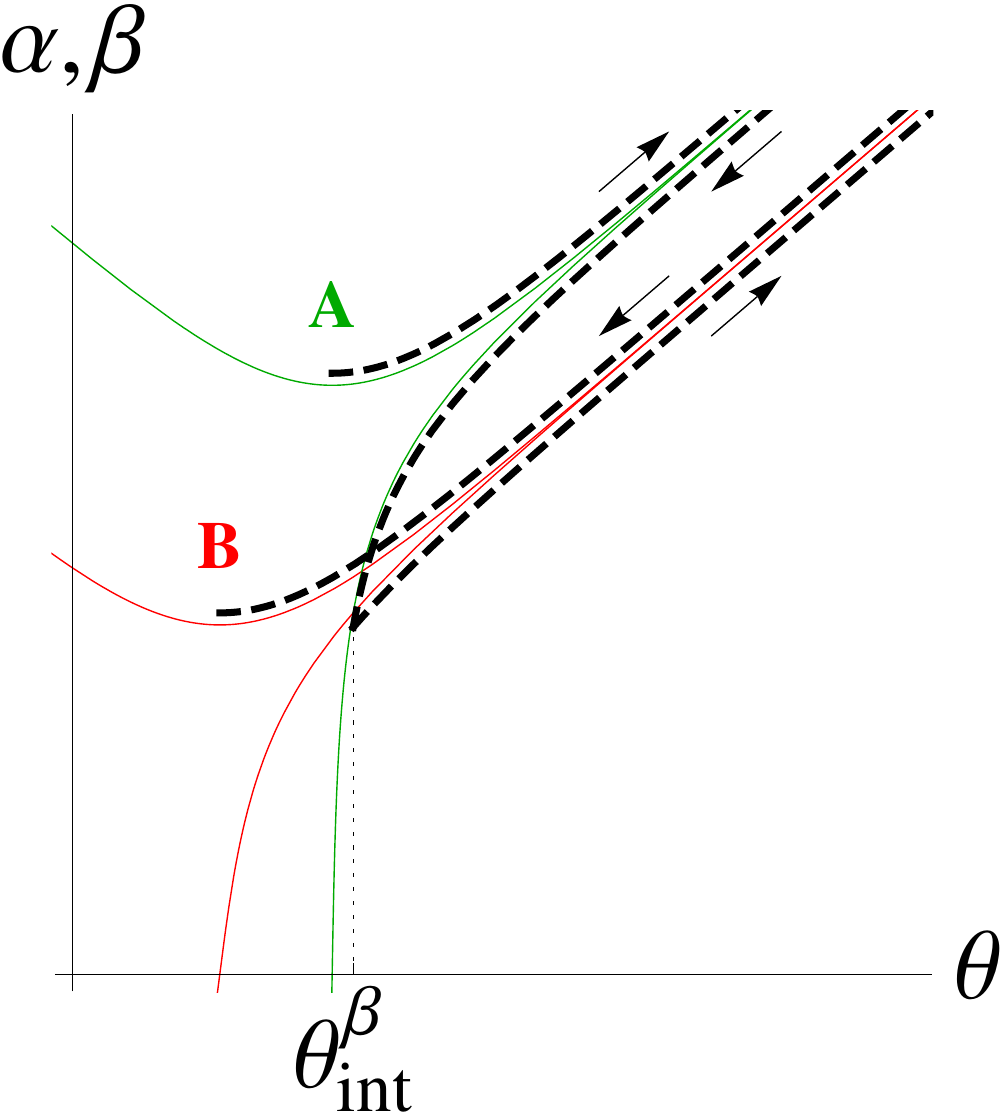}
\end{tabular}
\caption{The contours of integration in eqs.~(\ref{btzdist1}) and (\ref{btzdist2}), which calculate distances in the BTZ geometry. In the first two panels points $A$ and $B$ are connected by a geodesic of type~(\ref{btzgeod1}) centered at $\theta^\alpha_{\rm int}$; this distance is given by eq.~(\ref{btzgeod1}). In the left panel $\tilde\theta_A < \theta^\alpha_{\rm int} < \tilde\theta_B$ while in the center panel $\tilde\theta_B < \tilde\theta_A < \theta^\alpha_{\rm int}$. In the right panel, $A$ and $B$ are connected by a geodesic of type (\ref{btzgeod2}) centered at $\theta^\beta_{\rm int}$; this distance is given by eq.~(\ref{btzgeod2}).}
\label{BTZdistances}
\end{figure}

\paragraph{Distances} Our prescription for the geodesic distance given in Sec.~\ref{secdistance} employs the holographic construction of the closed convex cover of two points. In the BTZ geometry, a closed curve that does not encircle the horizon involves four transitions between the $\alpha$ and $\beta$ branches. At each transition, the differential entropy formula incurs mutually canceling infinities like those discussed in Appendix \ref{btzpointlength}. Because of this awkwardness of discussing contractible closed curves, we use the same trick as in eq.~(\ref{cddisttrick}) in the conical defect geometry. Rather than constructing a closed convex cover of points $A$ and $B$, we construct an open curve from $A$ to $B$. We then set the integration limits to be the minima of $\alpha_{A,B}(\theta)$ so that the boundary terms $f$ vanish. 
Denoting with $\theta^\alpha_{\rm int}$ the location where the graphs of $\alpha_{A,B}(\theta)$ intersect, we obtain:
\begin{equation}
d(A,B) = 
\frac{1}{2} \left| 
\int_{\tilde\theta_A}^{\theta^\alpha_{\rm int}} d\theta\, \frac{dS(\alpha)}{d\alpha}\Big|_{\alpha = \alpha_A(\theta)} 
+ 
\int_{\theta^\alpha_{\rm int}}^{\tilde\theta_B} d\theta\, \frac{dS(\alpha)}{d\alpha}\Big|_{\alpha = \alpha_B(\theta)} 
\right|
\label{btzdist1}
\end{equation}

Eq.~(\ref{btzdist1}) assumes that the graphs of $\alpha_A(\theta)$ and $\alpha_B(\theta)$ intersect. What happens if they do not? Because the intersection fixes the geodesic connecting the points, the absence of an intersection in the $\alpha$-branch means that the points are connected by a $\beta$-type geodesic instead. In this case, we must extend the region of integration to the $\beta$ regime. Denoting with $\theta^\beta_{\rm int}$ the intersection of the graphs of $\beta_{A,B}(\theta)$, we have $d(A,B)$ given by:
\begin{equation}
\frac{1}{2} \left| 
\int_{\tilde\theta_A}^\infty d\theta\, \frac{dS(\alpha)}{d\alpha}\Big|_{\alpha_A(\theta)} 
+
\int_\infty^{\theta^\beta_{\rm int}} d\theta\, \frac{dS(\beta)}{d\beta}\Big|_{\beta_A(\theta)} 
+ 
\int_{\theta^\beta_{\rm int}}^\infty d\theta\, \frac{dS(\beta)}{d\beta}\Big|_{\beta_B(\theta)} 
+
\int_\infty^{\tilde\theta_B} d\theta\, \frac{dS(\alpha)}{d\alpha}\Big|_{\alpha_B(\theta)} 
\right|
\label{btzdist2}
\end{equation}
Computations~(\ref{btzdist1}) and (\ref{btzdist2}) are illustrated in Fig.~(\ref{BTZdistances}). As in the conical defect geometry, in order to recover the full array of geodesic distances -- including ones along paths that wrap around the black hole -- we shift $\tilde\theta_B$ by a multiple of $2\pi$.


\section{Discussion}
\label{discussion}

In holographic duality, the conformal field theory is frequently taken as a definition of quantum gravity. In this view, the gravitational spacetime should emerge as an effective description of a set of quantities in the field theory. In a truly emergent spacetime every geometric construct -- including points and distances -- should arise from some field theory predecessors. Ideally, the field theory definitions of geometric objects should reflect some physical principle, which is important on the field theory side.

In the present work we took up the static slice of AdS$_3$ as a prime example of an emergent space. We have given {\it ab initio} field theory definitions of points (Sec.~\ref{pointsextrK}) and distances (Sec.~\ref{secdistance}) in AdS$_3$. Our construction uses vacuum entanglement entropies of intervals in CFT$_2$ as a basic input. A key physical principle underlying the success of our formalism is the strong subadditivity of entropy, which features prominently in proofs of many of our results, most conspicuously in establishing the triangle inequality. Our construction of points and distances does not assume the Ryu-Takayanagi proposal from the start, but recovers it in the limit where two bulk points are sent to the conformal boundary (Sec.~\ref{diamonds}). 

We know that other geometries cannot be reconstructed from the set of entanglement entropies alone, because Ryu-Takayanagi surfaces may not cover the whole space \cite{strongsub, fernando, plateau}. We have discussed two such examples in detail: the conical defect geometry (Sec.~\ref{condef}) and the nonrotating BTZ geometry (Sec.~\ref{btzthermal}). 
Since these spaces are quotients of AdS$_3$, it is possible to adapt our AdS$_3$ construction (Sec.~\ref{pointsent}) and ask what quantities supplant the information not contained in entanglement entropies.


\paragraph{Field theory input}
Broadly speaking, the Ryu-Takayanagi proposal suggests that the set of entanglement entropies is sufficient to recover the geometry on a coarse level, usually on the scale of $L_{\rm AdS}$. This is the intuition spelled out in the argument, which likens the AdS geometry to a tensor network \cite{briansessay, Swingle:2012wq, hartmanmaldacena, xiaoliang, complexity}. On the other hand, recovering bulk locality on sub-AdS scales is usually thought to depend on the matrix degrees of freedom \cite{lenny99, Bousso:2005ie, Berenstein:2008eg, Asplund:2008xd, idseetal}. Part of the motivation for the material in Secs.~\ref{condef} and \ref{btzthermal} is to emphasize that, at least in three bulk dimensions, both approaches to bulk reconstruction -- the macro-approach from entanglement and the micro-approach from matrix models -- should be unified. One heuristic way to envision such a unification is to imagine an RG flow taken to the extreme IR limit, where all the information about the spatial organization of the field theory has been coarse-grained away and all that remains is the matrix model as the s-wave sector of the theory \cite{Bousso:2005ie}. Secs.~\ref{condef} and \ref{btzthermal} contain clues for how to upgrade this heuristic scenario to a more realistic picture.

Sec.~\ref{condef} recovers the conical defect geometry from entwinement, a concept introduced in \cite{entwinement}. The field theory state dual to the conical defect geometry has a reduced gap and, in consequence, contains dynamical length scales that are larger than the size of the system \cite{longstring}. Entwinement $S(\alpha)$ is sensitive to such enlarged dynamical scales, but on scales smaller than system size it reduces to ordinary entanglement entropy. In the bulk, entwinement corresponds to geodesics that wind around the conical singularity. 

In Sec.~\ref{btzthermal} we considered the BTZ geometry. In the thermal state the energy gap is exponentially small and, semiclassically, we expect an infinite range of entwinement-like quantities $S(\alpha)$. In this work we assumed that such quantities can be extracted from the field theory -- either as entwinement per se or as subleading saddle points in semiclassical computations of 2-point functions. But, as we discuss in Sec.~\ref{btzthermal}, even an infinite real range of $S(\alpha)$ does not contain enough data to recover the full BTZ geometry. The missing information is the analytic continuation of $S(\alpha)$ to the complex domain: $S(\beta)$ in eq.~(\ref{btzlength2}). In terms of boundary 2-point functions, this continuation moves one of the insertion points from the original field theory to the thermofield double. In the bulk, this corresponds to spacelike geodesics in the two-sided BTZ black hole, which cross the horizon and connect distinct asymptotic boundaries \cite{excursions}. 

It is usually believed that the geometry in a single asymptotic region can be recovered entirely from data in a single boundary CFT, without recourse to two-sided quantities in the thermofield double state \cite{rhodual0, rhodual, chi, rqg, chi2, chiint, Kelly:2014owa, Headrick:2014cta}. From this point of view, the role played by $S(\beta)$ in our reconstruction of the BTZ geometry is puzzling. Perhaps $S(\beta)$ can secretly be recast in terms of single CFT quantities; another possibility is that its appearance is an artifact of working in the thermal state, which admits a purification through analytic continuation. To understand the role of two-sided quantities in bulk reconstruction, it would be useful to extend the results of the present paper to a broader class of field theory states and dual geometries. 

\paragraph{Generalizations}
We have defined points and distances on equal time slices of static geometries. Given the recent results on the hole-ography \cite{robproof} of time-dependent curves, it may be possible to generalize our construction away from static slices. Formulating such a generalization should be guided by the covariant version of the Ryu-Takayanagi proposal \cite{hrt}, which also satisfies the strong subadditivity of entropy \cite{maximin}. Perhaps this exercise can help us to recover the time component of the metric. If timelike distances can be constructed in a manner similar to the present paper, it would be interesting to compare their construction with the derivation of Einstein's equations from the ``first law of entanglement entropy'' \cite{einsteineq1, einsteineq2, einsteineq3}.
 
We have worked in the vacuum of a CFT$_2$ and in other highly symmetric states. A natural question to ask is whether our definition of points can be extended to less symmetric situations, for example to non-rotationally invariant states. If bulk points can then still be defined with a variational principle, action~(\ref{action}) will need generalizing. 


Another important extension would be to understand how bulk matter degrees of freedom can be described in a hole-ographic language. It is interesting to explore how the relation between local bulk fields and CFT operators \cite{hkll, hkll2, kll} translates into the auxiliary space (\ref{auxmetric}); see \cite{desitterhol} for early work in this direction. Further, combining the hole-ographic reconstruction of bulk geometry with a perturbative holographic representation of bulk fields may be useful for studying the dynamics of backreaction of matter on geometry.


Finally, it would be interesting to lift our construction to higher dimensions. The hole-ographic formula for areas of codimension-2 surfaces has been generalized to more than three bulk dimensions \cite{roblast, xijamieandi, robproof}, but the result is not well suited for constructing points. It appears that the best avenue toward higher-dimensional generalizations is the formal machinery of integral geometry.

\paragraph{Integral geometry} In Sec.~\ref{pointsent} we noted that our definition of points is equivalent to selecting spacelike geodesics in an auxiliary metric~(\ref{auxmetric}), which is 1+1-dimensional de Sitter space. Points in dS$_2$ represent geodesics on the hyperbolic plane. In Sec.~\ref{kinkina} we saw that cusps in dS$_2$ correspond to finite stretches of geodesics in $\mathbb{H}_2$; eqs.~(\ref{corner1}-\ref{corner2}) express the opposite relation. The duality between $\mathbb{H}_2$ and dS$_2$, which is reminiscent of twistor space \cite{twistors} (see \cite{twistorsrev} for an accessible review), is the subject of integral geometry \cite{santalo}. A wealth of results exist on the integral geometry of higher-dimensional symmetric spaces. It would be interesting to identify their applications to physics.

\paragraph{Other directions}
When the differential entropy formula involves only entanglement entropies, it can be interpreted in information theory \cite{patricknima}. It calculates the cost (measured in EPR pairs) of a restricted swapping protocol -- a task wherein Alice and Bob control complementary parts of a system and wish to exchange them. The shape of the curve is selected by a set of restrictions, which are imposed on Alice and Bob's operations. Our points in AdS$_3$ have vanishing differential entropy, so they correspond to restrictions, which allow for free exchange of states. An intriguing problem is to understand what this means from the viewpoint of the MERA network or the holographic renormalization group flow. 

\section*{Acknowledgments } We thank Michael Freedman, Matt Headrick, and Xi Dong for suggestions and help. We also thank Dio Anninos, Vijay Balasubramanian, Jan de Boer, Matthew Dodelson, Veronika Hubeny, Sam McCandlish, Rob Myers, Edgar Shaghoulian, Steve Shenker, James Sully and Leonard Susskind for useful conversations. LL would especially like to thank his adviser, Leonard Susskind, for his guidance, invaluable feedback and encouragement to explore the problems of his interest.
This work was supported in part by National Science Foundation Grant No. PHYS-1066293 and the hospitality of the Aspen Center for Physics during the ``Emergent Spacetime in String Theory'' workshop.

\appendix

\section{Extrinsic curvature of a curve in the hyperbolic plane}
\label{extrinsic}
We consider an arbitrary differentiable curve $R = r(\tilde\theta)$ in coordinates 
\begin{equation}
ds^2= \frac{dR^2}{1+R^2} + R^2 d\tilde\theta^2,
\label{h2metric}
\end{equation}
which cover the hyperbolic plane, that is a constant time slice of AdS$_3$ in coordinates (\ref{ads3metric}). 
Denoting $dr/d\tilde\theta = r'$, we can write the tangent vector and the induced metric as:
\myeq{\label{tangentvec} t^{\mu}=(r',\, 1) \qquad\Rightarrow\qquad 
h= g_{\mu\nu}t^{\mu}t^{\nu}= \frac{r'^2+R^2(1+R^2)}{1+R^2}}
The length element along the curve is then $d\tilde\theta\,\sqrt{h}$. The normal vector, which is
\begin{equation}
\label{normalvec} 
n_{\mu} = \frac{1}{\sqrt{1+R^2 +\frac{r'^2}{R^2}}}\, (1, -r')\,,
\end{equation}
can be used to rewrite metric (\ref{h2metric}) in the form $g_{\mu\nu}=h_{\mu\nu} +n_{\mu}n_{\nu}$, where:
\begin{equation}
h_{\mu\nu} = \frac{1}{1+R^2 +\frac{r'^2}{R^2}}\left(
\begin{array}{cc}
 \frac{r'^2}{R^2 (1+R^2)} & r' \\
 r' & R^2 (1+R^2)
\end{array}\right)
\end{equation}
The extrinsic curvature is:
\myeq{\label{Kdefinition} K_{\mu\nu}=\frac{1}{2}\mathcal{L}_{n}h_{\mu\nu}= \frac{1}{2}\left( n^{\kappa}\partial_{\kappa} h_{\mu\nu} +h_{\kappa\nu}\partial_{\mu}n^{\kappa} +h_{\kappa\mu}\partial_{\nu}n^{\kappa}\right), }
where the indices range over $R$ and $\tilde\theta$. Here $h_{\mu\nu}$ and $n^{\mu}$ are explicit functions of $R$ and implicit functions of $\tilde\theta$ (through their dependence on $r'(\tilde\theta)$). The scalar extrinsic curvature is calculated by contracting $K_{\mu\nu}$ with the full metric $g_{\mu\nu}$ and substituting $r(\tilde\theta) \to R$. The result is: 
\begin{equation}
\label{extrinsicK} 
K =\frac{(1 + R^2)(R^2 + R^4 + 2 R'^2 - R R'') + R^2 R'^2}{ (R^2+R^4+R'^2)^{3/2}}
\end{equation}
Combining eqs.~(\ref{tangentvec}) and (\ref{extrinsicK}), we obtain an expression, which reduces to eq.~(\ref{lagrangian}) after using eqs.~(\ref{tanalpha}-\ref{tantheta}). 

\section{Proof for the vanishing length of the BTZ points}
\label{btzpointlength}
To test our definition of a point, we should verify that its circumference vanishes. In the BTZ context this is an awkward task. Each time a bulk curve becomes tangent to the critical geodesic (\ref{btzgeodinf}), we encounter infinite integrals as in eq.~(\ref{openbtzgen}). If a closed curve does not enclose the horizon, it contains at least four such points, which makes the computation cumbersome. It is easier to compute one quarter of the circumference of a point ``curve'' -- from where its ``tangent'' is exactly angular all the way until it becomes exactly radial. This means integrating over $\theta$ from $\tilde\theta_A$ to infinity on the $\alpha$ branch and then from infinity back to $\tilde\theta_A$ on the $\beta$-branch:
\begin{equation}
\frac{1}{2} \int_{\tilde\theta_A}^\infty d\theta\, \frac{dS(\alpha)}{d\alpha}\Big|_{\alpha = \alpha_A(\theta)} +
\frac{1}{2} \int_{\infty}^{\tilde\theta_A} d\theta\, \frac{dS(\beta)}{d\beta}\Big|_{\beta = \beta_A(\theta)}
+ f_\beta(\tilde\theta_A)
= \frac{1}{4G} \cosh^{-1}(R_A/R_+) + f_\beta(\tilde\theta_A)
\label{btzcircum}
\end{equation}
There is no $f$-term from the first integral, because $\alpha_A'(\tilde\theta_A) = 0$. But on the $\beta$-branch we have to be more careful, because $\beta_A(\tilde\theta_A) = 0$. We can evaluate $f_\beta(\tilde\theta_A)$ as a limit, in which $\beta$ is sent to zero, but the radial position of the bulk point is kept constant at $R_A$. This means that $\theta$ must approach $\tilde\theta_A$ in the same limit according to eq.~(\ref{btzgeod2}):
\begin{equation}
\theta - \tilde\theta_A = \frac{1}{R_+} \sinh^{-1} \frac{\sinh\beta\, \sqrt{R_A^2 - R_+^2}}{R_A} \equiv \Delta(\beta)
\end{equation}
Substituting this into the definition of $f$ in eq.~(\ref{deffbtz}), we obtain
\begin{equation}
f_\beta(\tilde\theta_A) = 
\lim_{\beta \to 0}\,\frac{1}{8G}
\log\frac{\sinh R_+ \big( \beta - \Delta(\beta) \big)}{\sinh R_+ \big( \beta + \Delta(\beta) \big)}
= -\frac{1}{4G} \cosh^{-1} (R_A/R_+)\,,
\label{limitf}
\end{equation}
which is minus the geodesic distance from $A$ to the horizon. This confirms the fact that points have zero circumference, though the computation relies on a prior knowledge of the bulk. Luckily, knowing $f$ is not necessary to define a distance function.

\end{document}